\documentclass[useAMS,usenatbib,letterpaper]{mn2e}

\usepackage{url}
\usepackage{graphicx}
\usepackage{longtable}
\usepackage{lscape}
\usepackage{comment}
\usepackage{amssymb,amsmath}
\newcommand{\mm}{millimeter}
\newcommand{\revone}{ }

\def\aap{A\hbox{\rm \&}A} 
\def\aaps{A\hbox{\rm \&}A Suppl.}  \def\aj{AJ} 
  \def\apj{ApJ} \def\apjl{ApJ}
 \def\apjs{ApJS} 
\def\araa{ARA\hbox{\rm \&}A}

\def\mnras{MNRAS} \def\nat{Nat} \def\pasj{PASJ}

\title[AzTEC-SHADES: IDs, redshifts and large-scale structure]%
{\vspace{-0.5cm}AzTEC half square degree survey of the SHADES fields - II. Identifications, redshifts, and evidence for large-scale structure
\vspace{-0.5em}}
\author[M.~J.~Micha{\l}owski et al.]%
{Micha{\l}~J.~Micha{\l}owski$^{1}$\thanks{E-mail: mm@roe.ac.uk},
J.~S.~Dunlop$^{1}$,
R.~J.~Ivison$^{2,1}$,
M.~Cirasuolo$^{1,2}$,
\newauthor
K.~I.~Caputi$^{3,1}$,
I.~Aretxaga$^{4}$,
V.~Arumugam$^{1}$,
J.~E.~Austermann$^{5}$,
E.~L.~Chapin$^{6,7}$,
\newauthor 
S.~C.~Chapman$^{8}$, 
K.~E.~K.~Coppin$^{9}$,
E.~Egami$^{10}$,
D.~H.~Hughes$^{4}$, 
E.~Ibar$^{2}$,
\newauthor 
A.~M.~J.~Mortier$^{11,1}$,
A.~M.~Schael$^{12,1}$,  
K.~S.~Scott$^{13}$,
I.~Smail$^{14}$,
T.~A.~Targett$^{1}$, 
\newauthor 
J.~Wagg$^{15}$,
G.~W.~Wilson$^{16}$,
L.~Xu$^{10}$,
M.~Yun$^{16}$
\\
$^{1}$SUPA\thanks{Scottish Universities Physics Alliance}, Institute for Astronomy, University of Edinburgh, Royal Observatory, Edinburgh, EH9 3HJ\\
$^2$UK Astronomy Technology Centre, Royal Observatory, Edinburgh EH9 3HJ\\
$^3$Kapteyn Astronomical Institute, University of Groningen, P.O. Box 800, 9700 AV Groningen, The Netherlands\\
$^4$Instituto Nacional de Astrof\'{\i}sica, \'Optica y Electr\'onica (INAOE), Aptdo. Postal 51 y 216, 72000 Puebla, Pue., Mexico  \\
$^5$Center for Astrophysics and Space Astronomy, University of Colorado, Boulder, CO 80309, USA\\
$^{6}$Department of Physics \& Astronomy, University of British Columbia, 6224 Agricultural Road, Vancouver, BC V6T 1Z1, Canada\\
$^7$XMM SOC, ESAC, Apartado 78, 28691 Villanueva de la Ca\~nada, Madrid, Spain\\
$^8$Institute of Astronomy, University of Cambridge, Madingley Road, Cambridge CB3 0HA\\
$^9$Department of Physics, McGill University, 3600 Rue University, Montreal, QC H3A 2T8, Canada\\
$^{10}$Steward Observatory, University of Arizona, 933 N. Cherry Ave, Tucson, AZ 85721, USA \\
$^{11}$Astrophysics Group, Imperial College London, Blackett Laboratory, Prince Consort Road, London SW7 2AZ \\
$^{12}$Max Planck Institute for Extraterrestrial Physics, 85748 Garching bei M{\"u}nchen, Germany \\
$^{13}$North American ALMA Science Center, National Radio Astronomy Observatory, Charlottesville, VA 22903\\
$^{14}$Institute for Computational Cosmology, Durham University, South Road, Durham DH1 3LE\\
$^{15}$European Southern Observatory, Casilla 19001, Santiago, Chile\\
$^{16}$University of Massachusetts, Department of Astronomy, Amherst, MA01003, USA \\
\mbox{}
\vspace{-1.2cm}
}
\begin{document}

\date{Accepted 2012 July 30. Received 2012 July 27; in original form 2012 May 09}

\pagerange{\pageref{firstpage}--\pageref{lastpage}} \pubyear{2012}

\maketitle
\mbox{}\vspace{-1em}

\label{firstpage}

\begin{abstract}
The AzTEC $1.1$\,mm survey of 
the two SCUBA HAlf Degree Extragalactic Survey (SHADES) fields is the 
largest ($0.7$\,deg$^2$) blank-field millimetre-wavelength 
survey undertaken to date at a resolution of $\simeq18$\,arcsec and a depth of 
$\simeq 1$\,mJy. 
We have used the deep optical-to-radio multi-wavelength data in 
the SHADES Lockman Hole East and SXDF/UDS fields to obtain 
galaxy identifications for  $\revone\simeq64$\% ($\simeq80$\% {\revone including tentative identifications)} of the $148$ AzTEC-SHADES $1.1$\,mm 
sources reported by Austermann et al. (2010), exploiting 
deep radio and $24\,\mu$m data complemented 
by methods based on $8\,\mu$m flux-density and red optical-infrared 
($i-K$) colour.
This unusually high identification rate can be attributed to the 
relatively bright millimetre-wavelength flux-density
threshold, combined with the relatively deep supporting multi-frequency data
now available in these two well-studied fields.
We have further exploited the optical--mid-infrared--radio data to 
derive a $\revone\simeq60$\% ($\simeq75$\% {\revone including tentative identifications)} complete 
redshift distribution for the AzTEC-SHADES sources, 
yielding a median redshift of $z\simeq2.2$, with 
a high-redshift tail extending to at least $z \simeq 4$.
Despite the larger area probed by the AzTEC survey relative to the original SCUBA SHADES imaging, the redshift distribution 
of the AzTEC sources is consistent with that displayed by the SCUBA sources, and reinforces tentative evidence that 
the redshift distribution of mm/sub-mm sources in the Lockman Hole field is significantly different from that found in the 
SXDF/UDS field. Comparison with simulated surveys of similar scale extracted from
semi-analytic models based on the Millennium simulation
indicates that this is as expected if the mm/sub-mm sources are massive ($M > 10^{11}\,{\rm M_{\odot}}$)
star-forming galaxies tracing large-scale structures over scales of $10$--$20$\,Mpc. This 
confirms the importance of surveys covering several square degrees 
(as now underway with SCUBA2) to obtain representative samples of bright (sub)mm-selected galaxies.
This work provides a foundation for the further exploitation of the {\it Spitzer} and {\it Herschel} data in the SHADES fields in the study  of the stellar masses and specific star-formation rates of the most active star-forming galaxies in cosmic history.
\end{abstract}


\vspace{-3cm}

\newpage

\section{Introduction}

The objects uncovered by high galactic latitude surveys at (sub)millimetre wavelengths (generally termed `submillimetre galaxies'; SMGs),
are now understood to be distant ($z\simeq2$--$3$) and 
massive ($\simeq10^{11}\,$M$_\odot$) star-forming galaxies \citep{aretxaga03,aretxaga07,dannerbauer04,smail04, swinbank04, swinbank06,takagi04,borys05,chapman05,greve05,tacconi06,tacconi08,eales09,amblard10,dunlop10,engel10,michalowski10smg, michalowski10smg4,michalowski12mass,santini10,hainline11,hayward11b,targett11,wardlow11,bethermin12,bussmann12,shimizu12,yun12}. However, the role of these spectacular objects in the cosmic history 
of galaxy/star formation has yet to be properly established.

The majority of the submillimetre surveys undertaken with the Submillimetre Common-User Bolometer Array \citep[SCUBA;][]{hollandscuba}  
on the James Clerk Maxwell Telescope (JCMT) covered small ($<0.03$ deg$^2$) fields due to sensitivity and time constraints
\citep{smail97,hughes98,barger98,barger99,barger00,barger01,barger02,chapman01,chapman03c,scott,webb03,wang04,coppin05,pope06}.
Only two submillimetre surveys completed to date cover a significant fraction of a degree: the SCUBA HAlf Degree Extragalactic Survey \citep[SHADES; $0.2$ deg$^2$;][]{mortier05,coppin06} 
undertaken at the JCMT, and the $870\,\mu$m survey of the ECDFS field \citep[$0.25$ deg$^2$;][]{weiss09b} 
performed with the Large APEX BOlometer CAmera \citep[LABOCA;][]{laboca} on the Atacama Pathfinder Experiment (APEX). 

At slightly longer wavelengths ($1.1$--$1.2$\,mm) surveys with the Max-Planck Millimeter Bolometer array (MAMBO) on the IRAM 30-m telescope, and with 
the AzTEC \citep{aztec} on the JCMT and Atacama Submillimetre Telescope Experiment \citep[ASTE;][]{aste,aste2} have 
generally covered areas of $<0.25$ deg$^2$ \citep{greve04,greve08,bertoldi07, perera08,scott08,scott10,austermann09,hatsukade11,lindner11}. The 
notable exceptions are the MAMBO survey of the (possibly lensed) 
field of the cluster Abell 2125 totalling $0.44$ deg$^2$ \citep{wagg09}, and the AzTEC/ASTE survey of the COSMOS field 
with a $\simeq34$ arcsec beam covering $0.72$ deg$^2$ \citep{aretxaga11}. 

{\revone There are number of reasons why it is necessary to analyse SMG properties in fields with significant sizes (of the order of a deg$^2$ or more). Smaller fields miss the rare members of the SMG population, i.e.~both low- and high-redshift parts of the distribution as well as the bright-end of the luminosity function. Moreover, the clustering properties and large-scale structures traced by SMGs can only be studied in fields larger than these structures. Finally, significant field-to-field variation (cosmic variance) advocates the need of  observing multiple well-separated fields.}

Here we analyse the properties of the galaxies detected in the largest very deep (rms $0.9$--$1.7$\,mJy per beam) blank-field millimetre survey of the Lockman Hole East and the Subaru/{\it XMM-Newton} Deep Field (SXDF) 
(aka UKIDSS Ultra Deep Survey (UDS) field) conducted with AzTEC mounted on the JCMT (beam size $18$\,arcsec FWHM) which covers a total area of $0.7$\,deg$^2$ \citep{austermann10}.
At present, the only substantially larger millimetre-wavelength survey is the $87$\,deg$^2$ survey program undertaken with the South Pole Telescope \citep{viera10},
but this has a much larger beam size ($\simeq 60$\,arcsec) and much lower sensitivity ($3.4$\,mJy at $1.4$\,mm, which corresponds to $6.5$\,mJy at $1.1$\,mm 
assuming the average SMG spectral energy distribution (SED) from \citealt{michalowski10smg} at $z=2$--$3$).

The AzTEC survey considered here incorporates the two smaller fields (totalling $\simeq0.2$\,deg$^2$) covered by the 
original SCUBA-SHADES $850\,\mu$m imaging. The 850\,$\mu$m sources uncovered by SCUBA \citep{coppin06} have already 
been the subject of extensive multi-frequency analysis, including the identification of their galaxy counterparts 
via radio (VLA 1.4\,GHz) and mid-IR ({\it Spitzer} 24\,$\mu$m) imaging \citep{ivison07}, estimation of their
redshifts from the observed submillimetre-radio SEDs \citep{aretxaga07},
further redshift estimation and stellar mass determination based on the optical-infrared data \citep{takagi07,dye08,clements08},
follow-up imaging at $350\,\micron$ \citep{coppin08}, and studies of their environments and clustering \citep{serjeant08,vankampen05}.
Individual SHADES sources have also been the subject of detailed high-resolution follow-up with the SMA \citep{younger08,hatsukade10} and 
with {\it Spitzer} mid-infrared spectroscopy \citep{coppin10}. A final analysis of the redshift distribution and properties 
of the SCUBA-SHADES galaxies will be presented in Schael et al. (in preparation).

As is well-known, the large beams delivered by current single-dish (sub)millimetre facilities hampers 
the search for robust counterparts at other wavelengths. Usually the radio observations are used, 
both because the radio emission is believed to be related to the far-IR emission \citep[e.g.][]{condon}, and because 
the surface number density of bright radio sources is relatively low (and hence associations 
are often unique and statistically significant). Experience has shown that radio identifications can 
typically be secured for $\simeq30$--$70$\% of SMGs; the precise percentage depends 
on the relative depths of the (sub)millimetre and radio data 
\citep{ivison02,ivison07,dannerbauer04,dannerbauer10,pope06,chapin09,aretxaga11,chapin11,biggs11,wardlow11}, with 
completion rates of up to $80$\% being achieved for the relatively bright ($S_{1.1{\rm mm}}>4$\,mJy) 
and/or possibly lensed samples of \citet{ivison05} and \citet{wagg09}. 
Recently \citet{lindner11} obtained a 93\% identification (ID) rate for a sample of 41 SMGs using very deep radio data.
Attempts have been made to boost the identification 
fraction further using, for example, {\it Spitzer} MIPS $24\,\mu$m data, but these have generally 
resulted in only a modest increase in the number of secure galaxy counterparts \citep{ivison07}.
It has also been shown that for $30$--$60$\% of SMGs {\revone (depending on depth)} one can obtain 
$3$--$8\,\mu$m counterparts using the {\it Spitzer}/IRAC data \citep{ashby06,pope06,biggs11}.

The objective of this paper is to provide secure identifications and photometric redshifts for as many as possible 
of the $148$ AzTEC-SHADES $1.1$\,mm sources, exploiting not only the latest extremely-deep VLA/GMRT radio 
and {\it Spitzer} MIPS $24\,\mu$m data, but also the deep {\it Spitzer} IRAC $8\,\mu$m  maps, 
and the ever-improving optical--near-infrared imaging which has now 
been secured within both the UDS/SXDF and Lockman East fields. There are two reasons to expect that we should 
be able to secure a higher ID fraction for the AzTEC-SHADES sample than has been obtained for any previous
large ($>$100 source) and complete sample of (sub)millimetre sources. The first is that our $1.1$\,mm source sample is relatively 
bright \footnote{The rms of the AzTEC $1.1$\,mm data in the SHADES fields is  $0.9$--$1.7$\,mJy  \citep{austermann10}, 
i.e.~a factor of $\simeq2$ shallower than for the AzTEC survey of the GOODS-S field (rms $0.48$-$0.73$\,mJy \citep{scott10}) and for the LESS survey 
of the ECDFS (rms $1.2$\,mJy at $870\,\mu$m \citep{weiss09b}, corresponding to  $0.6$--$0.7$\,mJy at $1.1$\,mm 
assuming the average SMG spectral energy distribution from \citet{michalowski10smg} at $z=2$--$3$.}
and so we are not probing the faint part of the high-redshift 
(sub)millimetre luminosity function. The second is that, over the required
$\simeq 0.7$\,deg$^2$ area, the radio, mid-infrared and near-infrared/optical data are 
among some of the deepest currently available. These considerations have encouraged us to try to 
refine the use of mid-infrared data and optical-infrared colour information to maximise the completeness
of the galaxy identifications and hence the inferred redshift distribution of the AzTEC $1.1$\,mm sources. 

Here we present the final results of this ID analysis, and derive photometric redshifts for the
galaxies in the resulting near-complete SMG sample. We explore the implications of the inferred 
redshift distribution of this SMG sample, comparing it with redshift distributions previously 
reported from other, smaller, mm-wavelength surveys. We also explore the consistency of the redshift
distributions of the AzTEC sources found in the two separate SHADES survey fields, as well as 
comparing the AzTEC source redshift distribution with that already derived for the SCUBA 850\,$\mu$m
sources. Further analysis of the full SEDs of the AzTEC sources, 
including derived physical properties such as star formation rates and stellar masses, is deferred to a future paper. 

This paper is structured as follows.
In Section~\ref{sec:data} we summarize the available multi-wavelength data. 
Then, in Section~\ref{sec:idmeth}, we describe the methods used to identify potential
galaxy counterparts at other wavelengths, and to assess their statistical 
significance/robustness.
The resulting galaxy IDs, photometric redshifts, and optical-infrared SEDs  
are presented in Sections~\ref{sec:id}, \ref{sec:z} and \ref{sec:fluxr}, 
respectively, with notes on individual sources provided in Appendix A. In Section~\ref{sec:large} we explore the implications
of the derived redshift distributions, and assess the evidence for large-scale structure with the aid of simulated 
AzTEC surveys extracted from cosmological simulations.
Section~\ref{sec:conclusion} closes with our conclusions.
We use a cosmological model with $H_0=70$\,km\,s$^{-1}$\,Mpc$^{-1}$,  $\Omega_\Lambda=0.7$ and $\Omega_m=0.3$, and 
give all magnitudes in the AB system.

\section{Data}
\label{sec:data}

We utilised the JCMT/AzTEC $1.1$\,mm maps and catalogues from \citet{austermann10}\footnote{The fluxes have been recently revised by \citet{downes12}, but this change does not have any significant impact on our analysis.}. These data 
cover $0.7$\,deg$^2$ to an rms depth of $0.9$--$1.7$\,mJy per beam. 
We selected all {\revone 148} sources presented by \citet{austermann10} with signal-to-noise ratios (S/N)$\mbox{}>3.5$, and adopted the statistically 
deboosted $1.1$\,mm flux densities.

The VLA $1.4$\,GHz and GMRT $0.61$\,GHz radio data were taken from 
\citet{ivison05,ivison07} and \citet{ibar09,ibar10} respectively.
The $1\sigma$ rms depths at the centre of the radio images are $6$ and $9\,\mu$Jy beam$^{-1}$ for the $1.4$\,GHz maps in the Lockman Hole East 
and the UDS field respectively, 
and $15\,\mu$Jy beam$^{-1}$ for the $0.61$\,GHz map in the Lockman Hole. Both the VLA and GMRT radio imaging delivered a beam size of $5$\,arcsec (FWHM). 
The catalogues include sources for which $>3\sigma$ detections were obtained.

The mid-infrared {\it Spitzer} data in the Lockman Hole East are from programs 
PID 81 (PI: G.~Rieke) and PID 50249 (PI: E.~Egami), described in \citet{egami04} and \citet{dye08}, 
whereas in the UDS field the mid-infrared data are from the {\it Spitzer} Public Legacy Survey of the UKIDSS 
Ultra Deep Survey 
(SpUDS; PI: J.~Dunlop)\footnote{\url{http://ssc.spitzer.caltech.edu/spitzermission/} \url{observingprograms/legacy/spuds/}} described in \citet{caputi11}.

The optical data in both fields were obtained with Subaru/SuprimeCam \citep{suprimecam}, as described in 
\citet{dye06} and \citet{furusawa08}. The near-infrared data in both fields are provided by the UKIRT 
Infrared Deep Sky Survey  \citep[UKIDSS;][]{lawrence07} with the SXDF/UDS field  benefitting from the ultra-deep
$J,H,K$ coverage provided by the UDS survey \citep[e.g.][]{cirasuolo10}, while the Lockman Hole East field is part 
of the somewhat shallower UKIDSS DXS survey \citep{warren07}.

The depths of the data in both fields are summarized in Table~\ref{tab:data}.

\begin{table}
\caption{The $3\sigma$ depths of the multi-frequency data used in the Lockman Hole East and SXDF/UDS fields.}
\label{tab:data}
\begin{center}
\begin{tabular}{lccl}
\hline\hline
Filter			&LH 		& UDS	& Unit 		\\
\hline
$B$			& 27.4	& 28.8	& AB mag		\\
$R$			& 26.4	& 28.1	& AB mag		\\
$i$			& 26.3	& 27.8	& AB mag		\\
$z$			& 25.6	& 26.9	& AB mag		\\
$J$			& 22.9	& 25.5	& AB mag		\\
$H$			& $\cdots$& 24.8	& AB mag		\\
$K$			& 23.5	& 25.2	& AB mag		\\
$3.6\,\mu$m	& 0.8		& 1.4		& $\mu$Jy	\\
$4.5\,\mu$m	& 1.6		& 1.5		& $\mu$Jy	\\
$5.6\,\mu$m	& 11		& 19		& $\mu$Jy	\\
$8.0\,\mu$m	& 13		& 12		& $\mu$Jy	\\
$24\,\mu$m	& 25		& 30		& $\mu$Jy	\\
$1.1$ mm		& 0.9--1.3	& 1.0--1.7	& mJy		\\
$1.4$ GHz	& 18		& 27		& $\mu$Jy	\\
$0.6$ GHz	& 45		& $\cdots$& $\mu$Jy	\\
\hline 
\end{tabular}
\end{center}
\end{table}

\section[]{Identification Method}
\label{sec:idmeth}

\subsection{Radio and $24\,\mu$m IDs}
\label{sec:idrad24}

We obtained the radio and $24\,\mu$m counterparts applying the method outlined in 
\citet{downes86}, \citet{dunlop89} and \citet{ivison07}. The 2.5$\sigma$ search radius $r$ around each 
AzTEC position was determined on the basis of the deboosted signal-to-noise ratio (S/N): 
$r=2.5\times 0.6\times \mbox{FWHM}/(\mbox{S}/\mbox{N})$, where $\mbox{FWHM}=18$\,arcsec is the  size of the beam delivered by the 
JCMT at $1.1$\,mm. 
In order to account for systematic astrometry shifts \citep[due to either pointing inaccuracies or source blending; e.g.][]{dunlop10} 
we used a minimum search radius $r=8$ arcsec whenever the above formula produced $r < 8$\,arcsec (this proved necessary for only 20 AzTEC sources).

The statistical significance of each potential counterpart was assessed on the basis of the corrected Poisson probability $p$ that 
the chosen radio or $24\,\mu$m candidate could have been selected by chance.

\subsection{IRAC $8\,\mu$m IDs}
\label{sec:id80}

In order to maximize the fraction of AzTEC galaxies with identified counterparts we also explored other ways to select IDs. 

There is now a growing body of evidence indicating that SMGs 
are bright at the rest-frame near-IR wavelengths
\citep{ashby06,pope06,hainline09,biggs11,wardlow11}
and  hence are expected to be massive \citep{borys05,michalowski10smg, michalowski10smg4,michalowski12mass,hainline11,bussmann12,yun12}.
Similar to \citet{ashby06}, 
we thus explored the potential of using the available IRAC $8.0\,\mu$m imaging to search for IDs. We used the longest IRAC wavelength because 
the surface density of bright sources is lower than at shorter wavelengths.

We selected the $8\,\mu$m IDs in a similar way to the radio 
and $24\,\mu$m IDs; i.e. we searched for $8\,\mu$m objects within the same search radius, and estimated the probability of a 
chance association using analogous $p$-statistics. 
To do this we derived the cumulative number counts of $8.0\,\mu$m sources in our fields; we found that this had the 
form $N(>S_{8.0})\propto S_{8.0}^{-1.1}$ with a 
normalization given by $N(>20\,\mu\mbox{Jy})=4$ arcmin$^{-2}$. We did not select candidate identifications
at flux densities fainter than $20\,\mu$Jy.

\subsection{Red $i-K$ IDs}
\label{sec:idik}

It is now well known that SMGs generally 
exhibit red optical/near-IR colours \citep{smail99,smail02,smail04,ivison02,webb03,dannerbauer04,ashby06,yun12}. 
This is likely a consequence of two effects: SMGs are very dusty and they host a significant evolved stellar population 
\citep{michalowski10smg, michalowski12mass}.
Similarly, $\simeq60$\% of ULIRGs at $z\simeq2$--$3$ are very red with $i-K>2.5$ \citep{caputi06b}.

Therefore, we explored the potential of red $i-K$ colour to yield SMG galaxy counterparts, 
building on the work of Schael et al. (in preparation)\footnote{See also \citet{schael09phd}.}. 
Schael et al. showed, for SCUBA sources in the SHADES fields, that the reddest ($i-K>2$) source within $10$ arcsec of the SCUBA position usually corresponds 
to the radio ID. We therefore selected all sources with $i-K>2$ within the search radius defined above, 
and again estimated their reliability as AzTEC IDs by estimating the {\it a priori} probability $p$. 
The number density of \mbox{$K$-band} detected sources in the magnitude range of interest is well 
described by Euclidean counts, with $N(>S_K)\propto S_K^{-1.5}$. 
We estimated the number density at $K=23$ mag separately for sources with {\revone $i-K$ colours $2$--$2.5$, $2.5$--$3$ and $>3$} to be $N(K<23 \mbox{ mag})=8$ arcmin$^{-2}$,
$4$ arcmin$^{-2}$ and $2$ arcmin$^{-2}$, respectively. These number densities were used to calculate the $p$-values for sources with 
$i-K$ colours of $2$--$2.5$, $2.5$--$3$ and $>3$, respectively. We did not select sources fainter than $K = 23$ mag.

\section[]{Identifications}
\label{sec:id}
\label{sec:lens}

\begin{figure*}
\begin{center}
\includegraphics[width=0.59\textwidth]{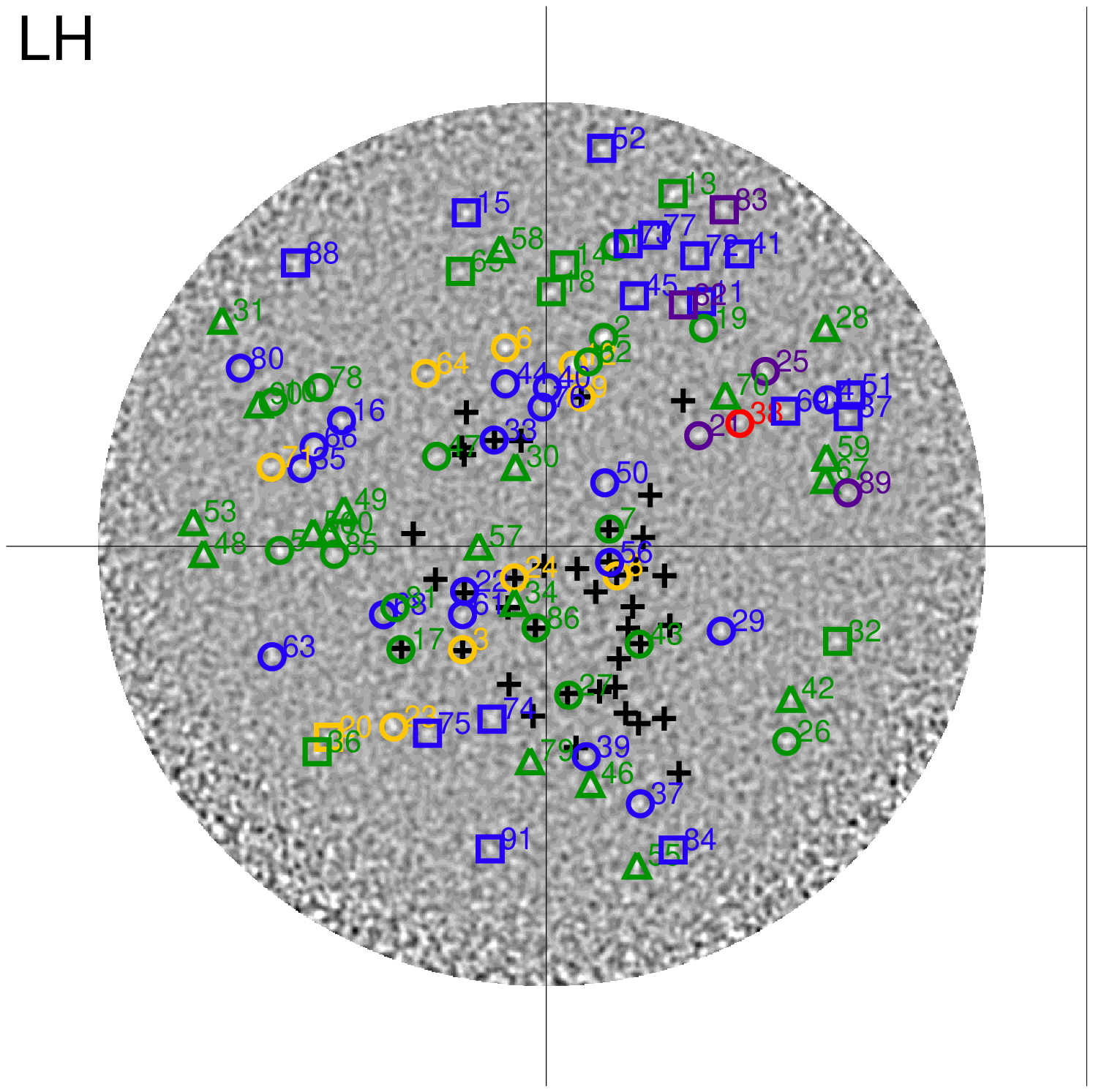}

\vspace{0.1em}

\includegraphics[width=0.59\textwidth]{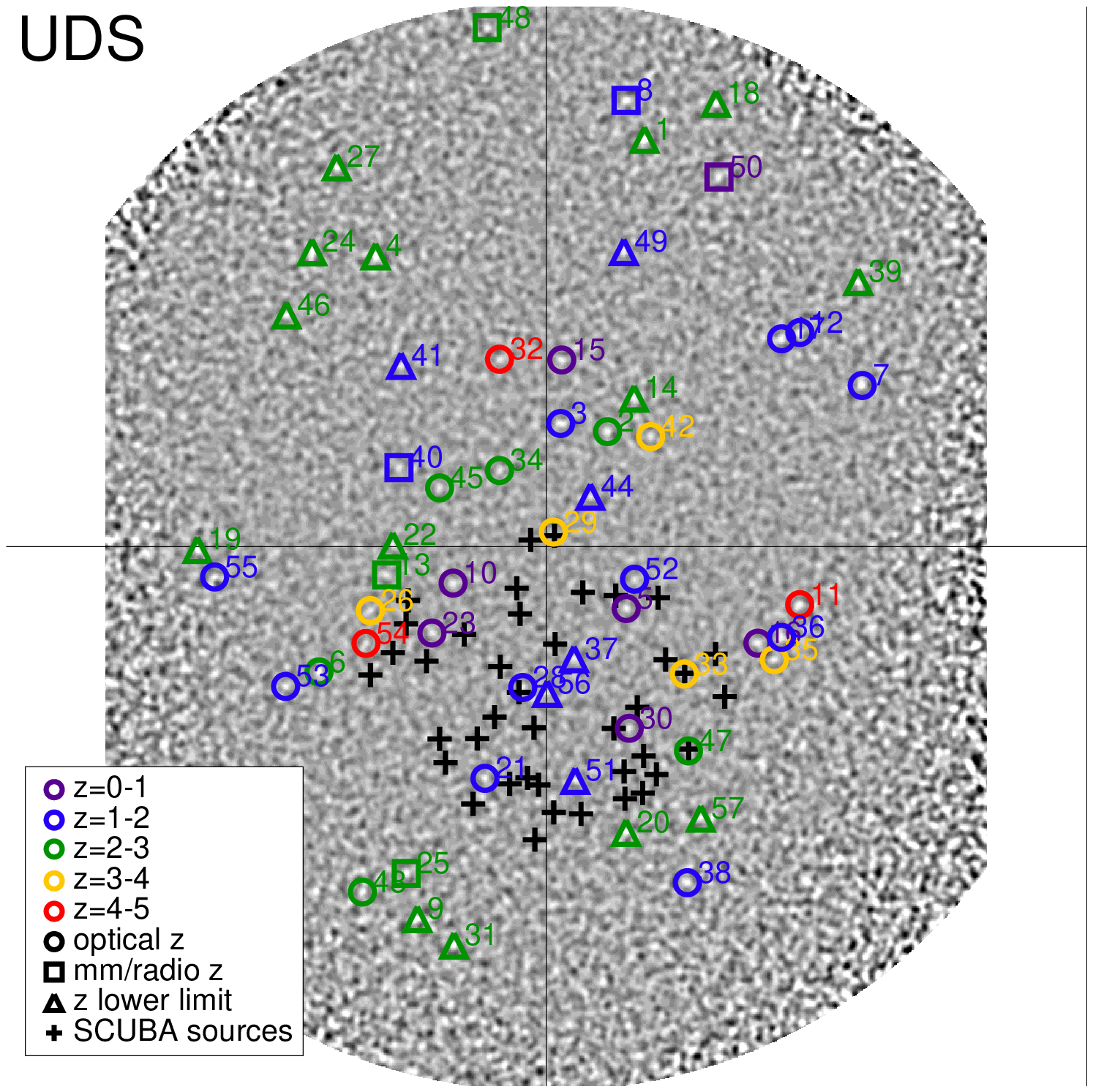}
\end{center}
 \caption{The AzTEC $1.1$\,mm maps of the Lockman Hole ({\it top}) and the SXDF/UDS field ({\it bottom}) from \citet{austermann10}. 
Both images are $0.88$\,deg on a side (the noisier edges have been removed). 
The sources analysed in this paper are marked and colour-coded according to their redshifts. {\it Circles} 
correspond to {\revone spectroscopic or photometric} optical or mid-infrared PAH redshifts, whereas {\it squares} correspond to redshifts derived from the $1.1$\,mm/$1.4$\,GHz flux-density ratio 
based on the average SED model of SMGs \citep{michalowski10smg}. In the case of a radio non-detection this method provides only a lower limit to the 
redshift and such cases are marked as {\it triangles}. {\it Black lines} divide both fields into four equal parts each with an area similar to that 
of the GOODS-N AzTEC survey studied by \citet{chapin09}. $50$\% (4/8) of these sub-fields do not contain any low-redshift SMGs (i.e. robust identifications
with $z < 1$).
{\it Crosses} denote $850\,\mu$m-selected galaxies  for which photometric redshifts have been derived by Schael et al. (in preparation).}
 \label{fig:map}
\end{figure*}

\begin{figure*}
\begin{center}
\includegraphics[width=0.59\textwidth]{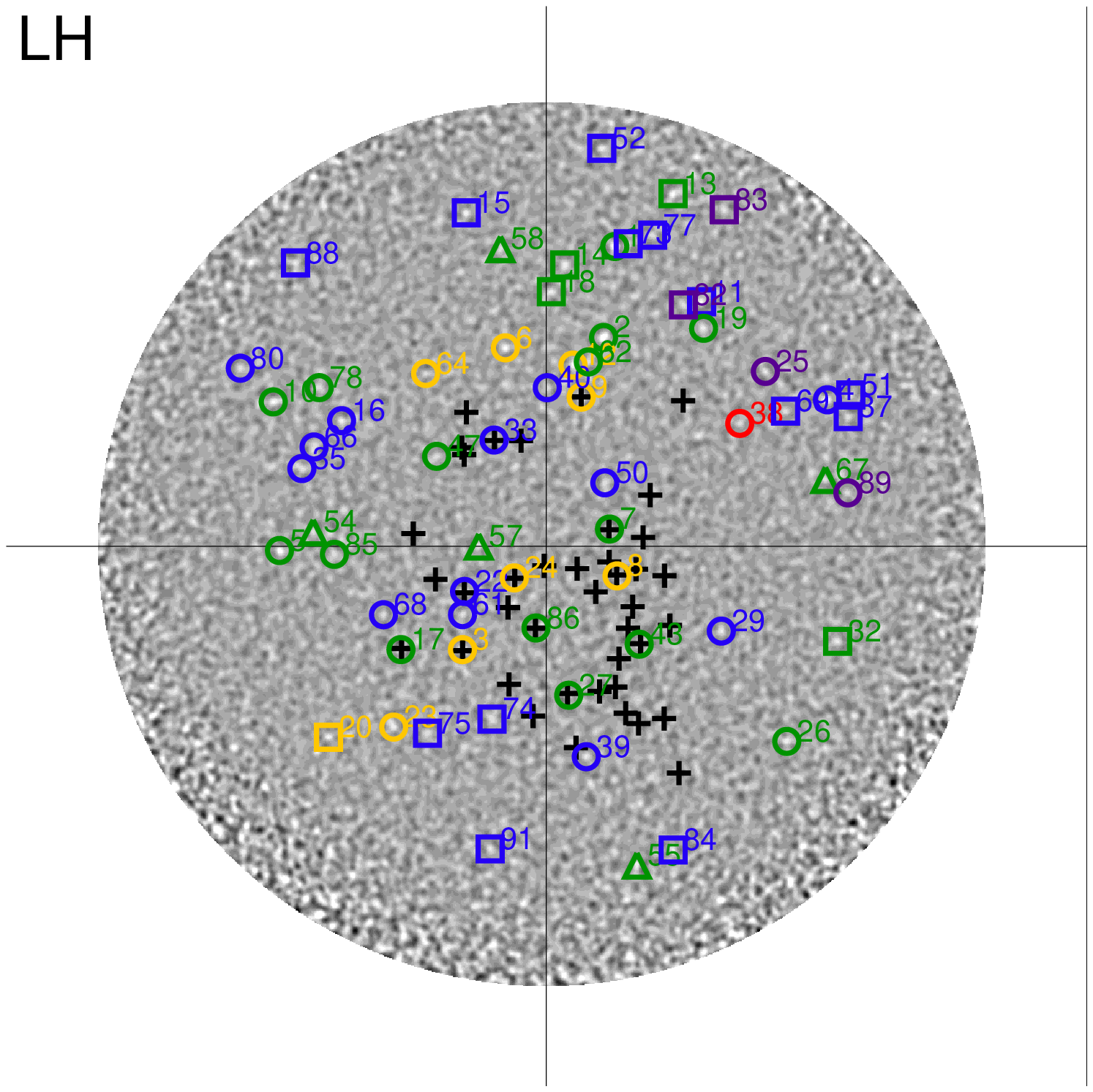}

\vspace{0.1em}

\includegraphics[width=0.59\textwidth]{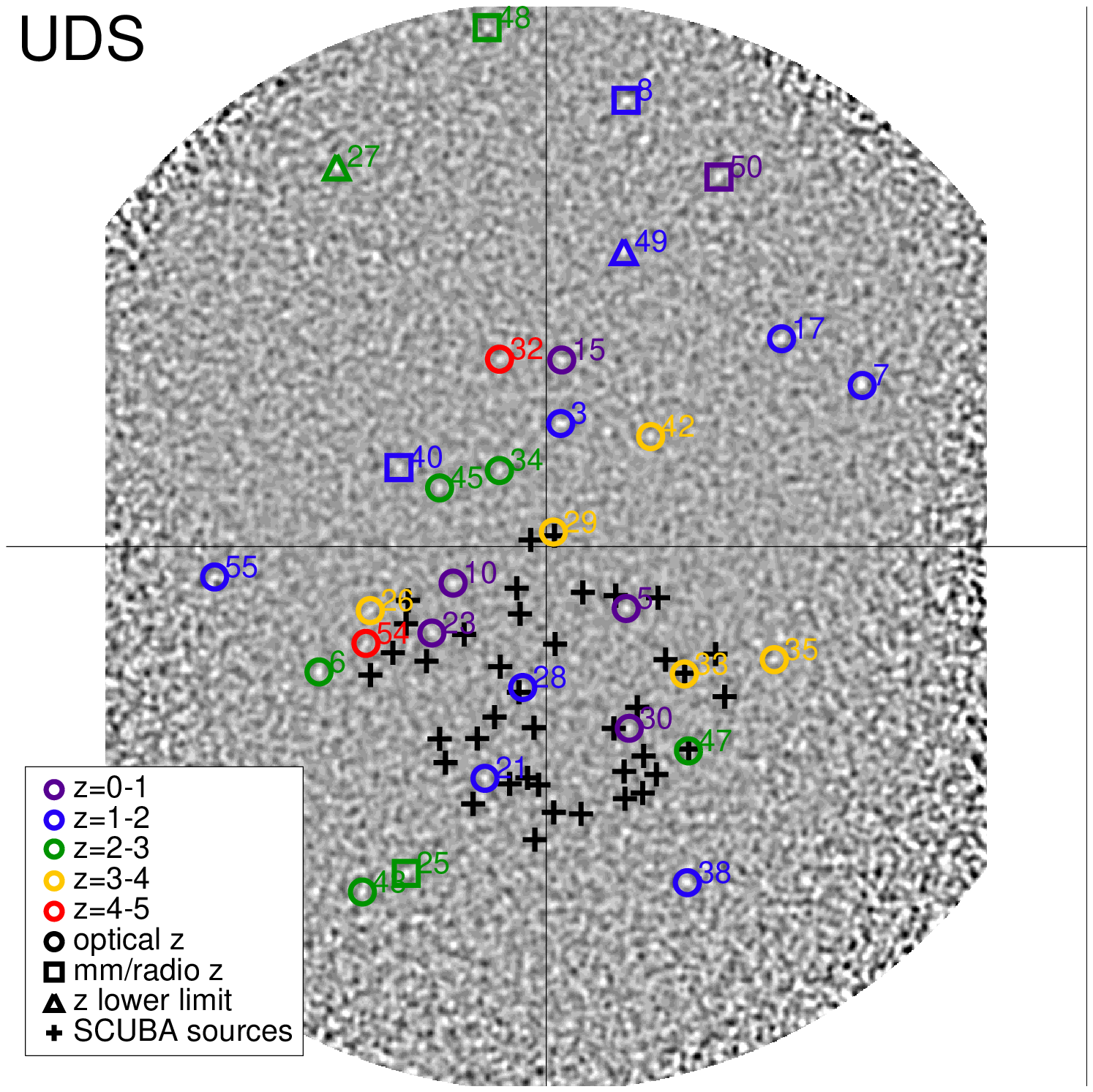}
\end{center}
 \caption{The same as Fig.~\ref{fig:map}, but this time only showing robust (category 1) IDs.
 }
 \label{fig:map1}
\end{figure*}




\begin{table*}
\caption{The success rate of the identification process. The columns show: (1) field name; (2) the total number of AzTEC sources, (3) the number of sources with IDs having at least one 
association with $p<0.05$ at radio, $24\,\mu$m, $8.0\,\mu$m or $i-K$; (4) the number of sources with IDs having at least two associations with 
$0.05<p<0.1$; (5), the number of sources with IDs having only one potential counterpart with $0.05<p<0.1$; (6) the number of sources with no IDs; 
(7) the total number of sources covered by the optical map (i.e.~those for which a photometric redshift could in principle be reliably estimated); (8) number (and percentage relative to number of sources covered b the optical map) of sources with an optical--near-infrared photometric redshift; (9) number of sources with category 1 and an optical--near-infrared photometric redshift.}
\label{tab:succ}
\begin{tabular}{lcccccccc}
\hline\hline
Field                  & N   & Cat 1 & Cat 2 & Cat 3 & No ID & N$_{\rm opt}$ & $z_{\rm opt}$ & Cat 1 with $z_{\rm opt}$\\
(1) & (2) & (3) & (4) & (5) & (6) & (7) & (8) & (9)\\
\hline
Lockman Hole & 91 & 64 (70\%) & 7 (8\%)&  7 (8\%)                         &  13 (14\%) &  62 & 47 (76\%) & 39 (63\%)\\
UDS                   & 57 & 31 (54\%) & 1 (2\%)& 8 (14\%)                         & 17 (30\%)  & 44 & 31 (70\%) & 24 (55\%)\\
Both                   & 148& 95 (64\%) & 8 (5\%)& 15 (10\%)                         & 30 (20\%) & 106 & 78 (74\%) & 63 (59\%)\\
\hline 
\end{tabular}
\end{table*}

\begin{table*}
\scriptsize
\caption{The success rate of the identification process for the five individual methods. The columns show: (1) field name; (2) the total number of AzTEC sources, (3) the number of sources with any $1.4$ GHz ID; (4) the number of sources with a category-1 $1.4$ GHz ID; (5), (6) any / category-1 $0.6$ GHz  IDs; (7), (8) any / category-1 $24\,\mu$m   IDs; 
(9) the number of sources covered by the IRAC map;  (10), (11)  any / category-1 $8.0\,\mu$m   IDs; (12)  the number of sources covered by the $i$- and $K$-band maps; (13), (14)  any / category-1 $i-K$   IDs.}
\label{tab:succsep}
\begin{tabular}{lccc@{}ccc@{}ccc@{}cccc@{}cccc}
\hline\hline
Field                  & N   & \multicolumn{2}{c}{$1.4$ GHz IDs} & & \multicolumn{2}{c}{$0.6$ GHz IDs} & & \multicolumn{2}{c}{$24\,\mu$m IDs} & & \multicolumn{3}{c}{$8.0\,\mu$m IDs} & & \multicolumn{3}{c}{$i-K$ IDs}  \\
\cline{3-4} \cline{6-7} \cline{9-10} \cline{12-14} \cline{16-18}
& & any & Cat 1 & & any & Cat 1 & & any & Cat 1 & & N & any & Cat 1 & & N & any & Cat 1 \\
(1) & (2) & (3) & (4) & & (5) & (6) & & (7) & (8) & & (9) & (10) & (11) & & (12) & (13) & (14)       \\
\hline
LH                     & 91 & 66 (73\%) & 50 (55\%)& & 57 (63\%) & 46 (51\%) & & 63 (69\%)  & 43 (47\%) & & 64 & 35 (55\%) & 24 (38\%) & & 54 & 26 (48\%) & 18 (33\%)\\
UDS                   & 57 & 25 (44\%) & 21 (37\%)& & N/A & N/A & & 22 (39\%)  & 16 (28\%) & & 51 & 14 (27\%) & 5 (10\%) & & 44 & 23 (52\%) & 13 (30\%)\\
Both                     & 148 & 91 (61\%) & 71 (49\%)& & N/A & N/A & & 85 (57\%)  & 59 (40\%) & & 115 & 49 (43\%) & 29 (25\%) & & 98 & 49 (50\%) & 31 (32\%)\\
\hline 
\end{tabular}
\end{table*}

\begin{figure*}
\includegraphics[width=0.8\textwidth]{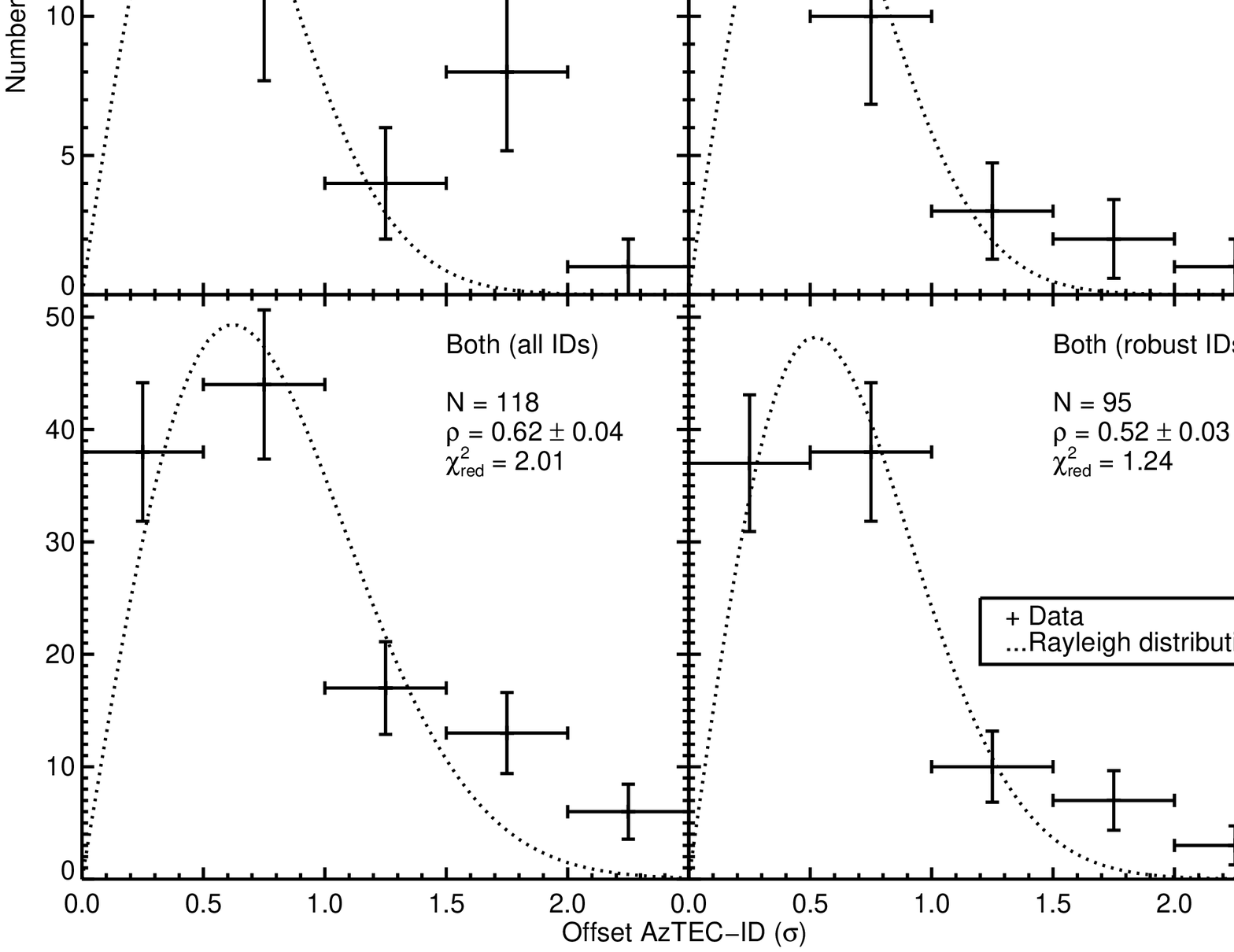}
 \caption{Distribution of the offsets  of IDs from the AzTEC positions ({\it crosses with Poissonian error bars}; only the best ID is taken into account for each source) and the Rayleigh distribution ({\it dotted lines}), $R(r)\propto r\exp(-r^2/2\rho^2)$, expected to explain the data if offsets originate from the statistical positional uncertainty of the AzTEC sources. Each panel represents all or robust (category-1 $p<0.05$) IDs, in the Lockman Hole, or the UDS field, or both. As in \citet{biggs11}, for each ID the value of the offset is normalized to the $1\sigma$ AzTEC positional uncertainty, $\sigma=0.6\times \mbox{FWHM}/(\mbox{S}/\mbox{N})$ (Sec.~\ref{sec:idrad24}). Hence, the value of $\rho$ corresponds to the $1\sigma$ positional uncertainty of IDs expressed in the units of such calculated $\sigma$ and hence should be equal to $1$. The numbers given in each panel are lower, so likely our estimates of the AzTEC positional uncertainties are overestimated. 
This is not because for 20 sources we reset the search radius  to $8$ arcsec ($2.5\sigma$; Sec.~\ref{sec:idrad24}), as after we remove this limit the values of $\rho$ increase only by $<0.05$.
 There is an indication that the distributions of all IDs exhibit a slight excess at higher offsets compared to the model and to the robust-only IDs, indicating that some of the category-2 and 3 IDs are not correct, but this difference is not statistically significant. For each panel the number of IDs and the reduced $\chi^2$ of the fit of the Rayleigh distribution to the data are shown indicating that the offset distributions are consistent with the Rayleigh distribution, so the number of wrongly assigned IDs is small (as they should manifest themselves as a significant signal at larger offsets).
}
 \label{fig:sep}
\end{figure*}

In Figs.~\ref{fig:AzLOCKthumb} and \ref{fig:AzSXDFthumb} we present thumbnail images for all of the AzTEC-SHADES sources, with the IDs indicated by coloured symbols.
For completeness and future reference, Tables \ref{tab:AzLOCKID} and \ref{tab:AzSXDFID} give the positions, 
relevant flux densities, angular offsets (from the original AzTEC $1.1$\,mm positions), and probabilities of chance associations ($p$) for 
all candidate radio and $24\,\mu$m IDs, regardless of their \mbox{$p$-values}. When IDs obtained at different wavelengths were separated by less than $2.5$\,arcsec 
they are listed as a single ID.
Tables~\ref{tab:AzLOCKgoodID} and \ref{tab:AzSXDFgoodID} present the relevant data only for the reliable IDs with $p<0.05$ (marked in bold) and tentative IDs with $0.05<p<0.1$ (marked in italics).  
The coordinates listed are those of the $1.4$\,GHz ID if present, or alternatively those of the appropriate $24\,\mu$m (or $0.61$ GHz), $8\,\mu$m, or $i-K$ selected ID. 

We matched these coordinates with the optical/near-infrared catalogues, using a matching search radius of $r=1.5$\,arcsec. 
The resulting multi-wavelength photometry was used both to derive photometric redshifts for the IDs (Section~\ref{sec:z}), and also to select additional IDs as described above.
Specifically, using the method presented in Section~\ref{sec:id80} we selected 9 (5) additional IDs in the  Lockman Hole field (UDS field) from the IRAC $8.0\,\mu$m imaging. 
For 2 (1) of the AzTEC-SHADES sources these are the only IDs, so this method adds a very small, but still useful set of extra identifications.
We checked the 22 AzTEC sources for which we found both robust radio ($p<0.05$) and $8.0\,\mu$m IDs, and found that they coincided in 20 cases ($91$\%), 
providing additional confidence that the $8.0\,\mu$m method can be reliably utilised when the radio data are not deep enough.

In addition, using the method presented in Section~\ref{sec:idik} we identified 16 (28) significant $i-K$ colour-selected IDs 
in the  Lockman Hole field (UDS field). 
For 1 (5) of the AzTEC-SHADES sources these are the only IDs, 
so again the result of this effort is another small, but helpful set of additional identifications, along with 
supporting evidence for several others.
Again we checked the AzTEC sources for which we found both robust radio and $i-K$ IDs, and found that they agreed in 19/23 
cases ($83$\%), providing reassurance that the $i-K$ method can be reliably used when the radio data are not deep enough.

The final success-rate of source  identification is summarised in Table~\ref{tab:succ}. The IDs have been divided into three categories. 
Category 1 is used for an ID which has a very low probability of chance association, $p<0.05$ ($p$ marked as bold in Tables~\ref{tab:AzLOCKgoodID} and \ref{tab:AzSXDFgoodID} 
and a big symbol on Figs.~\ref{fig:AzLOCKthumb} and \ref{fig:AzSXDFthumb}).
Category 2 denotes an ID  selected at least twice by one of the radio, $24\,\mu$m $8.0\,\mu$m or $i-K$ methods with $0.05<p<0.1$, while
category 3 indicates an ID that has been selected by only one of these methods with $0.05<p<0.1$. 

The higher category-1 rate for the Lockman Hole can be primarily attributed to the slightly deeper radio data 
available in this field. Indeed, if the radio depth was degraded to the same as that available in the 
UDS field ($\simeq 45\,\mu$Jy, $5\sigma$), then the category-1 ID percentage in the Lockman Hole would drop to $58$\%, very similar to that found 
within the UDS field.
The impact of the depth of the radio data on the ID rate is further demonstrated by the fact that the final total AzTEC-SHADES SMG ID rate of $\simeq80$\%
is surpassed only by the survey of \citet{lindner11}, who benefitted from the ultra-deep radio data available only in the Lockman
Owen field, which reaches down to an rms $\sigma_{1.4GHz} \simeq2.7\,\mu$Jy in the central regions.
 
 {\revone The ID success-rate separately for each method is shown in Table~\ref{tab:succsep}.  The radio and $24\,\mu$m methods in the Lockman Hole deliver higher success-rate than in the UDS field due to deeper data in the former field (see Table~\ref{tab:data}). On the other hand, deeper optical data in the UDS field do not help to increase the success-rate for the $i-K$ method. This is because no $i-K$ ID in the UDS field is fainter than the $3\sigma$ limit of $K=23.5$ mag in the Lockman Hole. However, deeper data in the UDS field help to increase the fraction of IDs with redshifts (see Table~\ref{tab:succ}).}
 
To determine the redshift distribution of the AzTEC-SHADES sources we decided to select only one ID for each source. 
For the vast majority of cases this is straightforward. However, for some sources there is more than one 
apparently significant galaxy counterpart. 
To deal with these cases we have adopted a policy of selecting the ID with the greatest number of high-significance ($p<0.05$) 
entries in Tables~\ref{tab:AzLOCKgoodID} and \ref{tab:AzSXDFgoodID} or, if IDs have only been 
uncovered at a single wavelength, the ID with the lowest value of $p$. 
If this procedure did not select a single ID, then we looked for the ID with the greatest number of moderate-significance ($0.05<p<0.1$) entries. 
In principle, such multiple IDs may indicate that the AzTEC source is a blend 
of a few galaxies \citep[e.g.][]{wang11}; higher-resolution submillimetre imagining is necessary to test this.

The frequency of multiple IDs is $\simeq21\pm5$\% (19/91) for the Lockman Hole 
and $\simeq11\pm4$\% (6/57) for the UDS field, where the errors reflect the Poissonian uncertainties. 
This is consistent with the rates found for SCUBA sources in the same fields ($\simeq9\pm5$\%; \citealt{clements08}; $\simeq19 \pm5$\%; \citealt{ivison07}) and for AzTEC sources in GOODS-N \citep[$\simeq18\pm8$\%][]{chapin09} and GOODS-S \citep[$\simeq10\pm5$\%][]{yun12}.

We identified two sources (AzUDS5 and 43) with statistically robust IDs corresponding to optically-bright galaxies with very low photometric redshifts ($z \simeq 0.45$ and $z \simeq 0.15$
respectively). Such low redshifts are completely inconsistent with their $1.1$\,mm/$24\,\mu$m flux-density ratios for any known long-wavelength SED (see Fig.~\ref{fig:fluxz}). In the case of AzUDS5
the angular offset between the radio position and the optical galaxy is 
$0.65$\,arcsec, suggesting that this is an example of galaxy lensing, and that the SMG lies at much higher redshift \citep[e.g.][]{dunlop04}. We note this, but leave the redshift as it is since we do not currently 
possess any CO spectroscopy of this source which might help to determine the true redshift of this SMG \citep[see e.g.][]{negrello10}. However, in the case of AzUDS43 we have 
identified a second, statistically-robust ID at $z\simeq2.43$, $3.5$ arcsec away from the bright low-redshift optical ID, and so we have adopted this ID 
and its photometric redshift in the determination of the redshift distribution.

{\revone To test the reliability of the IDs, in Fig.~\ref{fig:sep} we show the distribution of the offsets of IDs from the AzTEC positions. Similarly to \citet{biggs11}, we fitted the Rayleigh distribution, $R(r)\propto r\exp(-r^2/2\rho^2)$, expected to explain the data if offsets originate from the statistical positional uncertainty of the AzTEC sources. For each ID the value of the offset is normalized to the $1\sigma$ AzTEC positional uncertainty equal to $\sigma=0.6\times \mbox{FWHM}/(\mbox{S}/\mbox{N})$ (Sec.~\ref{sec:idrad24}). Hence the value of $\rho$ corresponds to the $1\sigma$ positional uncertainty of IDs expressed in the units of such calculated $\sigma$ and hence should be equal to $1$. The derived values are lower, so likely our estimates of the AzTEC positional uncertainties are overestimated. This is not because for 20 sources we reset the search radius  to $8$ arcsec ($2.5\sigma$; Sec.~\ref{sec:idrad24}), as after we remove this limit the values of $\rho$ increase only by $<0.05$. There is an indication that the distributions of all IDs exhibit a slight excess at higher offsets compared to the model and to the robust-only IDs, indicating that some of the category-2 and 3 IDs are not correct, but this difference is not statistically significant.  From the reduced $\chi^2$ values we conclude that the offset distributions are consistent with the Rayleigh distribution, so the number of wrongly assigned IDs is small (as they should manifest themselves as a significant signal at larger offsets). 

}

\section[]{Redshifts}
\label{sec:z}

\begin{figure}
\includegraphics[width=0.45\textwidth]{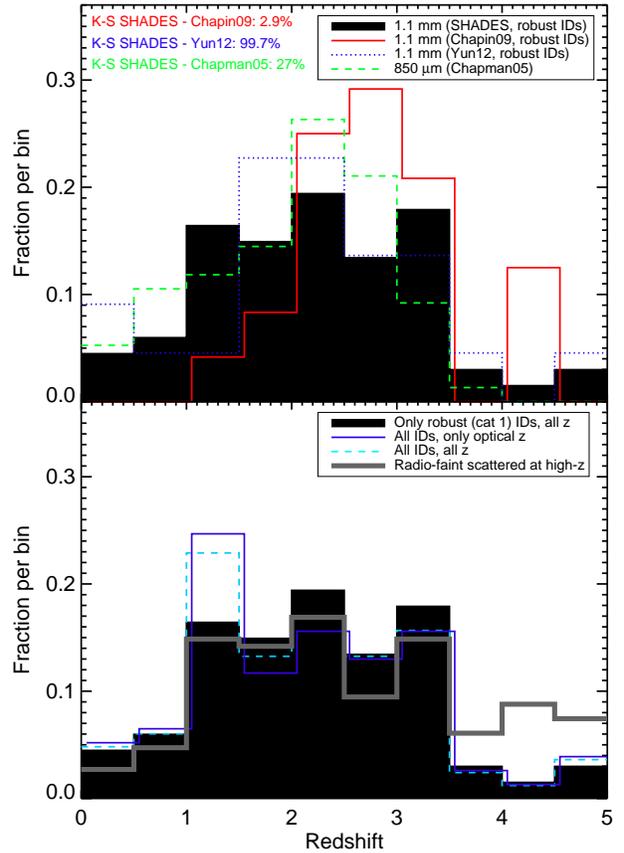}
 \caption{{\it Upper Panel}: the redshift distribution of $1.1$\,mm-selected galaxies in the SHADES fields with robust ($p<0.05$) IDs 
using optical--near-infrared and $1.1$\,mm/$1.4$\,GHz flux-density ratio redshifts ({\it solid black histogram}). In this plot any sources which only possess 
lower limits to their estimated redshifts have been excluded. 
Also shown are the redshift distributions of the $1.1$\,mm-selected galaxies in GOODS-N \citep[{\it solid red line};][]{chapin09} and in GOODS-S \citep[{\it dotted blue line};][]{yun12},
as well as the spectroscopically-determined redshift distribution of $850\,\mu$m-selected galaxies reported by \citet[][{\it green dashed line}]{chapman05}. 
The distributions peak at $z\simeq2$--$3$ but contain objects over the redshift range $z\simeq0$--$4$. 
The Kolmogorov-Smirnov test probabilities that samples are consistent with being drawn from the same parent population are indicated. 
The apparent difference at low redshifts between the SHADES and GOODS-N samples can be explained by the relatively small area of the latter survey (see Fig.~\ref{fig:map} and Section~\ref{sec:z}).
{\it Lower Panel}: An exploration of the robustness of the SHADES-AzTEC redshift distribution. The redshift distribution of the same set of robust IDs is again shown by the {\it solid black histogram}. 
The {\it solid blue line} represents only those IDs with optical--near-infrared
photometric redshifts, while the {\it dashed cyan line} includes all IDs, irrespective of robustness (i.e. including categories 1, 2, and 3) and type of redshift. 
The {\it grey thick line} shows the redshift distribution for all AzTEC-SHADES SMGs, this time including 
those not detected at radio or optical wavelengths, which were scattered between $z=5$ and the lower limit derived from the $1.1$\,mm/$1.4$\,GHz flux-density limit using the 
SMG SED template from \citet{michalowski10smg}.
}
 \label{fig:z}
\end{figure}

\begin{figure}
\includegraphics[width=0.45\textwidth]{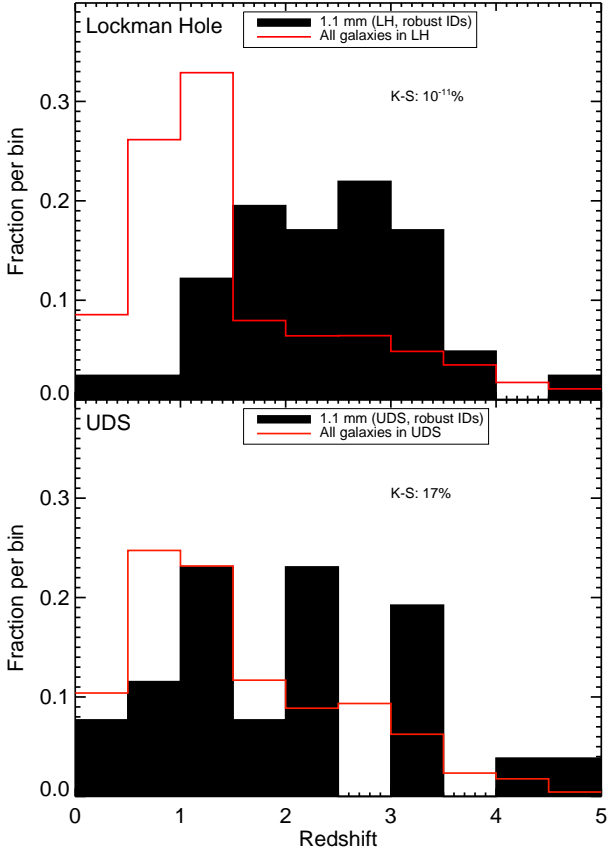}
 \caption{The redshift distribution of SHADES-AzTEC $1.1$\,mm-selected galaxies 
with robust ($p<0.05$) IDs and either optical--near-infrared or $1.1$\,mm/$1.4$ GHz redshifts ({\it solid black histograms}), this time shown separately for the 
Lockman Hole field ({\it upper panel}) and the UDS field ({\it lower panel}). Sources with only 
lower limits to their estimated redshifts were excluded.
Also shown for comparison are the redshift distributions of all the optical/near-infrared/{\it Spitzer} selected galaxies in the two survey fields ({\it solid red lines}). 
The Kolmogorov-Smirnov test probability that the samples on each panel are drawn from the same parent population are indicated. The AzTEC population in the Lockman Hole  
is not consistent with the general field galaxy population. 
The all galaxy samples (red lines) are different in both fields due to different optical depths, i.e.~the $i$-band data is $\simeq1.5$ mag deeper in the UDS field (Table~\ref{tab:data}), so the high redshift tail is much more pronounced in this field.
}
 \label{fig:z_all}
\end{figure}

\begin{figure}
\includegraphics[width=0.45\textwidth]{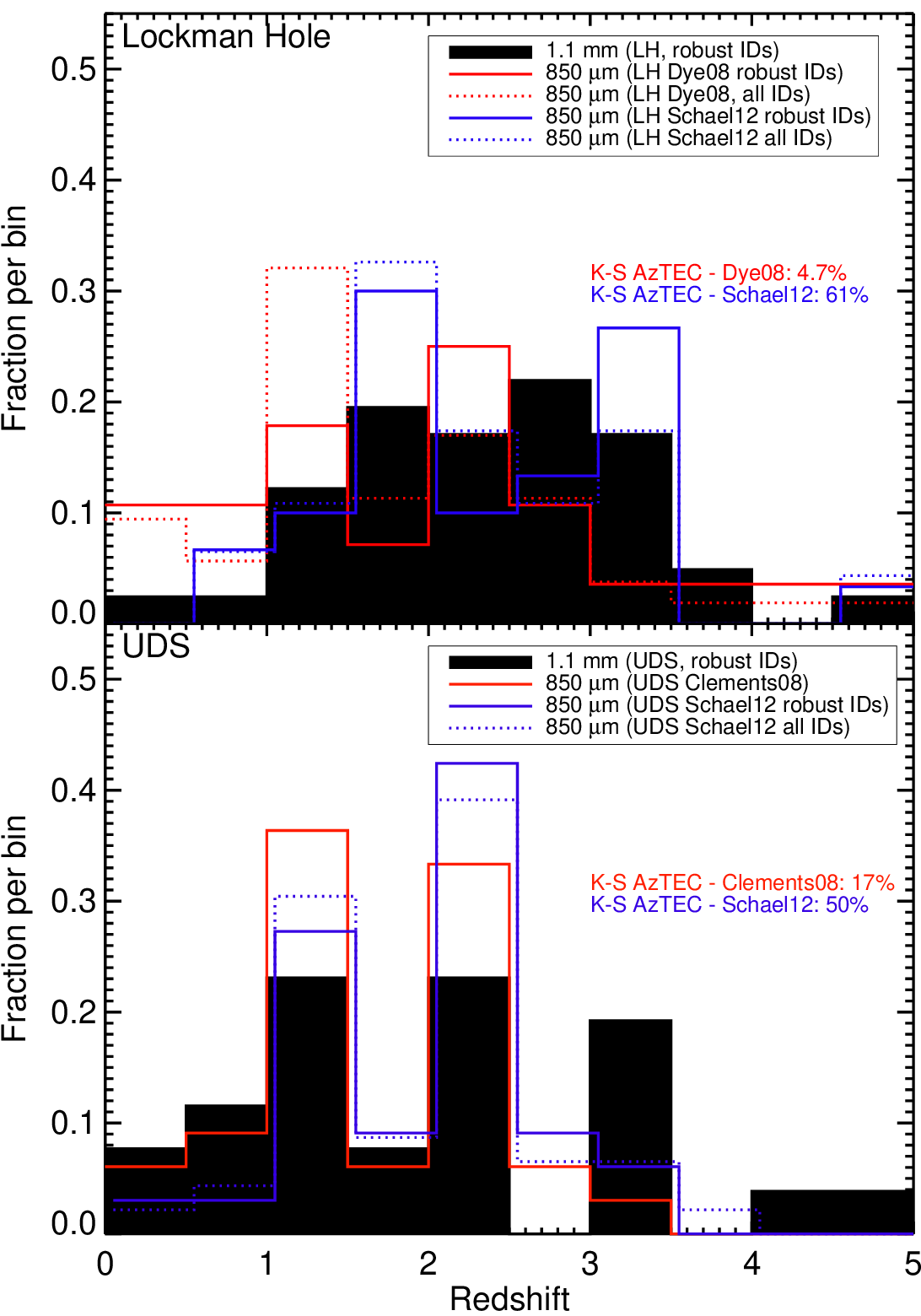}
 \caption{The redshift distribution of $1.1$\,mm-selected galaxies in the Lockman Hole ({\it upper panel}) and the UDS ({\it lower panel}) fields with robust ($p<0.05$) IDs 
and optical--near-infrared or $1.1$\,mm/$1.4$\,GHz flux-density ratio redshifts ({\it solid black histograms}). 
Sources with only 
lower limits to their estimated redshifts were excluded.
Also shown in this figure are the redshift distributions derived for the SHADES-SCUBA  $850\,\mu$m-selected galaxies in (sub regions of) the same fields 
\citep[robust IDs: {\it solid red lines}; all IDs: {\it dotted red line};][Schael et al. in prep.]{dye08,clements08}. 
The Kolmogorov-Smirnov test probability that the samples in each panel are drawn from the same populations are indicated. 
The distributions of the $1.1$\,mm- and $850\,\mu$m-selected populations are consistent, although there is an indication that the $1.1$\,mm selection results in slightly higher redshifts.
Combining the AzTEC and SCUBA samples in each field reveals a statistically significant difference between the inferred redshift distributions of SMGs in the Lockman Hole and UDS fields
(see Section~\ref{sec:large}).
 }
 \label{fig:z_sep}
\end{figure}

We have used the photometric redshift catalogues from  \citet{cirasuolo07,cirasuolo10} and \citet{mclure09} 
in the UDS field and have now produced an equivalent catalogue in the Lockman Hole East, 
albeit 
with larger uncertainties
due to the shallower near-infrared data. All optical, near-IR and IRAC data were used in the fits.
These catalogues have been produced using the {\sc HyperZ} package \citep{bolzonella00} 
with the stellar population models of \citet{bruzualcharlot03} and a \citet{chabrier03} initial 
mass function (IMF) with a mass range $0.1$--$100\,$M$_\odot$. 
A double-burst star-formation history was assumed, but this choice has little impact on derived redshifts \citep[as opposed to derived stellar masses;][]{michalowski12mass}. 
The metallicity was fixed at the solar value and 
reddening was calculated following the \citet{calzetti00} law within the range $0 \le A_V	\le 6$. 
The HI absorption along the line of sight was included according to the prescription of \citet{madau95}.

The resulting photometric redshifts are given in Tables \ref{tab:AzLOCKgoodID} and \ref{tab:AzSXDFgoodID}.
For sources with multiple IDs, the ID used in the determination of the redshift distribution is marked in bold (see Section~\ref{sec:id}). 
For sources with no optical-infrared photometric redshift we provide a redshift estimate 
based on the $1.1$\,mm / $1.4$\,GHz  ratio (shown in italics), adopting the average SED of SMGs from \citet{michalowski10smg}.

AzUDS48 is not covered by the radio map, so an estimate of its redshift has been deduced from its 
$1.1$\,mm/$24\,\mu$m flux-density ratio (see third panel of Fig.~\ref{fig:fluxz}).

The accuracy of the photometric catalogue of \citet{cirasuolo10} is excellent, 
with a mean $\Delta z/(1+z_{\rm spec})=0.008\pm0.034$. For 9 AzTEC sources with spectroscopic 
redshift (see appendix~\ref{sec:notes}) we derived a mean  $\Delta z/(1+z_{\rm spec})=-0.06\pm0.20$, also consistent with zero.

The median redshift (and 68\% bootstrap error) of AzTEC-SHADES sources in the Lockman Hole field 
is $z_{med}\simeq2.25^{+0.25}_{-0.21}$, in the UDS field it is $z_{med}\simeq1.64^{+0.63}_{-0.28}$, and for the full combined 
sample it is $z_{med} \simeq 2.19^{+ 0.10}_{-0.30}$. 


To date, only eleven SMGs have been shown to lie at $z>4$  \citep{coppin09,capak08,capak11,schinnerer08,daddi09,daddi09b,knudsen08b,knudsen09,riechers10,cox11,smolcic11,combes12,walter12b}. We 
have identified five AzTEC-SHADES sources ($\simeq5$\%) with photometric redshifts $z\gtrsim 4$ (AzLOCK6, AzLOCK38, AzUDS11, AzUDS32, AzUDS54). However, for all 
of these galaxies the uncertainties in the photometric redshifts are significant, and lower-redshift ($z\simeq2$) solutions cannot be excluded.
In addition, several AzTEC sources that are unambiguously 
detected in the radio (and in some cases also at $8.0\,\mu$m) remain undetected in the optical imaging
(AzLOCK36, AzLOCK38, AzLOCK71, AzLOCK81, AzLOCK91, AzUDS25, AzUDS40 and AzUDS57). These sources are thus also candidates for very high-redshift (or highly dust-obscured) galaxies.

The success rate of redshift determination is summarized in Table~\ref{tab:succ} (column 8).
Excluding AzTEC sources which are not covered by the necessary optical-infrared data, we obtained optical photometric redshift estimates for $\simeq75$\% of sources (i.e. out of the 106 AzTEC sources which lie within the area covered by the deep optical/near-infrared imaging, 
we obtained optical photometric redshift estimates for 78; out of them 63 have category-1 IDs)
{\revone or $\simeq60$\% if only category-1 IDs are taken into account}.
This is the most complete redshift information achieved to date for an unbiased sample of SMGs.
For the remaining 28 ($\simeq25$\%) SMGs no ID was selected, or the radio ID does not possess an optical counterpart, so the optical redshift could not be obtained.

The redshift distribution of the AzTEC-SHADES sources  is shown in Fig.~\ref{fig:z}, where it is 
compared with that displayed by similar (but smaller) samples of SMGs selected in GOODS-N \citep{chapin09} and in GOODS-S \citep{yun12}, as well as with 
the spectroscopically-determined redshift distribution of a somewhat heterogeneous selection of $850\,\mu$m-selected galaxies reported by \citet{chapman05}. 
The redshift distribution of the AzTEC-SHADES sources is peaked at $z\simeq2$--$2.5$, but has a high-redshift tail (extending to at least $z\simeq4$) as well as 
a significant intermediate-redshift population.

As perhaps expected, the redshift distribution of AzTEC-SHADES sources is similar to that previously derived for $850\,\mu$m-selected galaxies, 
and to that displayed by the AzTEC-ASTE sample in GOODS-S;
the Kolmogorov-Smirnov (K-S) test indicates that the redshift distribution of the AzTEC-SHADES 
sources and that reported by \citet{chapman05} and \citet{yun12} are consistent 
with the hypothesis that all three samples are drawn from the same parent population (Fig.~\ref{fig:z}).

However, our redshift distribution is slightly different than that reported by \citet{chapin09} in GOODS-N. The K-S test yields 
a probability of only $p \simeq2.9$\% that the AzTEC-SHADES and GOODS-N AzTEC samples are drawn from the same population (i.e.~the difference is at the $2.2\sigma$ level). 
The probability increases to $\simeq15$\% (no significant difference) when $z<1$ SMGs in the AzTEC-SHADES sample are removed.

To further explore the likelihood of this difference, we divided each of our fields into four equal parts, each similar in size to the GOODS-N field and  
analysed their redshift distributions separately (see Fig.~\ref{fig:map}). We found that in $50$\% (4/8) of such small fields we did not detect any $z<1$ source. 
To confirm this we performed a Monte-Carlo simulation by selecting in both of our fields 100\,000 randomly-located sub-fields with areas equal to that of  \citet{chapin09}. 
We found that $\simeq60$\% of these  sub-fields did not contain any $z<1$ object. This suggests that the sample of \citet{chapin09} may miss 
the lower-redshift population due to its small area and advocates for using larger fields to obtain a representative sample of {\mm}-selected galaxies.
Further support for field-to-field variation is provided by the differences in {\mm} number counts 
in different fields, in particular by the fact that GOODS-N field has systematically higher  number counts compared to other fields \citep{scott10,scott12}.

The lower panel of Fig.~\ref{fig:z} shows that the form of the derived redshift distribution is essentially unaffected by 
whether one restricts the redshift information to only optical-infrared photometric redshifts, or alternatively include all available redshift information 
(i.e. including estimates based on mm-radio colour). However, when we include the category 2 and 3 IDs, then a spike at $z\simeq1$--$1.5$ appears. 
At least some of these less-robust IDs are likely mis-identifications, because, as shown on Fig.~\ref{fig:z_all}, the redshift distribution of all 
optical/near-infrared galaxies in both the survey fields peaks at $z\simeq1$.

Fig.~\ref{fig:z_sep} shows the redshift distribution of the robust ($p<0.05$) IDs of the AzTEC sources 
separately in the Lockman Hole and the UDS fields compared with that of the $850\,\mu$m-selected 
SMGs in (smaller sub-regions of) the same fields \citep[][Schael et al. in prep.]{dye08,clements08}. Applying 
the K-S test we found that, within each field, the redshift distributions of the $1.1$\,mm- and $850\,\mu$m-selected populations are completely 
consistent (the significance values, $p$, are indicated in Fig.~\ref{fig:z_sep}), although (unsurprisingly) there is a tendency for the $1.1$\,mm 
selected SMGs to lie at slightly higher redshifts. 
This consistency is not driven by the fact that both populations are composed of the same galaxies; the AzTEC images cover 
significantly larger areas than the SCUBA maps, to somewhat shallower effective depths (and different completeness), and consequently there 
are only 13 and 6 sources in common between 
the $1.1$\,mm and $850\,\mu$m-selected samples in the Lockman Hole and UDS fields, respectively 
(see Fig.~\ref{fig:map}, Fig.~\ref{fig:map1} and appendix~\ref{sec:notes}).
The implications of this result are explored further below in Section~\ref{sec:large}.

In principle AGN contamination may influence our photometric redshift estimates, which are based on purely star-forming templates. 
\citet{hainline09,hainline11} claimed a significant AGN contribution to the IRAC fluxes of SMGs, 
however \citet{michalowski12mass} found that 
the AGN contamination is unlikely to influence the SED modelling. 
Additionally, the photometric errors of IRAC fluxes are typically $\sim10$\%, 
significantly larger than that of our deep optical and near-IR data, so even if the AGN contamination was significant at 
IRAC wavelengths, these data should not influence the derived photometric redshifts significantly.
Similarly, \citet{wardlow11} found that the exclusion of the $8.0\,\mu$m data does not significantly change the derived photometric redshifts of SMGs.

\section[]{Flux ratios}
\label{sec:fluxr}

\begin{figure}
\includegraphics[width=0.4\textwidth]{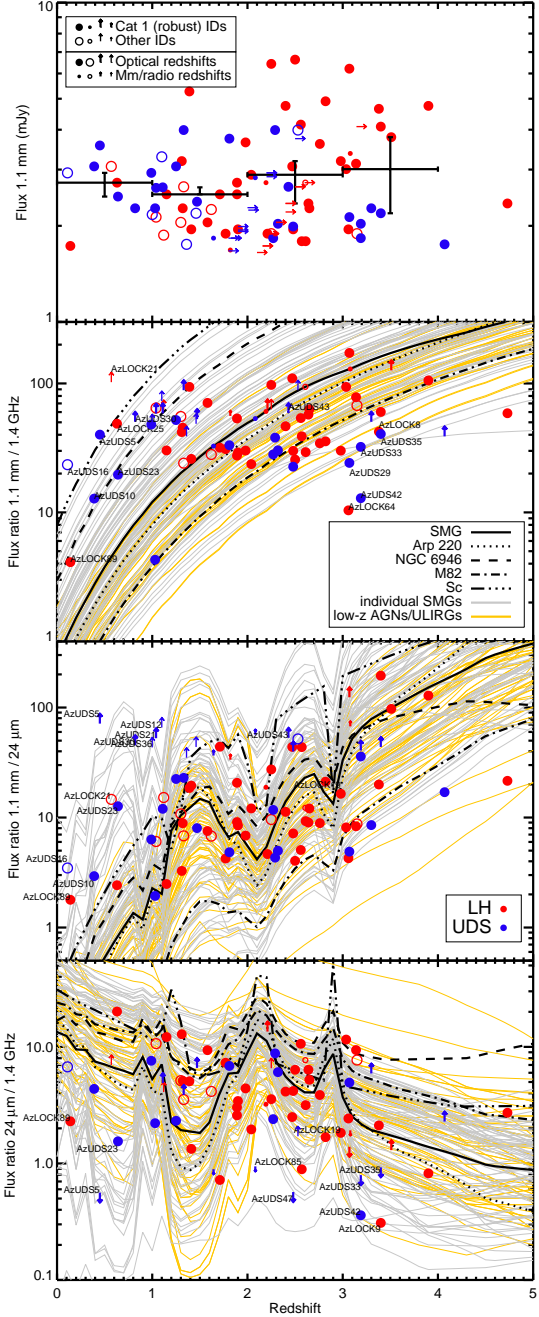}
 \caption{$1.1$\,mm flux densities (medians in four redshift bins with $68$\% bootstrap errors are shown as {\it thick crosses}) and the $1.1$\,mm/$1.4$\,GHz, $1.1$\,mm/$24\,\mu$m and $24\,\mu$m/1.4\,GHz flux-density ratios plotted against redshift for all the AzTEC-SHADES sources 
(the best ID is shown for each source) presented in Tables \ref{tab:AzLOCKgoodID} and \ref{tab:AzSXDFgoodID} ({\it colour symbols}; 
{\it red}: Lockman Hole; {\it blue}: UDS field). {\it Circles} denote detections at both bands, whereas {\it upward and downward arrows} denote non-detections at one of the wavelengths.
 {\it Filled circles and thick arrows} indicate category 1 (robust; $p<0.05$) IDs, whereas {\it open circles and thin arrows} indicate other IDs.
 {\it Big symbols} indicate sources with optical--near-infrared photometric redshifts, whereas {\it small symbols} indicate that $1.1$\,mm/1.4\,GHz redshift estimator 
was used due to a lack of optical/infrared photometry. 
 For comparison, the predicted redshift evolution of the flux-density ratios 
calculated using the SEDs of an average SMG and individual SMGs \citep{michalowski10smg}, of local galaxies \citep{silva98} and local ULIRGs/AGNs \citep{vega08} are shown ({\it lines}).
 }
 \label{fig:fluxz}
\end{figure}

In order to verify that these redshift estimates are consistent with the available photometry, in 
Fig.~\ref{fig:fluxz} we show $1.1$\,mm flux density, as well as the three flux-density ratios 
$1.1$\,mm/$1.4$\,GHz, $1.1$\,mm/$24\,\mu$m, and $24\,\mu$m/$1.4$\,GHz plotted against redshift
for all the identified AzTEC-SHADES sources (utilising the best ID for each source, determined as described above). 
The lower 3 panels also show the expected redshift dependence 
of the appropriate flux-density ratio as predicted by 
the SED of an average  SMG and individual SMGs \citep{michalowski10smg}, the SEDs of local galaxies \citep[a ULIRG, a starburst  and spirals;][]{silva98} and the SEDs of local ULIRGs/AGNs \citep{vega08}. 
Filled/thick symbols indicate category 1 (robust, $p<0.05$) IDs, and the smaller symbols indicate that the $1.1$\,mm/$1.4$\,GHz flux-ratio redshift estimator 
was used due to a lack of optical-infrared photometry. 

The top panel shows that the brightest AzTEC sources preferentially lie at 
higher redshifts \citep[consistent with ][]{ivison02,wall08,marsden11}; 
all but one of the sources brighter than $4$\,mJy are at $z>2$ (i.e. the upper-left corner of this panel is almost empty). 
However, this does not mean that the median flux densities of AzTEC sources is higher at higher redshifts (see thick crosses on this panel).

For sources with no radio detections the derived $1.1$\,mm/$1.4$\,GHz redshifts are lower limits.
As can be seen from the arrows in the top panel of Fig.~\ref{fig:fluxz}, 
the derived limits do not allow us to place very strong redshift constraints on these sources (typically $z > 2$).
Hence, in general, the lack of a radio detection does not imply that such sources lie at substantially higher redshifts than the radio-detected subset; given the $1.1$\,mm flux densities of the 
sources, and the depth of the available radio data, a radio-blank SMG could be at $z<3$ or even at $z < 1.5$ if its dust is very cold (similar to spiral galaxies).

The second panel of Fig.~\ref{fig:fluxz} reveals the redshift dependence of the implied dust temperature of the galaxies within the AzTEC-SHADES sample (assuming the photometric redshifts are broadly correct). Namely, the $1.1$ mm / $1.4$ GHz flux ratios of the AzTEC sources do not change with redshift, implying that  
those at $z<1.5$ are cooler (consistent with the SEDs of Sc spiral and NGC 6946), whereas those at higher redshifts are hotter (consistent with the SEDs of an average SMG, Arp 220 and M82).

Finally, the bottom two panels of Fig.~\ref{fig:fluxz} show that the vast majority of the AzTEC-SHADES sources lie within the regions spanned by the range of SED models considered (only for the sources denoted as small symbols were the redshifts in fact derived from the fluxes plotted in this figure). 
This indicates that our optical photometric redshift estimates are reasonable, in the sense that they are generally 
consistent with the anticipated range of $1.1$\,mm/$24$\,$\mu$m and $24\,\mu$m/1.4\,GHz flux-density ratios.
We note that radio-loud AGN from \citet{shang11} have $1.1$\,mm/1.4\,GHz and $24\,\mu$m/1.4\,GHz flux-density ratios of $0.01$--$0.03$  
in the redshift range $0$--$5$. Hence the sources which lie below the SED locus in panels 2 {\revone (AzLOCK64 and AzUDS42, both category-1 IDs)} may be readily explained by some contribution from AGN synchrotron radiation at radio wavelengths.

\section[]{Large-scale structures}
\label{sec:large}

\begin{figure}
\includegraphics[width=0.45\textwidth]{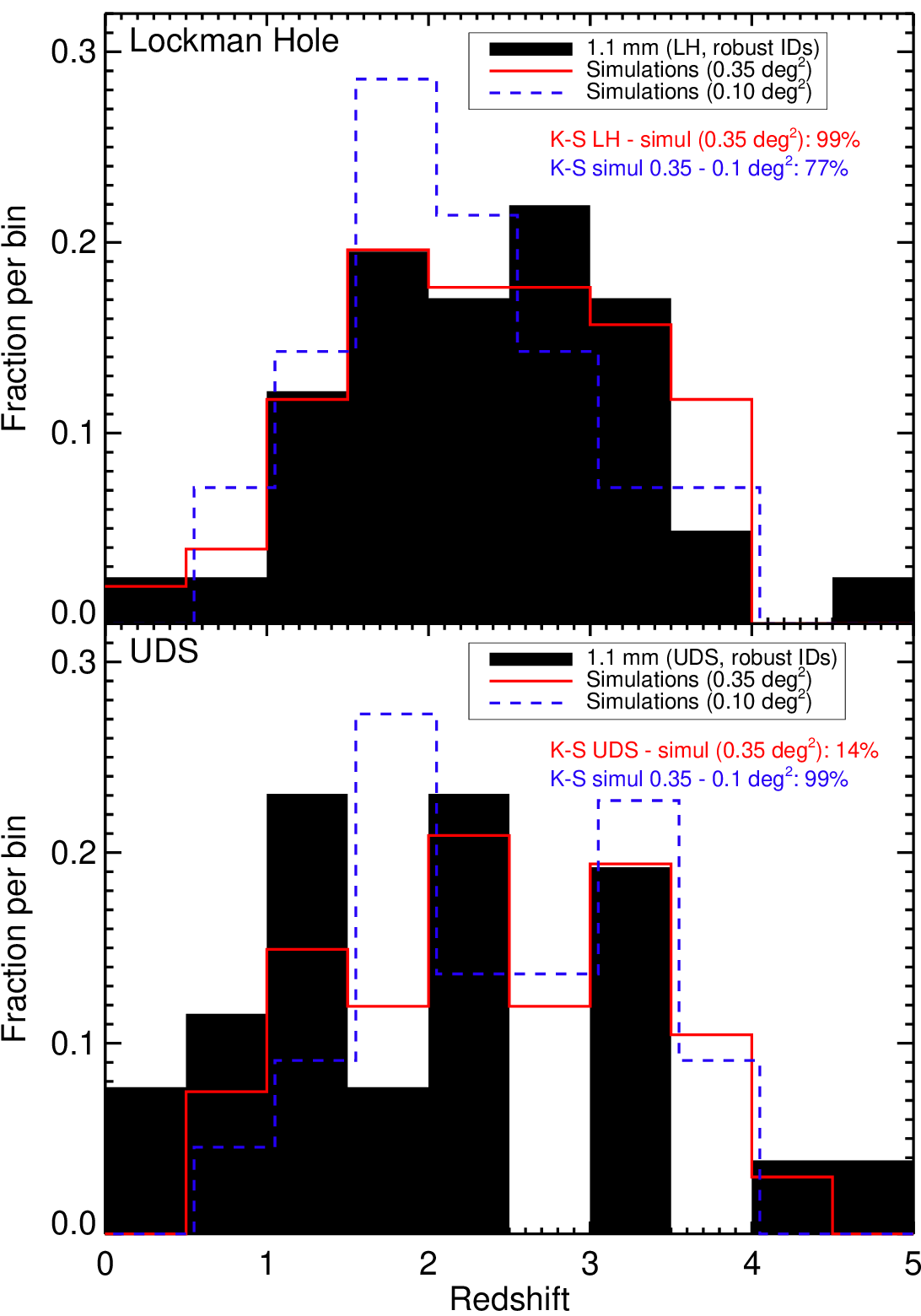}
 \caption{The redshift distribution of $1.1$\,mm-selected galaxies in the Lockman Hole ({\it upper panel}) and the UDS ({\it lower panel}) fields with robust ($p<0.05$) IDs 
and optical--near-infrared or $1.1$\,mm/$1.4$\,GHz flux-density ratio redshifts ({\it solid black histograms}). 
Sources with only  lower limits to their estimated redshifts were excluded.
Also shown in this figure are the redshift distributions of simulated galaxies \citep{croton06} in random $0.35$ (AzTEC-like, {\it red solid lines}) and $0.1$ deg$^{-2}$ (SCUBA-like, {\it blue dotted lines}) fields. For each panel these simulated fields were chosen out of 1000 random fields to best represent the distributions of real SMGs.
The Kolmogorov-Smirnov test probability that the real and simulated galaxies (as well as simulated galaxies in fields with different sizes) in each panel are drawn from the same populations are indicated. 
 }
 \label{fig:zcrot}
\end{figure}

With only photometric redshifts it is difficult to study the large-scale structures traced by SMGs. However, as is evident from Fig.~\ref{fig:z_sep}, 
the redshift distributions of the AzTEC sources in both the Lockman Hole and UDS fields are strikingly similar to that displayed 
by the SCUBA sources in the same fields, 
even though SCUBA covered much smaller sub-regions within these fields 
(see Fig.~\ref{fig:map} and Fig.~\ref{fig:map1}) 
and there are only 13 and 6 sources in 
common in the Lockman Hole and UDS samples respectively.  
Indeed, the K-S test shows that the AzTEC and SCUBA sources in each 
field are fully consistent with being drawn from the same population ($p = 61$\% in the Lockman Hole, and 
$p=50$\% in UDS field). Interestingly this consistency between the redshift distributions of the AzTEC and SCUBA sources {\it within} each field is 
better than the consistency of the redshift distributions {\it between} the two 
SHADES fields (for AzTEC Lockman Hole vs AzTEC UDS $p=16$\%, while for SCUBA Lockman Hole vs SCUBA UDS $p=14$\%). 
This difference starts to become statistically significant if we combine the AzTEC and SCUBA samples in each field (i.e.~ if for each field 
we construct a combined catalogue containing both AzTEC and SCUBA sources in that field), and then repeat the redshift distribution comparison.
This yields a probability that the redshift distributions of the combined (AzTEC+SCUBA) Lockman Hole and UDS samples are drawn from the same population of 
only $2$\% (a statistical difference between the AzTEC source population in the Lockman Hole and 
UDS fields was also noted by \citealt{austermann10}, who showed that the $1.1$\,mm number counts at high fluxes in the Lockman Hole 
are higher than in the UDS field; see their Fig. 8 as well as the top panel of Fig.~\ref{fig:fluxz} in this paper). 

This result suggests that the large-scale structures traced by the 
AzTEC survey are significantly different between the Lockman Hole and UDS fields, 
and the AzTEC and SCUBA SMG samples are tracing {\it the same} large-scale structures within each field.
This latter result implies that these structures extend from the $\simeq0.3$ deg areas sampled by 
SCUBA to at least the $\simeq0.7$ deg scales traced by each AzTEC-SHADES survey field. 
These angular scales correspond to $10 - 20$\,Mpc at $z=1.25$ and $2.25$, 
(the redshifts of the prominent peaks seen in the redshift distribution of the UDS SMGs). 
Alternatively, large-scale structure at lower redshifts may also contribute to the appearance of this effect, 
as  \citet{aretxaga11} showed that the clustering of the foreground galaxies at $z\simeq0.7$  is correlated with the position of bright SMGs in the COSMOS field. 

In order to investigate whether large-scale structure is in fact expected to produce the 
variations we see between the Lockman Hole and UDS fields, we have analysed a 900\,deg$^2$ 
light cone of simulated galaxies at $z=0$--$5$ with $\mbox{SFR}>100\,$M$_\odot$\,yr$^{-1}$
\citep{croton06}\footnote{\url{http://tao.it.swin.edu.au/mock-galaxy-factory/}}, produced in a semi-analytic model 
based on the Millennium simulation \citep{springel05}. Interestingly, to match the AzTEC number counts and produce a 
redshift distribution which peaks at $z\sim2$-$3$ rather than rising monotonically out to $z > 5$, we had to apply a stellar 
mass cut to the simulation to confine our analysis to galaxies with $M_*>10^{11}\,{\rm M_\odot}$. 
Confining our attention to this high-mass regime, we then randomly chose 1000 $0.35$\,deg$^2$ fields from within this light-cone and, 
for each pair of fields, calculated the K-S probability that the galaxy redshift distributions are drawn from the same population.  
We found that $\simeq10$\% of field pairs were inconsistent at a K-S probability level of $<5$\% and that 
$\simeq25$\% of pairs were inconsistent at a probability level $<15$\%. 
Hence, the level of inconsistency we found between the Lockman Hole and the UDS AzTEC source populations is 
broadly as predicted by the simulation, providing some support for the interpretation that it is indeed due to large-scale structure.  

For illustrative purposes, Fig.~\ref{fig:zcrot} shows the redshift distributions of the random fields 
which yielded the highest K-S probability of being consistent with the Lockman Hole and UDS fields ($99$ and $14$\% respectively). 
Thus, the light cones produced using this particular simulation can certainly yield SMG number densities and redshift distributions
consistent with those seen in both AzTEC-SHADES fields. However, it is clear that the SMG redshift distribution in the Lockman Hole 
is much more typical of what is found within the simulations than is the redshift distribution in the UDS field.
Specifically, $\simeq70$\% of the simulated $0.35$\,deg$^2$ AzTEC surveys 
yield a K-S probability $>5$\% that their SMG redshift distribution is consistent with that of the AzTEC sources in the Lockman Hole, 
but this is true for only $\simeq1$\% of the simulated fields when compared to the UDS SMG redshift distribution. Thus, {\it for this particular 
simulation} it would appear the the observed redshift distribution in the UDS is relatively unusual, albeit not unfeasibly so.

Our analysis of this simulation also indicates that it is unsurprising that the redshift distributions of the SHADES-SCUBA and SHADES-AzTEC 
sources within a given field should be in good agreement (Fig.~\ref{fig:z_sep}), as the overlapping 1.1\,mm and 850\,$\mu$m surveys, while differing in size,
wavelength and depth, are tracing the same large scale structures (see Fig.~\ref{fig:zcrot}). Namely, virtually all random $0.35$ deg$^2$ fields have  the K-S probability above $5$\% that their redshift distribution is consistent with that of overlapping $0.1$ deg$^2$ fields.

Finally we note that the fact the two well-separated SHADES fields, despite each covering $\simeq 0.35$\,deg$^2$, still yield significantly different redshift SMG distributions
serves to reinforce the importance of completing SMG surveys 
over several square degrees in order to overcome cosmic variance and obtain a complete and representative view of the number density, redshift distribution and evolution 
of SMGs in the context of the general high-redshift galaxy population.

\section{Conclusions}
\label{sec:conclusion}
We have used the deep optical-to-radio multi-wavelength data in 
the SHADES Lockman East and SXDF/UDS and fields to obtain 
galaxy identifications for $\revone\simeq64$\% ($\simeq80$\% {\revone including tentative identifications)}  of the $148$ AzTEC-SHADES $1.1$\,mm 
sources reported by \citet{austermann10}, exploiting 
deep radio and $24\,\mu$m data complemented 
by methods based on $8\,\mu$m flux-density and red optical-infrared 
($i-K$) colour.
This unusually high identification rate can be attributed to the 
relatively bright millimetre-wavelength flux-density
threshold, combined with the relatively deep supporting multi-frequency data
now available in these two well-studied fields.
We have further exploited the optical--mid-infrared--radio data to 
derive a $\revone\simeq60$\% ($\simeq75$\% {\revone including tentative identifications)} complete 
redshift distribution for the AzTEC-SHADES sources, 
yielding a median redshift of $z\simeq2.2$, with 
a high-redshift tail extending to at least $z \simeq 4$.

Despite the larger area probed by the AzTEC survey relative to the original SCUBA SHADES imaging, the redshift distribution 
of the AzTEC sources is consistent with that displayed by the SCUBA sources, and reinforces tentative evidence that 
the redshift distribution of mm/sub-mm sources in the Lockman Hole field is significantly different from that found in the 
SXDF/UDS field. Comparison with simulated surveys of similar scale extracted from
semi-analytic models based on the Millennium simulation
indicates that this is as expected if the mm/sub-mm sources are massive ($M > 10^{11}\,{\rm M_{\odot}}$)
star-forming galaxies tracing large-scale structures over scales of $10$--$20$\,Mpc. This 
confirms the importance of surveys covering several square degrees 
(as now underway with SCUBA2) to obtain representative samples of bright (sub)mm-selected galaxies.

\section*{Acknowledgments}

We  thank Joanna Baradziej, Helmut Dannerbauer and our anonymous referee for comments and suggestions and Darren Croton and Pratika Dayal for help with numerical simulations.

MJM acknowledges the support of the Science and Technology Facilities Council. JSD acknowledges the support of the Royal Society via a Wolfson Research Merit award, and also the support of the European Research Council via the award of an Advanced Grant. KIC acknowledges the Leverhulme Trust for support through the award of an Early Career Fellowship. 
IRS acknowledges support from STFC and the Leverhulme Trust.
 KSS is supported by the National Radio Astronomy Observatory, which is a facility of the National Science Foundation operated under cooperative agreement by Associated Universities, Inc.
Support for this work was provided in part by NSF grant AST 05-40852 and a grant from the Korea Science \& Engineering Foundation (KOSEF) under a cooperative Astrophysical Research Center of the Structure and Evolution of the Cosmos (ARCSEC).

 This research has made use of the Tool for OPerations on Catalogues And Tables \citep[TOPCAT;][]{topcat}: \url{www.starlink.ac.uk/topcat/ };
 SAOImage DS9, developed by Smithsonian Astrophysical Observatory \citep{ds9}; SExtractor: Software for source extraction \citep{sextractor},
 and NASA's Astrophysics Data System Bibliographic Services.



\appendix

\section{Notes on individual sources}
\label{sec:notes}


{\bf AzLOCK1}: A robust single ID (category 1). The spectroscopic {\revone mid-infrared PAH} redshift from \citet{coppin10} was adopted.

\noindent
{\bf AzLOCK2}: A robust single ID (category 1).

\noindent
{\bf AzLOCK3}: This source has two category 1 radio IDs, but the high-resolution submillimetre imaging 
obtained by \citet{younger08} revealed that the first one (the northern, $24\,\mu$m-faint ID) is the correct ID. This is the SCUBA source LOCK850.02 \citep{ivison07}. 
The same IDs were selected.

\noindent
{\bf AzLOCK4}: Three candidate category 1 radio identifications. The one with the lowest $p$ value, and confirmation at low radio frequency and 24\,$\mu$m was adopted.

\noindent
{\bf AzLOCK5}: A robust single ID (category 1). A possible blend of weak $24\,\mu$m sources. The spectroscopic {\revone mid-infrared PAH} redshift from \citet{coppin10} was adopted.

\noindent
{\bf AzLOCK6}: A robust single radio and 8\,$\mu$m ID (category 1).

\noindent
{\bf AzLOCK7}: A possible blend of two radio/$24\,\mu$m sources and an additional red $i-K$ source (category 1) at the same redshift as the first ID.  
This is the SCUBA source LOCK850.04 \citep{ivison07}. The faintest radio ID considered by \citet{ivison07} is not in the catalogue used here, 
as it is blended with the brighter object. The spectroscopic redshift from \citet{ivison07} was adopted for the second ID, but the first ID is the one adopted here.

\noindent
{\bf AzLOCK8}: A robust single ID (category 1). This is the SCUBA source LOCK850.01 \citep{ivison07}. The same ID was selected. The spectroscopic {\revone mid-infrared PAH} redshift from \citet{coppin10} was adopted.

\noindent
{\bf AzLOCK9}: A robust ID (category 1) with two possible weaker IDs (category 3). This is the SCUBA source LOCK850.34 \citep{ivison07}. 
We did not select their fainter radio counterpart due to its low significance and we also considered  one additional $24\,\mu$m candidate. 

\noindent
{\bf AzLOCK10}: A robust ID (category 1) with a possible weaker ID (category 3). The spectroscopic {\revone mid-infrared PAH} redshift from \citet{coppin10} was adopted for the robust ID.

\noindent
{\bf AzLOCK11}: A robust ID (category 1) with a possible radio companion (category 1).

\noindent
{\bf AzLOCK12}: A robust ID (category 1) with a possible radio companion (category 3).

\noindent
{\bf AzLOCK13}: A robust single ID (category 1).

\noindent
{\bf AzLOCK14}: A robust single ID (category 1).

\noindent
{\bf AzLOCK15}: A robust single ID (category 1).

\noindent
{\bf AzLOCK16}: A robust single ID (category 1) with a possible $8.0\,\mu$m companion (category 3). A possible blend of weak $24\,\mu$m sources.

\noindent
{\bf AzLOCK17}: A robust single ID (category 1). A possible blend of weak $24\,\mu$m sources.  This is the SCUBA source LOCK850.15 \citep{ivison07}. We selected only their second radio IDs as the remaining two are too far away from the AzTEC position.

\noindent
{\bf AzLOCK18}: A robust single ID (category 1). 

\noindent
{\bf AzLOCK19}: A robust ID (category 1) with a possible red $i-K$ companion (category 3). A possible blend of weak $24\,\mu$m sources.

\noindent
{\bf AzLOCK20}: Two robust ID (category 1) with a possible weaker radio companion (category 3). A possible blend of weak $24\,\mu$m sources.

\noindent
{\bf AzLOCK21}: A single ID (category 2). A possible blend of weak $24\,\mu$m sources. This is the SCUBA source LOCK850.13 \citep{ivison07}. We only selected the brighter (southern) out of their two $24\,\mu$m IDs, because of low significance of the other one.

\noindent
{\bf AzLOCK22}: A blend of four radio sources (category 1 and 3). This is the SCUBA source LOCK850.43 \citep{ivison07}.  We selected two out of three of their radio IDs and selected two additional IDs. The inconsistency can be explained by the fact that the radio image is severely blended at this position.

\noindent
{\bf AzLOCK23}: A robust single ID (category 1) with a possible weaker red $i-K$ companion (category 3). 
 
\noindent
{\bf AzLOCK24}: A blend of three radio sources (category 1 and 2). This is the SCUBA source LOCK850.03 \citep{ivison07}. We selected both of their radio IDs and an additional faint radio ID. The spectroscopic redshift from \citet{ivison07} was adopted.

\noindent
{\bf AzLOCK25}: A robust ID (category 1) with a possible radio companion (category 3).

\noindent
{\bf AzLOCK26}: A robust single ID (category 1).

\noindent
{\bf AzLOCK27}: Three robust IDs (category 1). A possible blend of weak $24\,\mu$m sources. This is the SCUBA source LOCK850.71 \citep{ivison07}. The same IDs were selected.

\noindent
{\bf AzLOCK28}: No IDs obtained (but only the radio selection could yield IDs due to lack of coverage at other wavelengths).

\noindent
{\bf AzLOCK29}: A robust single ID (category 1).

\noindent
{\bf AzLOCK30}: No IDs obtained.

\noindent
{\bf AzLOCK31}: No IDs obtained. A possible blend of weak $24\,\mu$m sources.

\noindent
{\bf AzLOCK32}: A robust single ID (category 1). A possible blend of weak $24\,\mu$m sources.

\noindent
{\bf AzLOCK33}: Two robust IDs (category 1). This is the SCUBA source LOCK850.52 \citep{ivison07}. The same ID was selected.

\noindent
{\bf AzLOCK34}: No IDs obtained.

\noindent
{\bf AzLOCK35}:  Three IDs (category 1 and 3).

\noindent
{\bf AzLOCK36}: A single ID (category 2).

\noindent
{\bf AzLOCK37}: Three IDs (category 2 and 3).

\noindent
{\bf AzLOCK38}: A robust single ID (category 1).

\noindent
{\bf AzLOCK39}: Three IDs (category 1 and 3). A possible blend of weak $24\,\mu$m sources.

\noindent
{\bf AzLOCK40}: Three IDs (category 1 and 3).

\noindent
{\bf AzLOCK41}: A single ID (category 2). A possible blend of weak $24\,\mu$m sources.

\noindent
{\bf AzLOCK42}: No IDs obtained. A possible blend of weak $24\,\mu$m sources.

\noindent
{\bf AzLOCK43}: Three  IDs (category 1 and 3). A possible blend of weak $24\,\mu$m sources. This is the SCUBA source LOCK850.79 \citep{ivison07}. We selected the same IDs.

\noindent
{\bf AzLOCK44}: A tentative single ID (category 3). A possible blend of weak $24\,\mu$m sources.

\noindent
{\bf AzLOCK45}: A single ID (category 2). A possible blend of weak $24\,\mu$m sources.

\noindent
{\bf AzLOCK46}: No IDs obtained. A possible blend of weak $24\,\mu$m sources.

\noindent
{\bf AzLOCK47}: A robust single ID (category 1).

\noindent
{\bf AzLOCK48}: No IDs obtained (but the $8.0\,\mu$m and $i-K$ methods could not be used due to lack of coverage at these wavelengths).

\noindent
{\bf AzLOCK49}: No IDs obtained.

\noindent
{\bf AzLOCK50}: A robust single ID (category 1). A possible blend of weak $24\,\mu$m sources.

\noindent
{\bf AzLOCK51}: Three IDs (category 1 and 3).

\noindent
{\bf AzLOCK52}: A robust single ID (category 1).

\noindent
{\bf AzLOCK53}: No IDs obtained.

\noindent
{\bf AzLOCK54}: Four IDs (category 1, 2 and 3).

\noindent
{\bf AzLOCK55}: Three IDs (category 1 and 3).

\noindent
{\bf AzLOCK56}: Three IDs (category 2). This is the SCUBA source LOCK850.06 \citep{ivison07}. The same IDs were selected.

\noindent
{\bf AzLOCK57}: A single ID (category 1).  A possible blend of weak $24\,\mu$m sources.

\noindent
{\bf AzLOCK58}: Two IDs (category 1).

\noindent
{\bf AzLOCK59}: No IDs obtained (but  the $8.0\,\mu$m and $i-K$ methods could not be used due to lack of coverage at these wavelengths).

\noindent
{\bf AzLOCK60}: No IDs obtained. A possible blend of weak $24\,\mu$m sources.

\noindent
{\bf AzLOCK61}: Two IDs (category 1 and 3). A possible blend of weak $24\,\mu$m sources.

\noindent
{\bf AzLOCK62}: A robust single ID (category 1). The spectroscopic {\revone mid-infrared PAH} redshift from \citet{coppin10} was adopted.

\noindent
{\bf AzLOCK63}: Two IDs (category 2 and 3).

\noindent
{\bf AzLOCK64}: A robust single ID (category 1). A possible blend of weak $24\,\mu$m sources.

\noindent
{\bf AzLOCK65}: A tentative single ID (category 3). A possible blend of weak $24\,\mu$m sources.

\noindent
{\bf AzLOCK66}: A robust single ID (category 1). A possible blend of weak $24\,\mu$m sources.

\noindent
{\bf AzLOCK67}: Two IDs (category 1 and 3).

\noindent
{\bf AzLOCK68}: A robust single ID (category 1). A possible blend of weak $24\,\mu$m sources.

\noindent
{\bf AzLOCK69}: A robust single ID (category 1).

\noindent
{\bf AzLOCK70}: No IDs obtained. A possible blend of weak $24\,\mu$m sources.

\noindent
{\bf AzLOCK71}: A tentative single ID (category 3).

\noindent
{\bf AzLOCK72}: A tentative single ID (category 3). A possible blend of weak $24\,\mu$m sources.

\noindent
{\bf AzLOCK73}: A robust single ID (category 1).

\noindent
{\bf AzLOCK74}: Two IDs (category 1 and 3). A possible blend of weak $24\,\mu$m sources.

\noindent
{\bf AzLOCK75}: Two IDs (category 1 and 3). A possible blend of weak $24\,\mu$m sources.

\noindent
{\bf AzLOCK76}: A tentative single ID (category 3). A possible blend of weak $24\,\mu$m sources.

\noindent
{\bf AzLOCK77}: Two IDs (category 1 and 3). A possible blend of weak $24\,\mu$m sources.

\noindent
{\bf AzLOCK78}: A robust single ID (category 1). A possible blend of weak $24\,\mu$m sources.

\noindent
{\bf AzLOCK79}: A tentative single ID (category 3). A possible blend of weak $24\,\mu$m sources.

\noindent
{\bf AzLOCK80}: Two IDs (category 1) at similar redshifts. 

\noindent
{\bf AzLOCK81}: Two IDs (category 3). A possible blend of weak $24\,\mu$m sources.

\noindent
{\bf AzLOCK82}: Five IDs (category 1 and 3).

\noindent
{\bf AzLOCK83}: A robust single ID (category 1). A possible blend of weak $24\,\mu$m sources.

\noindent
{\bf AzLOCK84}: A robust single ID (category 1). A possible blend of weak $24\,\mu$m sources.

\noindent
{\bf AzLOCK85}: Four IDs  (category 1 and 2).

\noindent
{\bf AzLOCK86}: Three IDs (category 1 and 3). A possible blend of weak $24\,\mu$m sources. This is the SCUBA source LOCK850.14 \citep{ivison07}. We only selected the closer out of their two radio IDs, because of low significance of the other one. The spectroscopic redshift from \citet{ivison07} was adopted.

\noindent
{\bf AzLOCK87}: A robust single ID (category 1).

\noindent
{\bf AzLOCK88}: A robust single ID (category 1). A possible blend of weak $24\,\mu$m sources.

\noindent
{\bf AzLOCK89}: Two robust IDs (category 1) at low redshift ($0.14$). A possible blend of weak $24\,\mu$m sources. Its $1.1$\,mm / $1.4$ GHz flux ratio is also consistent with $z\lesssim0.5$.

\noindent
{\bf AzLOCK90}: No IDs obtained.

\noindent
{\bf AzLOCK91}: A robust single ID (category 1).

\mbox{}

\noindent
{\bf AzUDS1}: No IDs obtained (but the $8.0\,\mu$m and $i-K$ methods could not be used due to lack of coverage at these wavelengths).

\noindent
{\bf AzUDS2}: A tentative single ID (category 3).

\noindent
{\bf AzUDS3}: Two robust IDs (category 1).

\noindent
{\bf AzUDS4}: No IDs obtained (but the $i-K$ method could not be used due to lack of coverage at these wavelengths). A possible blend of weak $24\,\mu$m sources.

\noindent
{\bf AzUDS5}: A robust single ID (category 1). This is the SCUBA source SXDF850.03 \citep{ivison07}. The same ID was selected. The optical counterpart and the radio ID may  correspond to a possible lensing galaxy and a lensed source (see Section~\ref{sec:lens}).

\noindent
{\bf AzUDS6}: Two IDs (category 1 and 3).

\noindent
{\bf AzUDS7}:  Two IDs (category 1 and 3).

\noindent
{\bf AzUDS8}: A robust single ID (category 1).

\noindent
{\bf AzUDS9}: No IDs obtained. A possible blend of weak $24\,\mu$m sources.

\noindent
{\bf AzUDS10}: A robust single ID (category 1). A very close star makes the photometry difficult to obtain. This is the SCUBA source SXDF850.29 \citep{ivison07}. The same ID was selected.

\noindent
{\bf AzUDS11}: A tentative single ID (category 3). A possible blend of weak $24\,\mu$m sources.

\noindent
{\bf AzUDS12}: A tentative single ID (category 3).

\noindent
{\bf AzUDS13}: A tentative single ID (category 3).

\noindent
{\bf AzUDS14}: No IDs obtained.

\noindent
{\bf AzUDS15}: Two IDs (category 1 and 3).

\noindent
{\bf AzUDS16}: A single ID (category 2).

\noindent
{\bf AzUDS17}: Two IDs (category 1 and 3).

\noindent
{\bf AzUDS18}: No IDs obtained (but the $8\,\mu$m and $i-K$ methods could not be used due to lack of coverage at these wavelengths).

\noindent
{\bf AzUDS19}: No IDs obtained (but the $i-K$ methods could not be used due to lack of coverage at these wavelengths).

\noindent
{\bf AzUDS20}: No IDs obtained.

\noindent
{\bf AzUDS21}: Two IDs (category 1 and 3).

\noindent
{\bf AzUDS22}: No IDs obtained (but the $i-K$ methods could not be used due to lack of coverage at these wavelengths).

\noindent
{\bf AzUDS23}: A robust  single ID (category 1).

\noindent
{\bf AzUDS24}: No IDs obtained (but the $i-K$ methods could not be used due to lack of coverage at these wavelengths).

\noindent
{\bf AzUDS25}: A robust  single ID (category 1).

\noindent
{\bf AzUDS26}: Two IDs (category 1 and 3).

\noindent
{\bf AzUDS27}: Three IDs (category 1 and 3). A possible blend of weak $24\,\mu$m sources.

\noindent
{\bf AzUDS28}: Two IDs (category 1 and 3) at similar redshifts. A possible blend of weak $24\,\mu$m sources.

\noindent
{\bf AzUDS29}: Two IDs (category 1 and 3). A possible blend of weak $24\,\mu$m sources.

\noindent
{\bf AzUDS30}: Two IDs (category 1 and 3).  A possible blend of weak $24\,\mu$m sources.

\noindent
{\bf AzUDS31}: No IDs obtained.

\noindent
{\bf AzUDS32}: A robust  single ID (category 1).

\noindent
{\bf AzUDS33}: Two IDs (category 1 and 3). This is the SCUBA source SXDF850.01 \citep{ivison07}. We selected an additional category 3 ID.

\noindent
{\bf AzUDS34}: A robust  single ID (category 1).

\noindent
{\bf AzUDS35}: A robust  single ID (category 1).

\noindent
{\bf AzUDS36}: A single ID (category 3).

\noindent
{\bf AzUDS37}: No IDs obtained.


\noindent
{\bf AzUDS38}: Five IDs (category 1 and 3). A possible blend of weak $24\,\mu$m sources.

\noindent
{\bf AzUDS39}: No IDs obtained. A possible blend of weak $24\,\mu$m sources.

\noindent
{\bf AzUDS40}: Two IDs (category 1 and 3).

\noindent
{\bf AzUDS41}:  No IDs obtained.

\noindent
{\bf AzUDS42}:  Five IDs (category 1 and 3). A possible blend of weak $24\,\mu$m sources.

\noindent
{\bf AzUDS43}: Two robust  IDs (category 1). A possible blend of weak $24\,\mu$m sources. The first and second IDs may  correspond to a possible lensing galaxy and a lensed source (see Section~\ref{sec:lens}). The redshift of the second ID was used in the analysis of the redshift distribution.

\noindent
{\bf AzUDS44}: No IDs obtained.

\noindent
{\bf AzUDS45}: A robust  single ID (category 1). A possible blend of weak $24\,\mu$m sources.

\noindent
{\bf AzUDS46}: No IDs obtained.

\noindent
{\bf AzUDS47}:   Five IDs (category 1). The high-resolution submillimetre imaging revealed that the second one (the most northern with $z\simeq2.48$) is the correct ID \citep{hatsukade10}. This is the SCUBA source SXDF850.06 \citep{ivison07}. The same radio IDs were selected.

\noindent
{\bf AzUDS48}: A robust  single ID (category 1). A possible blend of weak $24\,\mu$m sources.

\noindent
{\bf AzUDS49}: A robust  single ID (category 1).

\noindent
{\bf AzUDS50}: Two IDs (category 1 and 3). A possible blend of weak $24\,\mu$m sources.

\noindent
{\bf AzUDS51}: No IDs obtained. This is the SCUBA source SXDF850.5 \citep{ivison07}. Their $3\sigma$ radio ID is not selected as it is too far away from the AzTEC position, which is offset from the SCUBA position by $\simeq5$ arcsec.

\noindent
{\bf AzUDS52}: Four tentative IDs (category 3).

\noindent
{\bf AzUDS53}: A tentative ID (category 3).

\noindent
{\bf AzUDS54}: A robust  single ID (category 1).

\noindent
{\bf AzUDS55}: A robust  single ID (category 1). A possible blend of weak $24\,\mu$m sources.

\noindent
{\bf AzUDS56}: No IDs obtained. A possible blend of weak $24\,\mu$m sources. This is the SCUBA source SXDF850.18 \citep{ivison07}. Their radio ID was not selected due to its low significance.

\noindent
{\bf AzUDS57}: A tentative single ID (category 3).

\section[]{Figures and Tables}

\voffset=0in

\onecolumn
\begin{landscape}
\begin{scriptsize}
\begin{longtable}{lccccccccccccccccc}
\caption{Reliable and tentative radio, $24\,\mu$m, $8.0\,\mu$m and $i-K$ identifications in the Lockman Hole field. $S_{1.1\/\micron}$ is deboosted millimetre flux from \citet{austermann10}. The coordinates and the distances to the AzTEC sources are that of the $1.4$ GHz ID if present and $24\,\mu$m (or $0.61$ GHz) otherwise. The probabilities of a counterpart to be a chance superposition, $p$, are bold if $p<0.05$ and italic if $0.05<p<0.1$. In the last column the photometric redshifts based on optcial-near-IR photometry is given. Optical spectroscopic redshift from \citet{ivison07} are shown with no errors and with three significant digits. Mid-IR spectroscopic redshift from \citet{coppin10} are shown with symetric errors. Italicized redshift were calculated from the 1.1 mm / 1.4 GHz flux ratio using the average SMG SED from \citet{michalowski10smg}. Bold redshifts indicate an ID used for redshift histogram analysis. N/A is given if an object is not covered by the IRAC map (column $S_{8.0}$ and $p_{8.0}$); or by the $i$- or $K$-band map (column $i-K$ and $p_{i-K}$). Moreover, N/A is given for $p$-values if $S_{8.0}<20\,\mu$Jy, or $i-K<2$ or the object is not detected at the $K$-band as such objects were not taken into account for the calculations of $p$-values. Dots represent a non-detection at $8.0\,\mu$m (column $S_{8.0}$); a nondetection at both $i$- and $K$-band (column $i-K$).}
\label{tab:AzLOCKgoodID}\\
\hline\hline
No. & $S_{1.1\/\micron}$ & RA$_{\rm ID}$ & DEC$_{\rm ID}$ & Dist & $S_{1.4}$ & $p_{1.4}$ & $S_{0.6}$ & $p_{0.6}$ & $S_{24}$ & $p_{24}$ & $S_{8.0}$ & $p_{8.0}$ & $K$ & $i-K$ & $p_{i-K}$ & $z$ \\
 & (mJy) & (deg) & (deg) & ($''$) & ($\mu$Jy) & & ($\mu$Jy) & & ($\mu$Jy) & & ($\mu$Jy) & & (mag) & (mag) & & \\
\hline
\endfirsthead
\caption{continued.}\\
\hline\hline
No. & $S_{1.1\/\micron}$ & RA$_{\rm ID}$ & DEC$_{\rm ID}$ & Dist & $S_{1.4}$ & $p_{1.4}$ & $S_{0.6}$ & $p_{0.6}$ & $S_{24}$ & $p_{24}$ & $S_{8.0}$ & $p_{8.0}$ & $K$ & $i-K$ & $p_{i-K}$ & $z$ \\
 & (mJy) & (deg) & (deg) & ($''$) & ($\mu$Jy) & & ($\mu$Jy) & & ($\mu$Jy) & & ($\mu$Jy) & & (mag) & (mag) & & \\
\hline
\endhead
\hline
\endfoot
\hline
\multicolumn{16}{l}{$^a$The optical photometric redshift could not been measured because this object is outside of the optical map (or close to a bright star)}
\endlastfoot
       1 & $6.63^{+0.85}_{-0.97}$ & 163.008040 & 57.681007 & 2.3 & $258 \pm 11$ & {\bf 0.002 } & $596 \pm 24$ & {\bf 0.001 } & $1642 \pm 11$ & {\bf 0.002 } & 215 $\pm$ 19 & {\bf 0.009 } & N/A & N/A & N/A & {\boldmath $2.50\pm0.02$}\\
       2 & $6.43^{+0.90}_{-0.89}$ & 163.022900 & 57.607548 & 6.5 & $66 \pm 8$ & {\bf 0.038 } & $\cdots$ & $\cdots$ & $235 \pm 6$ & {\it 0.070 } & 31 $\pm$ 3 & 0.136 & 21.7 & 3.3 & {\bf 0.035 } & {\boldmath $2.25^{+0.13}_{-0.13}$}\\
       3 & $6.21^{+1.05}_{-0.92}$ & 163.237710 & 57.350926 & 2.8 & $67 \pm 7$ & {\bf 0.011 } & $149 \pm 16$ & {\bf 0.004 } & $636 \pm 13$ & {\bf 0.008 } & 28 $\pm$ 2 & {\it 0.069 } & 23.2 & $<$2.0 & $\cdots$ & $2.58^{+0.12}_{-0.23}$\\
$\cdots$ & $\cdots$ & 163.237530 & 57.352273 & 2.6 & $36 \pm 7$ & {\bf 0.019 } & $\cdots$ & $\cdots$ & $\cdots$ & $\cdots$ & 30 $\pm$ 3 & {\bf 0.045 } & 24.1 & $<$2.9 & $\cdots$ & {\boldmath $3.07^{+0.11}_{-0.64}$}\\
       4 & $5.27^{+0.91}_{-0.97}$ & 162.686880 & 57.556959 & 7.4 & $150 \pm 12$ & {\bf 0.021 } & $121 \pm 17$ & {\bf 0.045 } & $\cdots$ & $\cdots$ & N/A & N/A & N/A & N/A & N/A & $2.43^{+0.10}_{-0.08}$\\
$\cdots$ & $\cdots$ & 162.686160 & 57.554737 & 2.1 & $56 \pm 11$ & {\bf 0.009 } & $99 \pm 17$ & {\bf 0.025 } & $284 \pm 13$ & {\bf 0.013 } & N/A & N/A & N/A & N/A & N/A & {\boldmath $1.39^{+0.58}_{-0.44}$}\\
$\cdots$ & $\cdots$ & 162.682660 & 57.555667 & 5.5 & $43 \pm 11$ & {\bf 0.044 } & $\cdots$ & $\cdots$ & $\cdots$ & $\cdots$ & N/A & N/A & N/A & N/A & N/A & $\it 3.0\pm0.5$\\
       5 & $4.91^{+1.02}_{-0.90}$ & 163.515860 & 57.431498 & 0.5 & $138 \pm 9$ & {\bf 0.000 } & $277 \pm 19$ & {\bf 0.000 } & $232 \pm 6$ & {\bf 0.000 } & 31 $\pm$ 3 & {\bf 0.002 } & 23.9 & $<$2.9 & $\cdots$ & {\boldmath $2.82\pm0.02$}\\
\hline
\end{longtable}

\end{scriptsize}
\newpage
\begin{longtable}{lccccccccccccccc}
\caption{Reliable and tentative radio, $24\,\mu$m, $8.0\,\mu$m and $i-K$ identifications in the UDS field. Columns are described in Table~\ref{tab:AzLOCKgoodID}}
\label{tab:AzSXDFgoodID}\\
\hline\hline
No. & $S_{1.1\/\micron}$ & RA$_{\rm ID}$ & DEC$_{\rm ID}$ & Dist & $S_{1.4}$ & $p_{1.4}$ & $S_{24}$ & $p_{24}$ & $S_{8.0}$ & $p_{8.0}$ & $K$ & $i-K$ & $p_{i-K}$ & $z$ \\
 & (mJy) & (deg) & (deg) & ($''$) & ($\mu$Jy) & & ($\mu$Jy) & & ($\mu$Jy) & & (mag) & (mag) & & \\
\hline
\endfirsthead
\caption{continued.}\\
\hline\hline
No. & $S_{1.1\/\micron}$ & RA$_{\rm ID}$ & DEC$_{\rm ID}$ & Dist & $S_{1.4}$ & $p_{1.4}$ & $S_{24}$ & $p_{24}$ & $S_{8.0}$ & $p_{8.0}$ & $K$ & $i-K$ & $p_{i-K}$ & $z$ \\
 & (mJy) & (deg) & (deg) & ($''$) & ($\mu$Jy) & & ($\mu$Jy) & & ($\mu$Jy) & & (mag) & (mag) & & \\
\hline
\endhead
\hline
\endfoot
\hline
\multicolumn{16}{l}{$^a$The optical photometric redshift could not been measured because this object is outside of the optical map (or close to a bright star)}
\endlastfoot
       1 & $5.29^{+1.40}_{-1.68}$ & $\cdots$ & $\cdots$ & $\cdots$ & $\cdots$ & $\cdots$ & $\cdots$ & $\cdots$ & N/A & N/A & N/A & N/A & N/A & {\boldmath $\it >2.9$}$^a$\\
       2 & $3.99^{+1.10}_{-1.34}$ & 34.442284 & -4.796439 & 5.7 & $\cdots$ & $\cdots$ & $77 \pm 7$ & 0.199 & 21 $\pm$ 2 & 0.162 & 22.8 & $>$3.6 & {\it 0.084 } & {\boldmath $2.53^{+0.47}_{-0.28}$}\\
       3 & $3.75^{+1.22}_{-1.18}$ & 34.479333 & -4.789667 & 1.5 & $113 \pm 20$ & {\bf 0.002 } & $776 \pm 11$ & {\bf 0.002 } & 78 $\pm$ 7 & {\bf 0.014 } & 21.6 & 3.8 & {\bf 0.005 } & {\boldmath $1.81^{+0.04}_{-0.06}$}\\
$\cdots$ & $\cdots$ & 34.479412 & -4.789859 & 1.4 & $\cdots$ & $\cdots$ & $\cdots$ & $\cdots$ & $\cdots$ & $\cdots$ & 21.6 & 4.6 & {\bf 0.003 } & $3.13^{+0.07}_{-0.10}$\\
       4 & $4.39^{+1.67}_{-1.58}$ & $\cdots$ & $\cdots$ & $\cdots$ & $\cdots$ & $\cdots$ & $\cdots$ & $\cdots$ & $\cdots$ & $\cdots$ & N/A & N/A & N/A & {\boldmath $\it >2.7$}$^a$\\
       5 & $3.57^{+1.19}_{-1.21}$ & 34.425542 & -4.941028 & 1.0 & $89 \pm 17$ & {\bf 0.002 } & $\cdots$ & $\cdots$ & 40 $\pm$ 4 & {\bf 0.022 } & 19.3 & 1.3 & N/A & {\boldmath $0.45^{+0.05}_{-0.05}$}\\
       6 & $3.99^{+1.27}_{-1.57}$ & 34.677833 & -4.992167 & 4.3 & $105 \pm 16$ & {\bf 0.014 } & $924 \pm 37$ & {\bf 0.009 } & 77 $\pm$ 7 & {\bf 0.047 } & 21.2 & 3.6 & {\bf 0.011 } & {\boldmath $2.29^{+0.09}_{-0.07}$}\\
$\cdots$ & $\cdots$ & 34.677021 & -4.991021 & 5.2 & $\cdots$ & $\cdots$ & $\cdots$ & $\cdots$ & 73 $\pm$ 6 & {\it 0.075 } & 19.8 & 1.4 & N/A & $0.45^{+0.08}_{-0.05}$\\
\hline
\end{longtable}

\end{landscape}


\voffset=-0.8in

\twocolumn


\begin{figure*}
\begin{center}
\includegraphics[width=0.86\textwidth,clip]{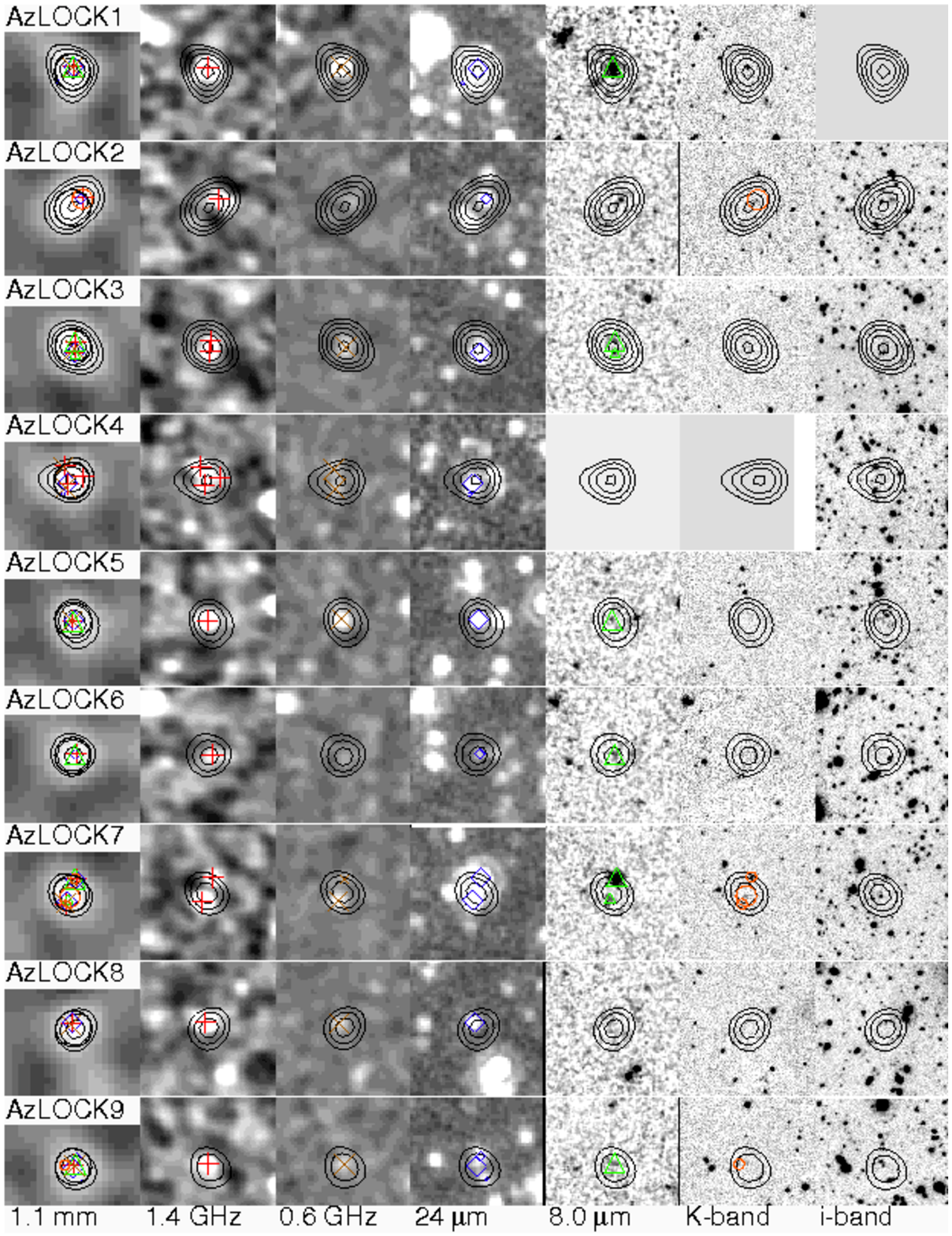}
\end{center}
\caption{Thumbnail images of AzTEC sources in the Lockman Hole. Each panel is $60$ arcsec on a side and centred on the AzTEC position. From left to right: $1.1$\,mm,  $1.4$ GHz, $0.61$ GHz, $24\,\mu$m,  $8.0\,\mu$m, $K$-band and $i$-band. The IDs are marked on the relevant images: {\it red pluses}: $1.4$ GHz IDs, {\it blue diamonds}: $24\,\mu$m IDs, {\it brown crosses}: $0.61$ GHz IDs, {\it green triangles}: the $8.0\,\mu$m IDs, {\it orange circles}: the $i-K>2$.  {\it Big symbols}: reliable IDs ($p<0.05$), {\it medium symbols}: tentative IDs ($0.05<P<0.1$), {\it small symbols}: bad IDs ($p>0.1$). On the $1.1$\,mm images all the ID symbols as well as a {\it thick circle} corresponding to the search radius (Sec.~\ref{sec:idmeth}) are shown. The contours represent $1.1$ mm flux and start at $3$ mJy with a $1$ mJy increment.
The rest of the images is available in the electronic edition of the Journal.}
\label{fig:AzLOCKthumb}
\end{figure*}



\addtocounter{figure}{-1}
\begin{figure*}
\begin{center}
\includegraphics[width=0.95\textwidth,clip]{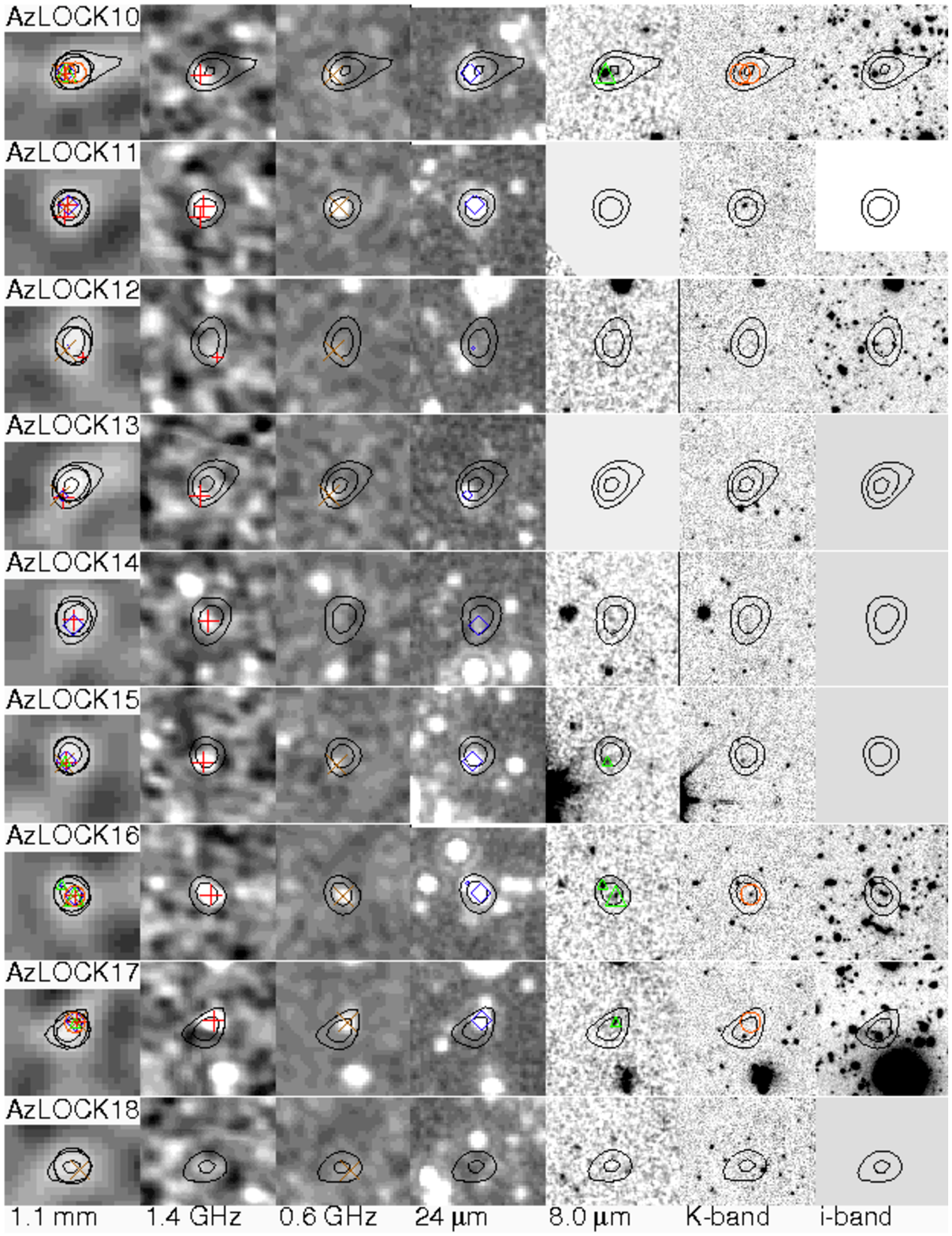}
\end{center}
\caption{(continued).}
\end{figure*}


\addtocounter{figure}{-1}
\begin{figure*}
\begin{center}
\includegraphics[width=0.95\textwidth,clip]{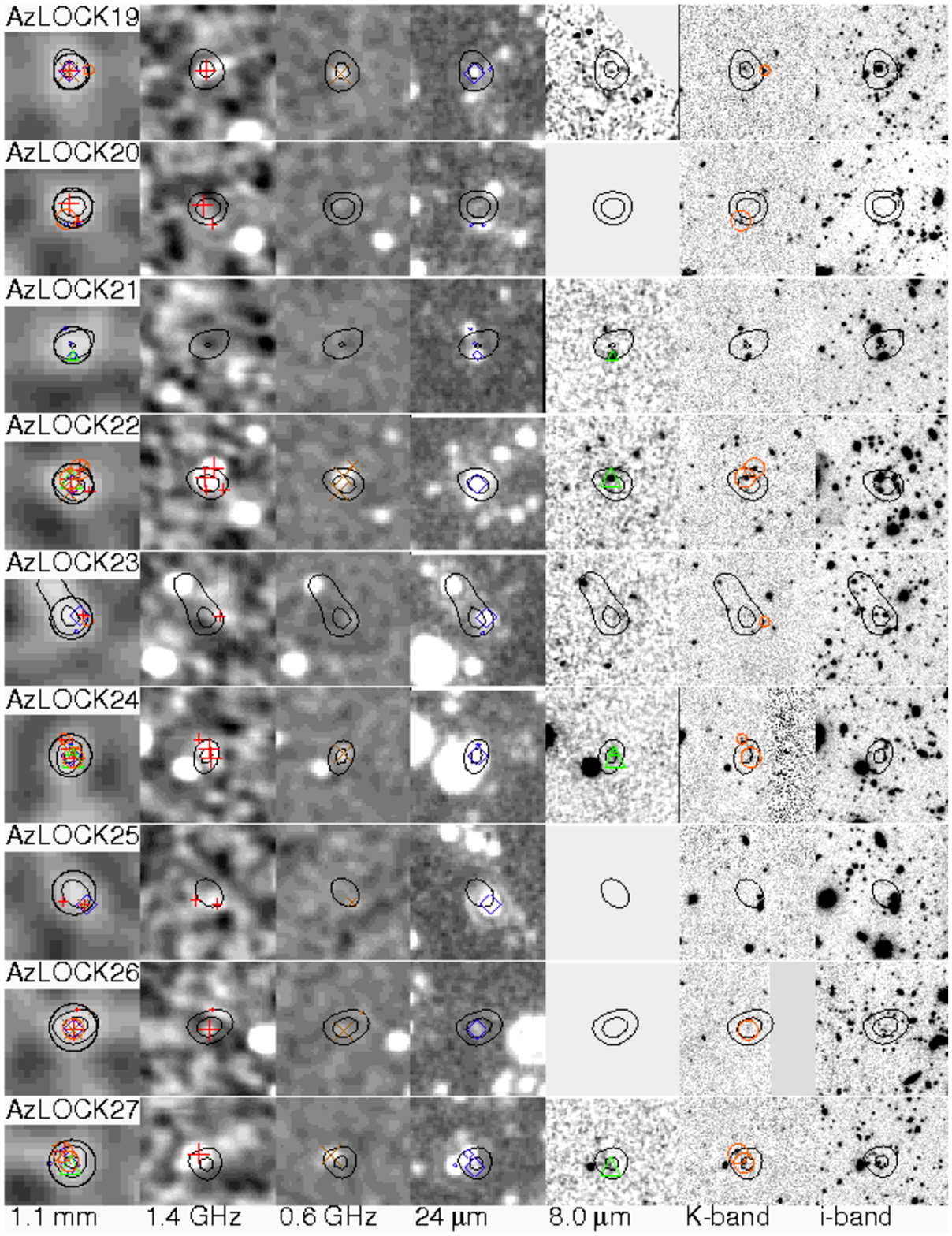}
\end{center}
\caption{(continued).}
\end{figure*}

\addtocounter{figure}{-1}
\begin{figure*}
\begin{center}
\includegraphics[width=0.95\textwidth,clip]{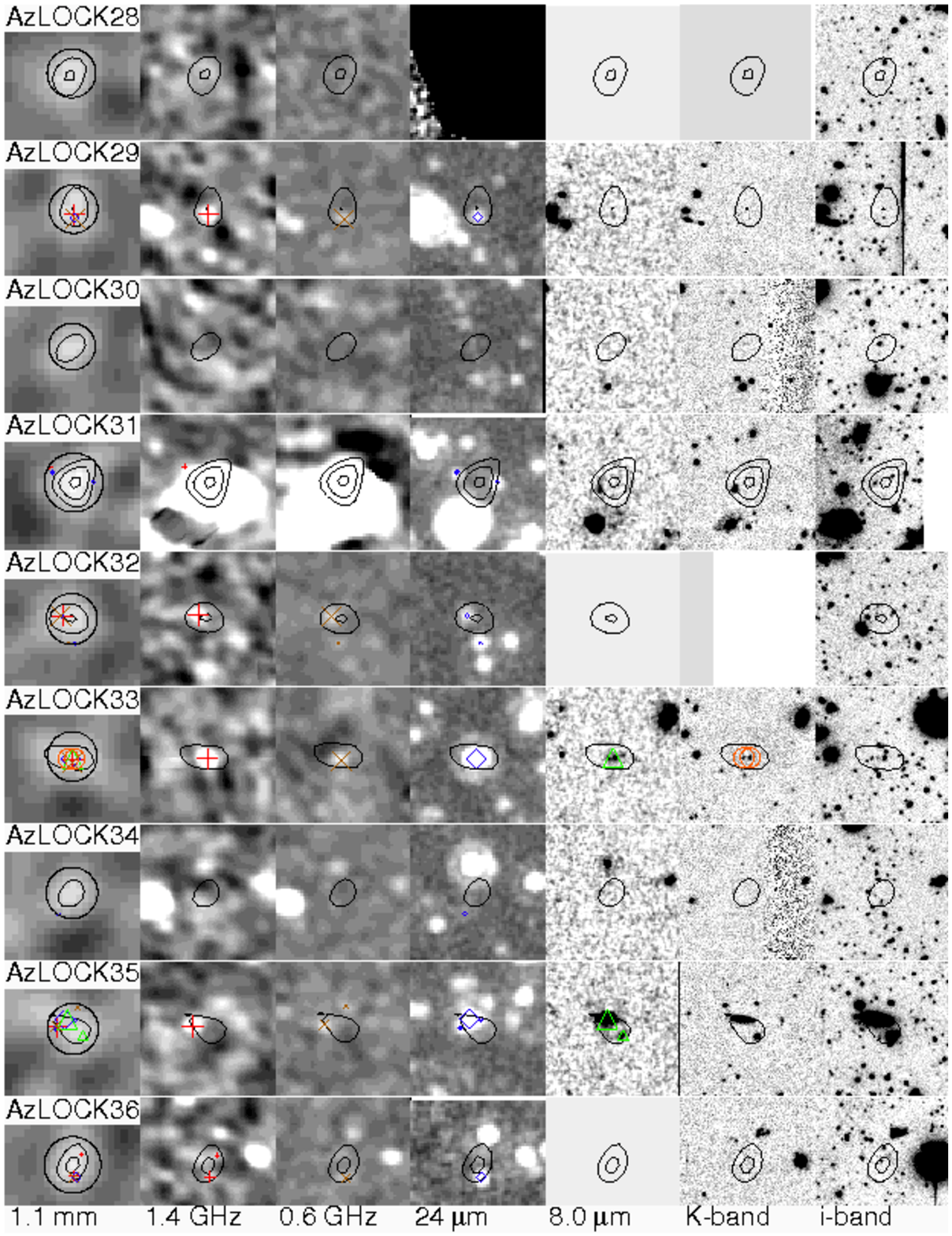}
\end{center}
\caption{(continued).}
\end{figure*}

\addtocounter{figure}{-1}
\begin{figure*}
\begin{center}
\includegraphics[width=0.95\textwidth,clip]{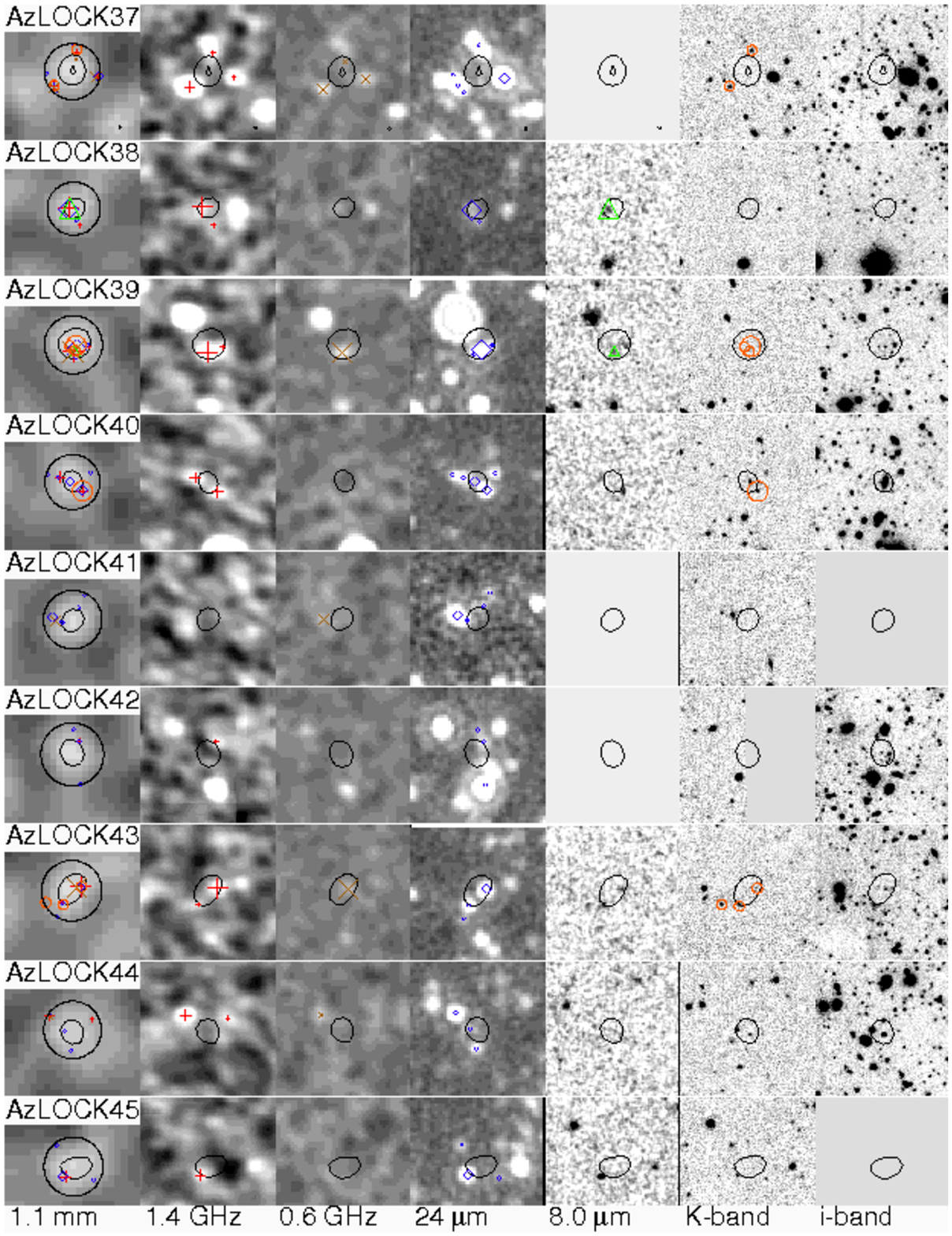}
\end{center}
\caption{(continued).}
\end{figure*}

\addtocounter{figure}{-1}
\begin{figure*}
\begin{center}
\includegraphics[width=0.95\textwidth,clip]{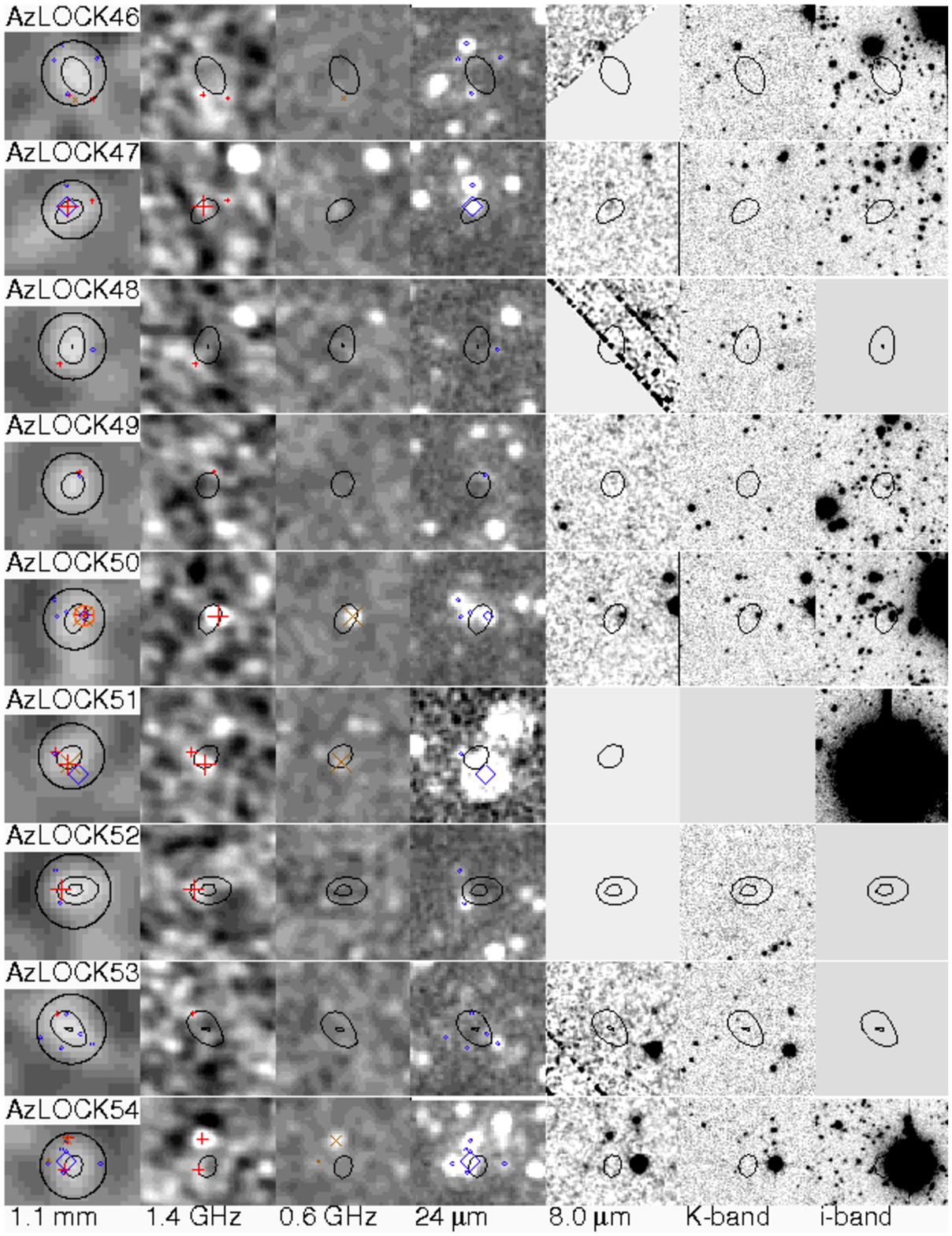}
\end{center}
\caption{(continued).}
\end{figure*}

\addtocounter{figure}{-1}
\begin{figure*}
\begin{center}
\includegraphics[width=0.95\textwidth,clip]{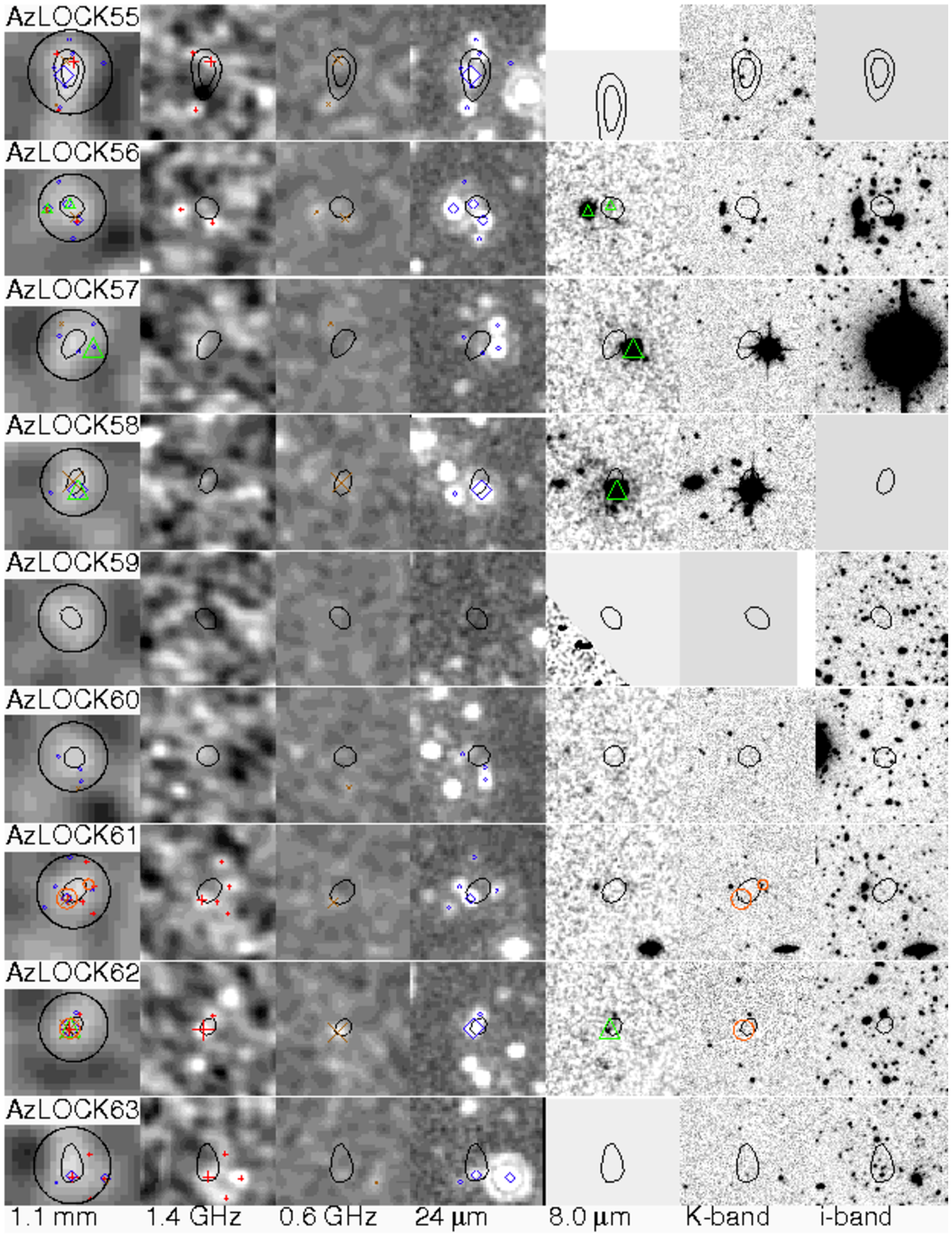}
\end{center}
\caption{(continued).}
\end{figure*}

\addtocounter{figure}{-1}
\begin{figure*}
\begin{center}
\includegraphics[width=0.95\textwidth,clip]{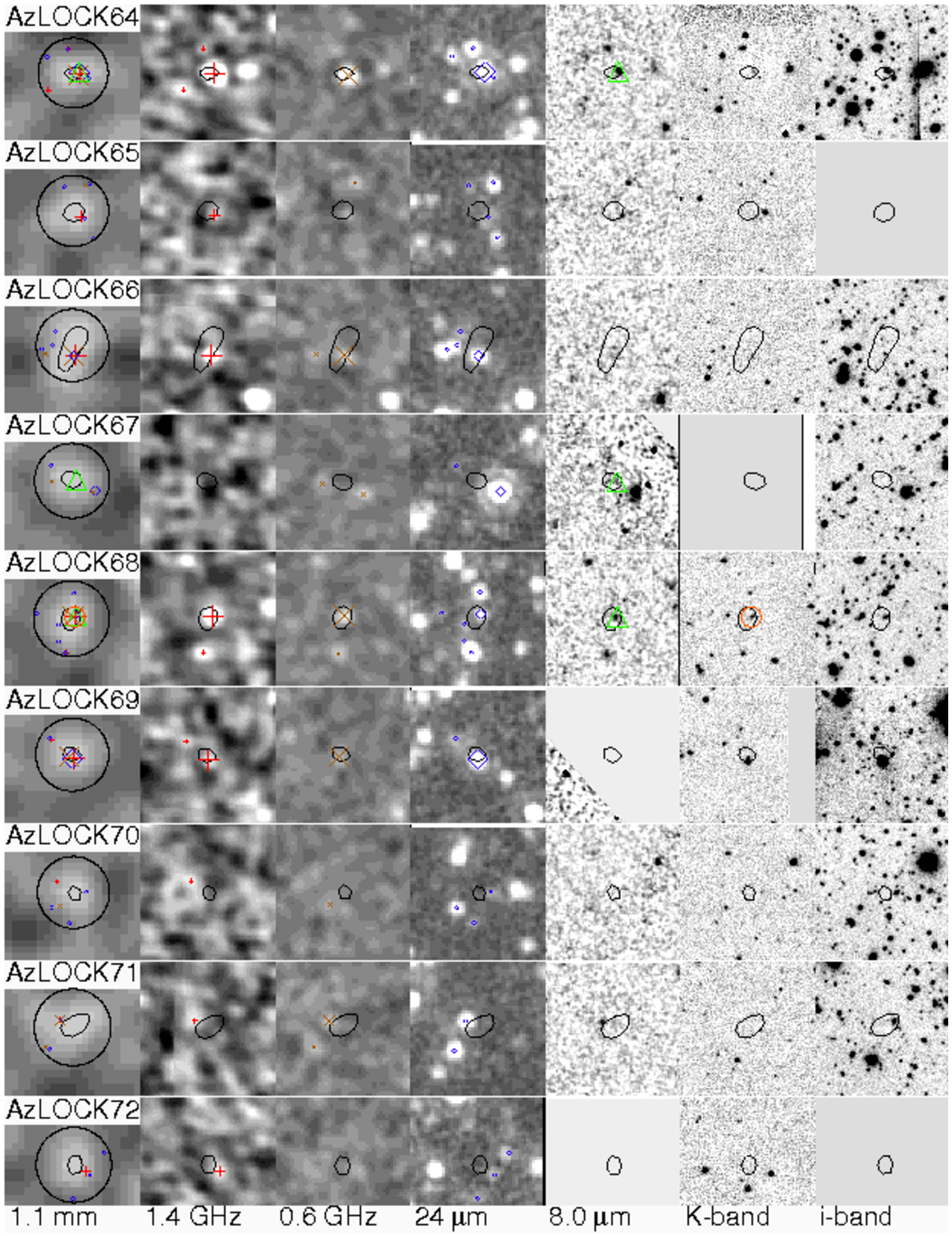}
\end{center}
\caption{(continued).}
\end{figure*}

\addtocounter{figure}{-1}
\begin{figure*}
\begin{center}
\includegraphics[width=0.95\textwidth,clip]{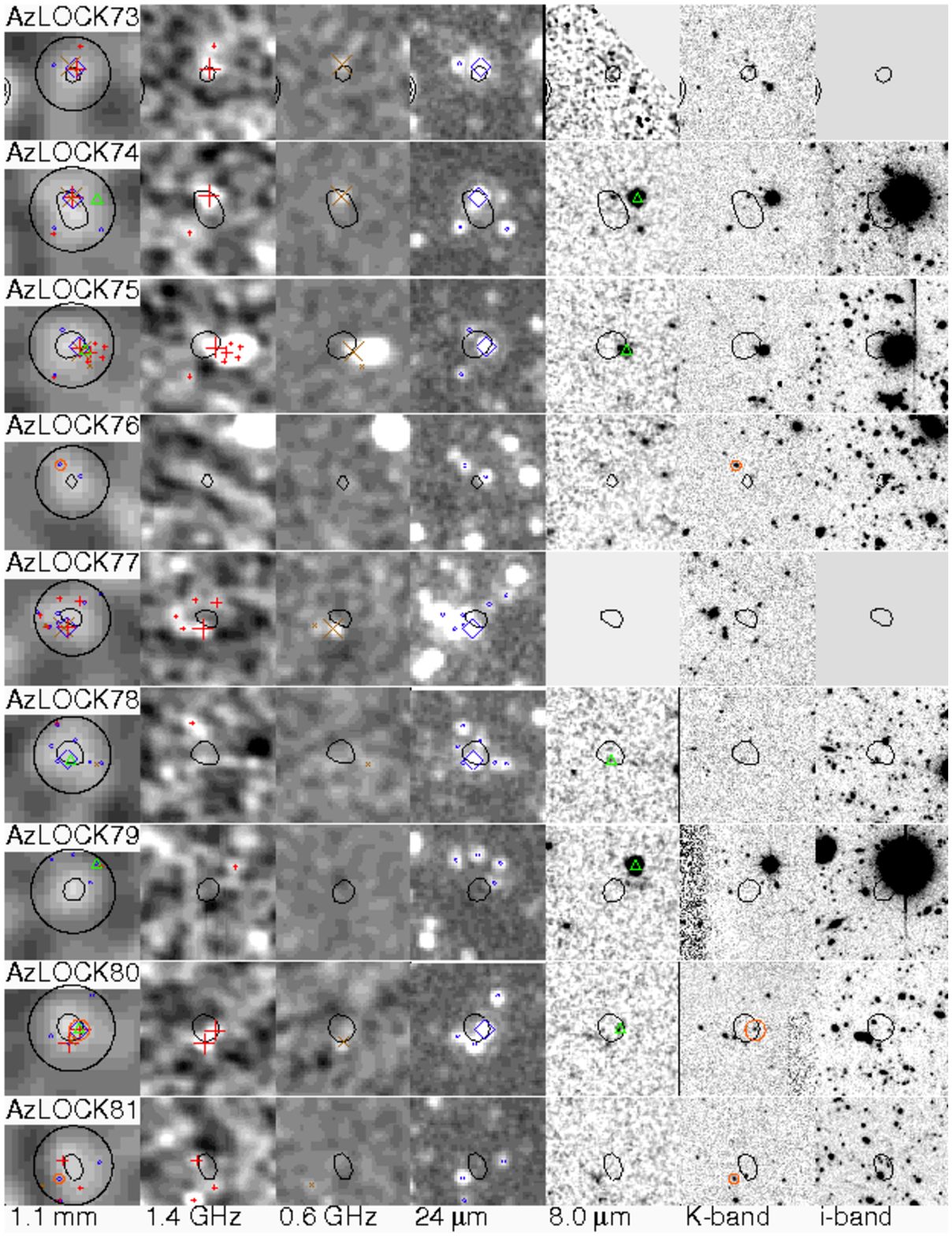}
\end{center}
\caption{(continued).}
\end{figure*}

\addtocounter{figure}{-1}
\begin{figure*}
\begin{center}
\includegraphics[width=0.95\textwidth,clip]{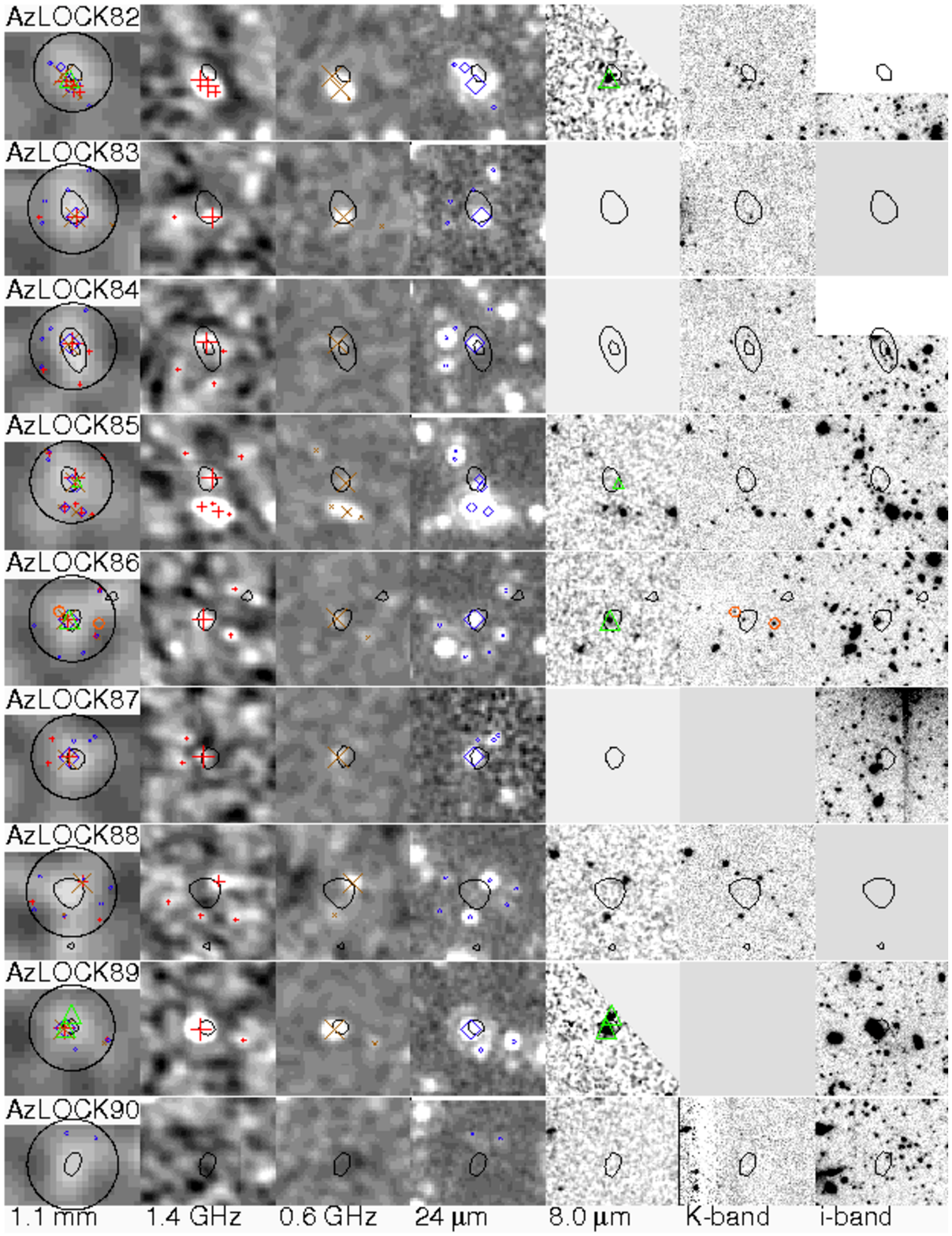}
\end{center}
\caption{(continued).}
\end{figure*}

\addtocounter{figure}{-1}
\begin{figure*}
\begin{center}
\includegraphics[width=0.95\textwidth,clip]{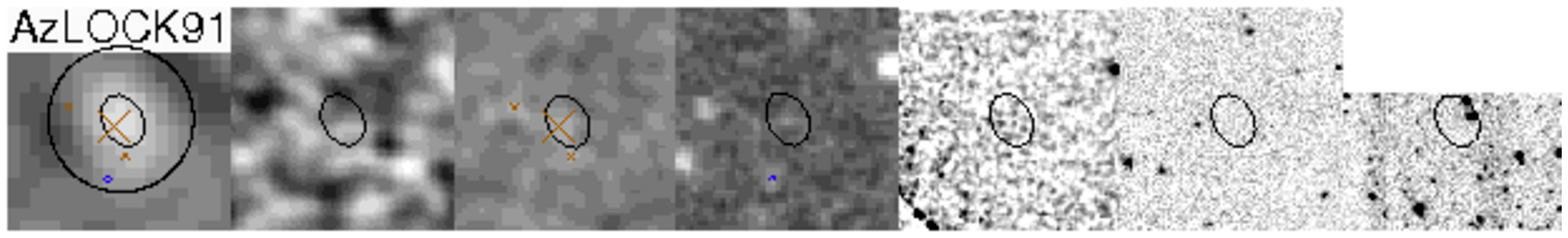}
\end{center}
\caption{(continued).}
\end{figure*}



\begin{figure*}
\begin{center}
\includegraphics[width=0.93\textwidth,clip]{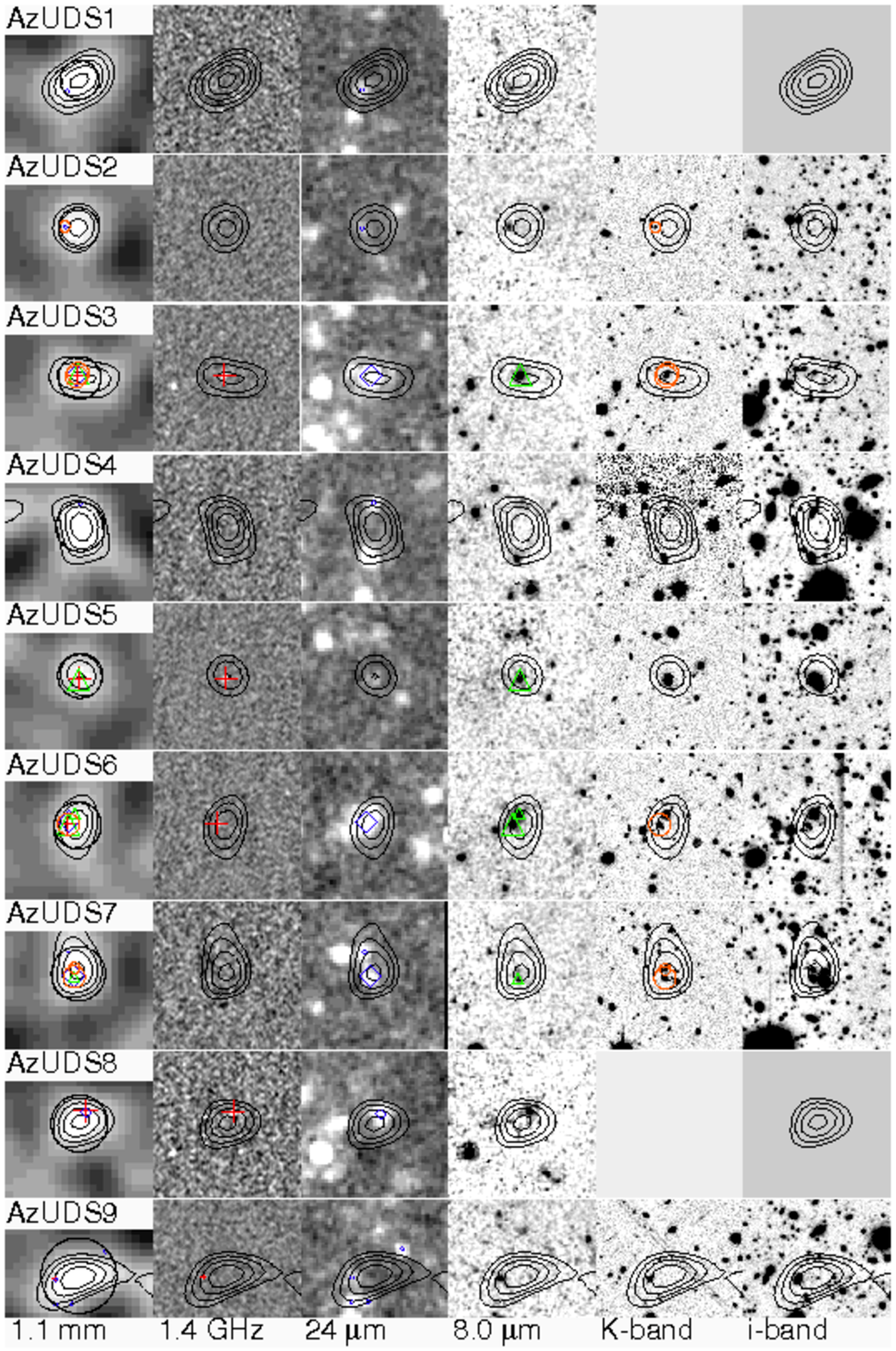}
\end{center}
\caption{Thumbnail images of AzTEC sources in the UDS field. Symbols are the same as in Fig.~\ref{fig:AzLOCKthumb}. 
The rest of the images is available in the electronic edition of the Journal.
}
\label{fig:AzSXDFthumb}
\end{figure*}


\addtocounter{figure}{-1}
\begin{figure*}
\begin{center}
\includegraphics[width=0.95\textwidth,clip]{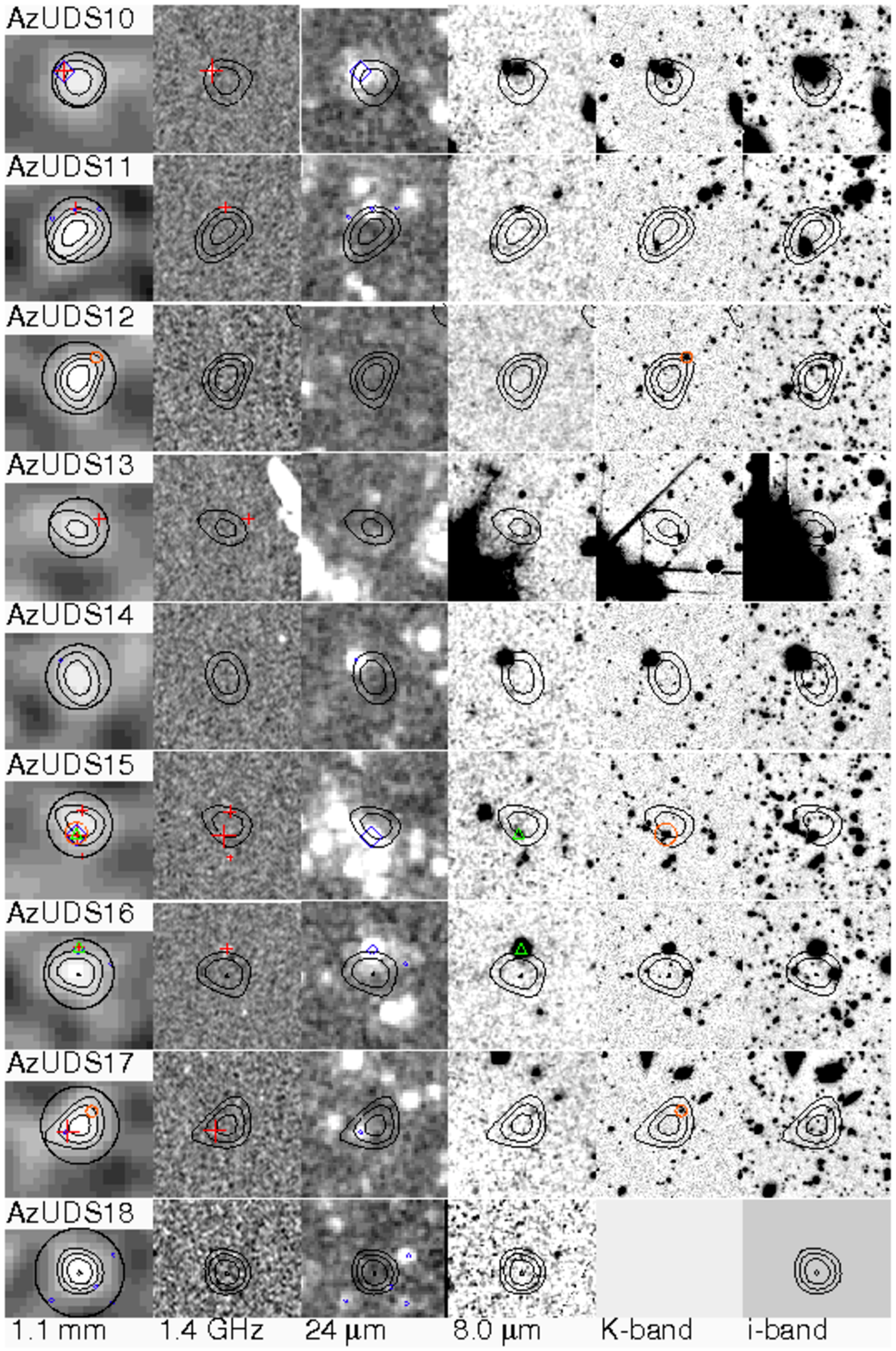}
\end{center}
\caption{(continued).}
\end{figure*}

\addtocounter{figure}{-1}
\begin{figure*}
\begin{center}
\includegraphics[width=0.95\textwidth,clip]{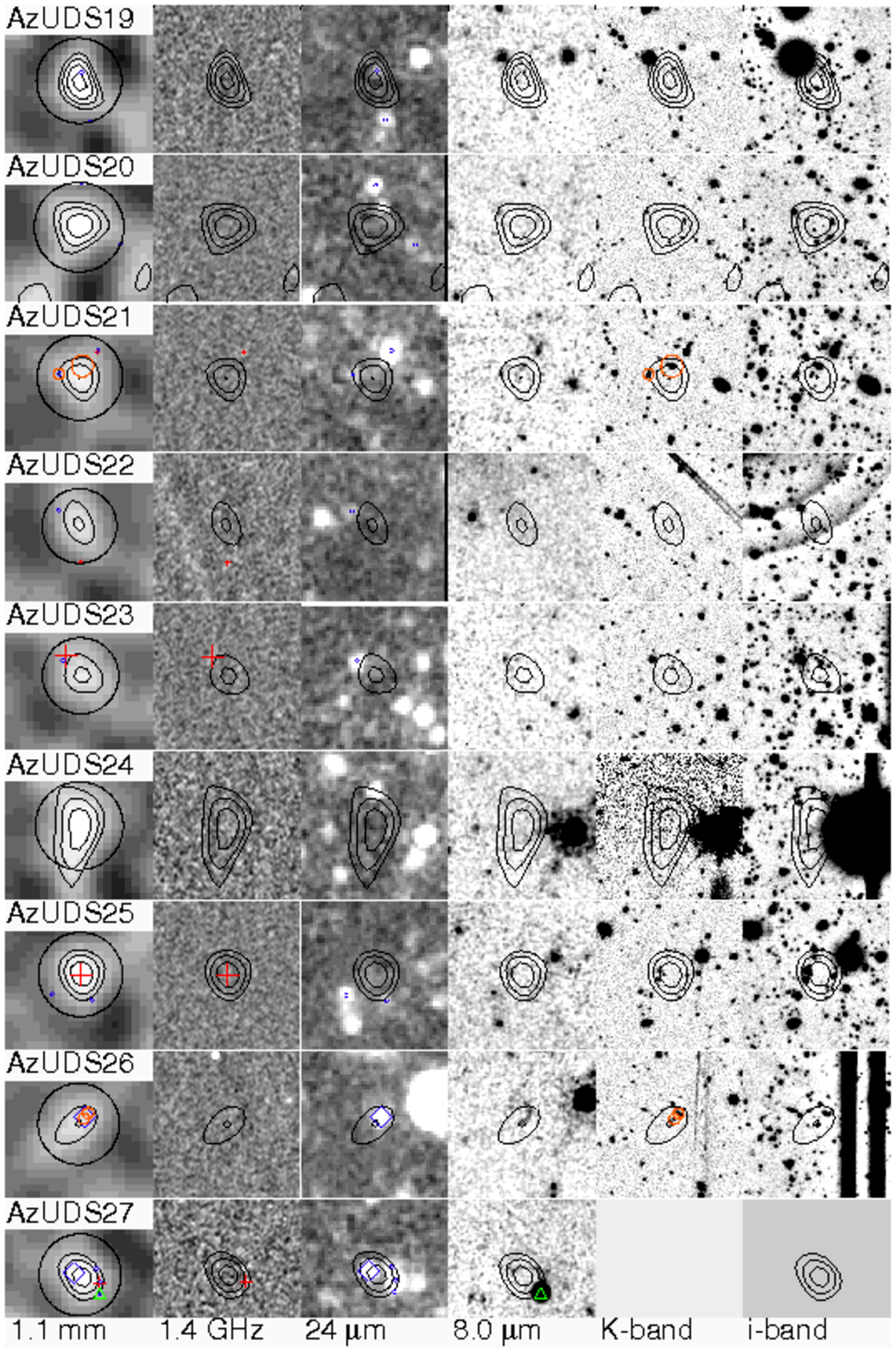}
\end{center}
\caption{(continued).}
\end{figure*}

\addtocounter{figure}{-1}
\begin{figure*}
\begin{center}
\includegraphics[width=0.95\textwidth,clip]{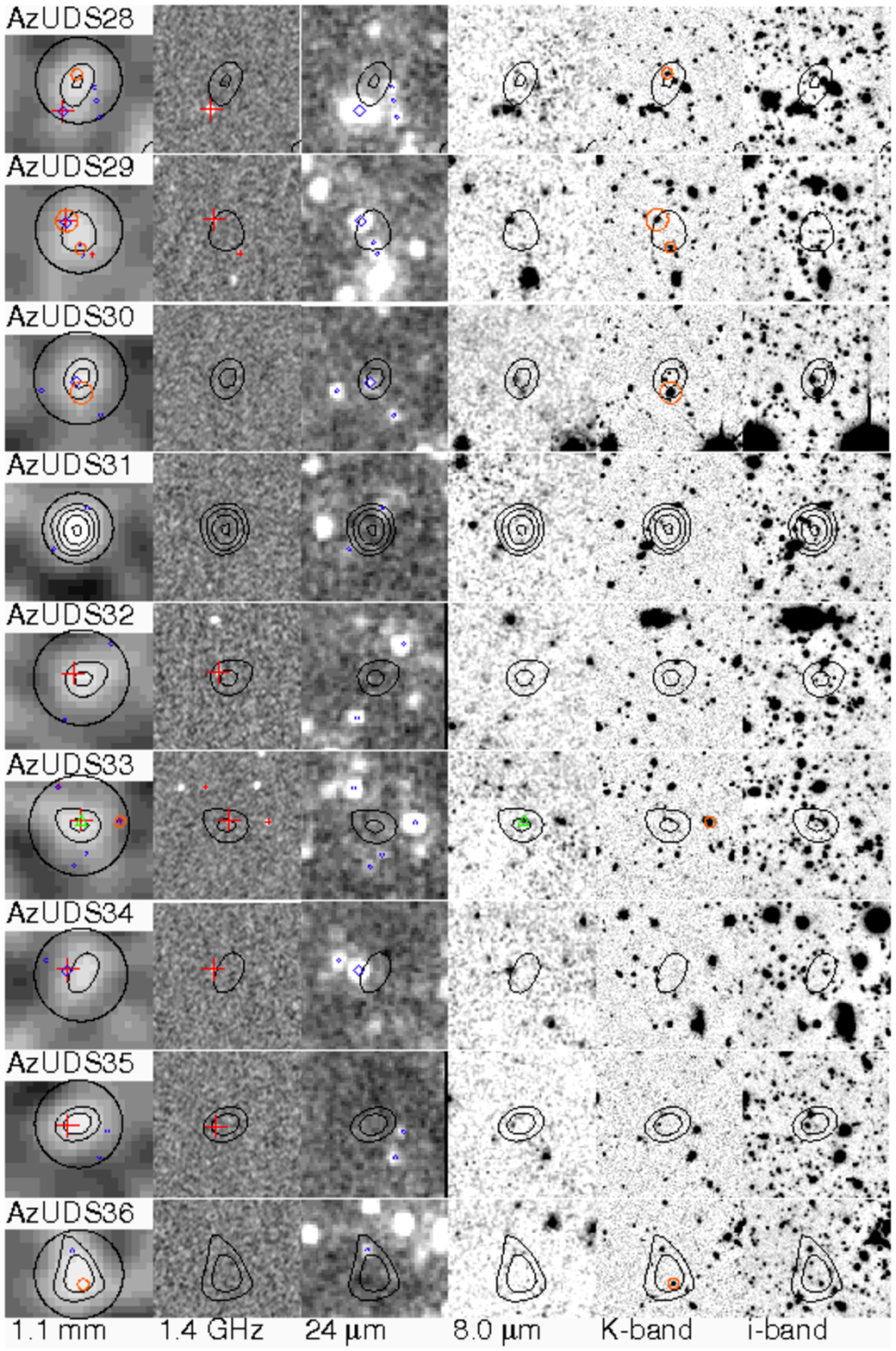}
\end{center}
\caption{(continued).}
\end{figure*}

\addtocounter{figure}{-1}
\begin{figure*}
\begin{center}
\includegraphics[width=0.95\textwidth,clip]{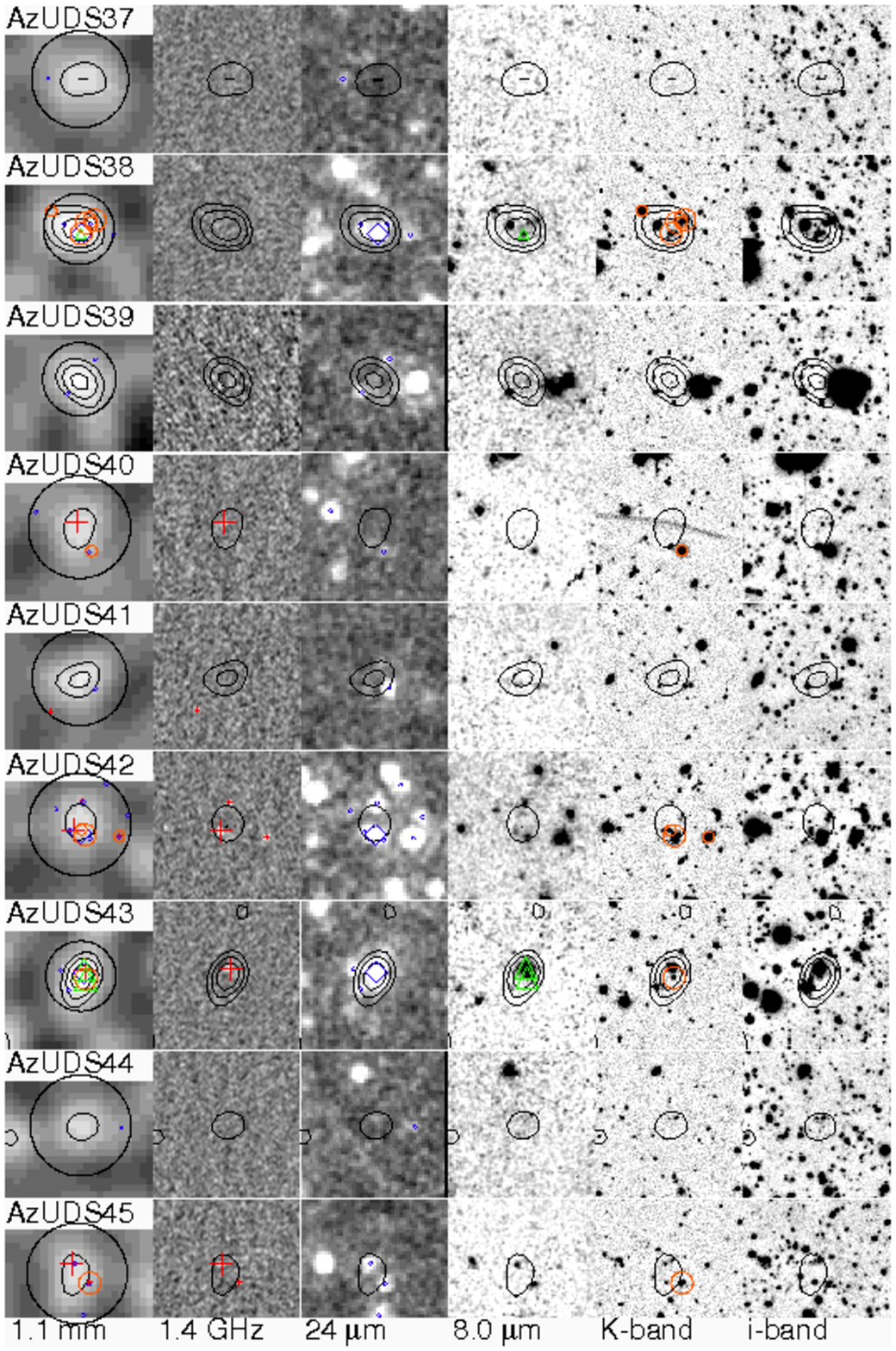}
\end{center}
\caption{(continued).}
\end{figure*}

\addtocounter{figure}{-1}
\begin{figure*}
\begin{center}
\includegraphics[width=0.95\textwidth,clip]{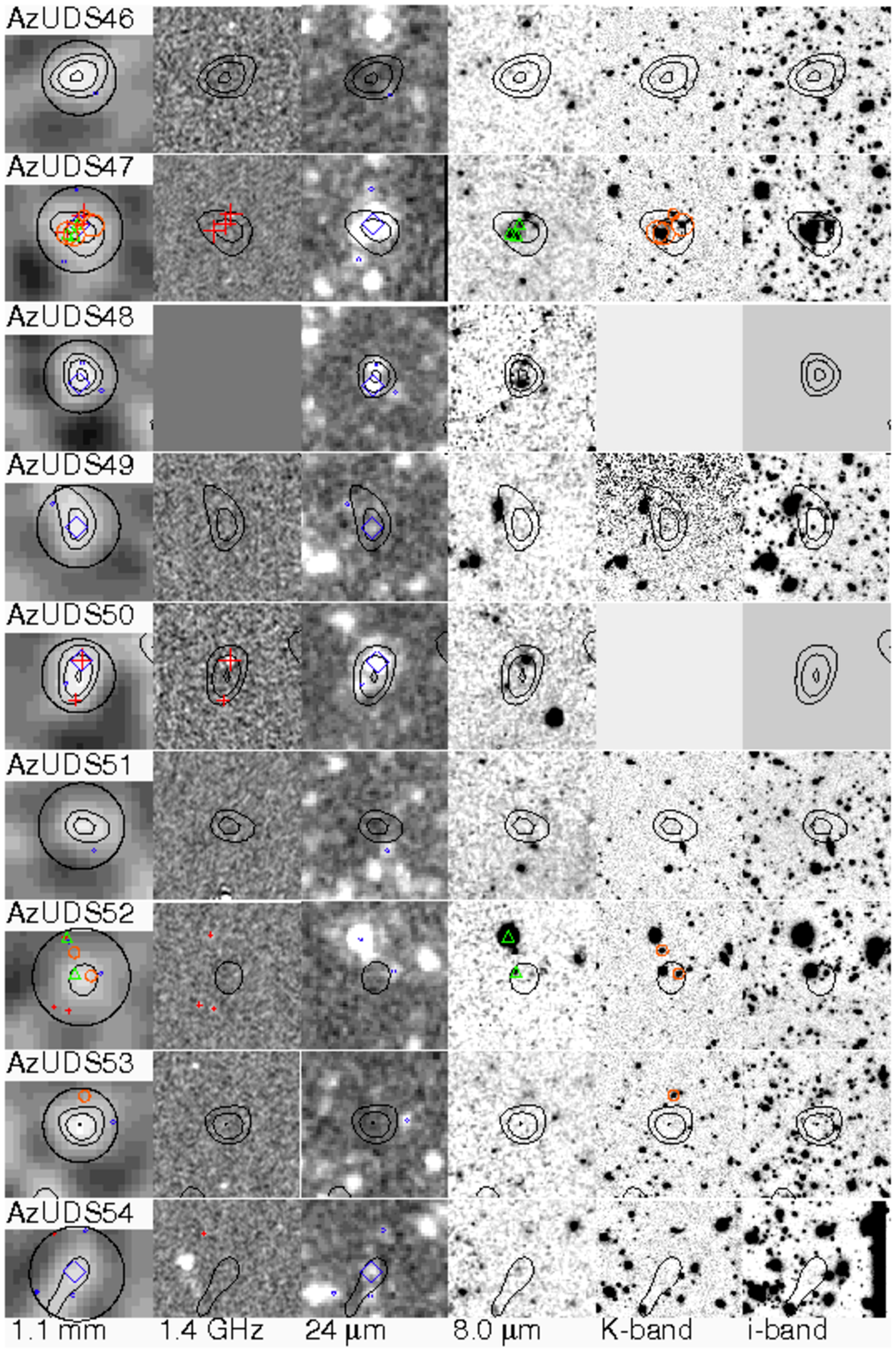}
\end{center}
\caption{(continued).}
\end{figure*}

\addtocounter{figure}{-1}
\begin{figure*}
\begin{center}
\includegraphics[width=0.95\textwidth,clip]{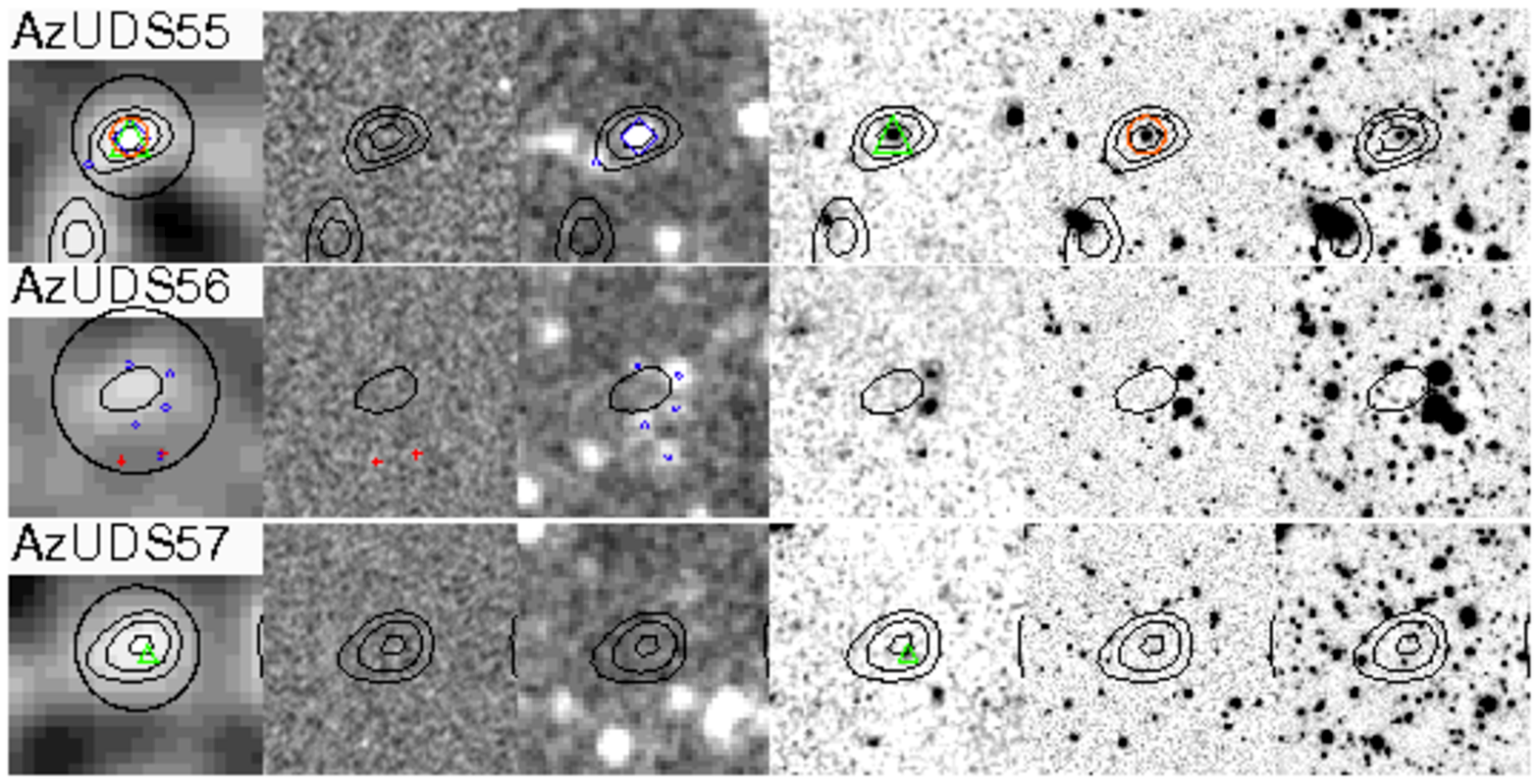}
\end{center}
\caption{(continued).}
\end{figure*}


\voffset=0in

\onecolumn
\begin{landscape}
\begin{scriptsize}
\begin{longtable}{lcccccccccccccccc}
\caption{Radio and $24\,\mu$m identifications in the Lockman Hole field.}
\label{tab:AzLOCKID}\\
\hline\hline
No. & $S_{1.1\/\micron}$ & RA$_{1.4}$ & DEC$_{1.4}$ & $S_{1.4}$ & Dist. & $p$ &  RA$_{24}$ & DEC$_{24}$ & $S_{24}$ & Dist. & $p$ & RA$_{0.61}$ & DEC$_{0.61}$ & $S_{0.61}$ & Dist. & $p$  \\
 & (mJy) & (deg) & (deg) & ($\mu$Jy) & ($''$) & & (deg) & (deg) & ($\mu$Jy) & ($''$) & & (deg) & (deg) & ($\mu$Jy) & ($''$) & \\
\hline
 
1 & $6.63^{+0.85}_{-0.97}$ & 163.008040 & 57.681007 & $258\pm11$ & 2.3 & 0.002
 & 163.008072 & 57.681003 & $1642\pm11$ & 2.3 & 0.002 & 163.008390 & 57.680893
 & $596\pm24$ & 1.9 & 0.001 \\
$\cdots$ & $\cdots$ & $\cdots$ & $\cdots$ & $\cdots$ & $\cdots$ & $\cdots$ & 
163.011618 & 57.679160 & $123\pm14$ & 7.8 & 0.206 & $\cdots$ & $\cdots$ & 
$\cdots$ & $\cdots$ & $\cdots$ \\
2 & $6.43^{+0.90}_{-0.89}$ & 163.022900 & 57.607548 & $66\pm8$ & 6.5 & 0.038 & 
163.023461 & 57.607408 & $235\pm6$ & 5.4 & 0.070 & $\cdots$ & $\cdots$ & 
$\cdots$ & $\cdots$ & $\cdots$ \\
3 & $6.21^{+1.05}_{-0.92}$ & 163.237710 & 57.350926 & $67\pm7$ & 2.8 & 0.011 & 
163.237751 & 57.350931 & $636\pm13$ & 2.8 & 0.008 & 163.237780 & 57.351320 & 
$149\pm16$ & 1.5 & 0.004 \\
$\cdots$ & $\cdots$ & 163.237530 & 57.352273 & $36\pm7$ & 2.6 & 0.019 & $\cdots$
 & $\cdots$ & $\cdots$ & $\cdots$ & $\cdots$ & $\cdots$ & $\cdots$ & $\cdots$ & 
$\cdots$ & $\cdots$ \\
\hline
\end{longtable}

\end{scriptsize}
\begin{longtable}{lccccccccccc}
\caption{All radio and $24\,\mu$m identifications in the UDS field.}
\label{tab:AzSXDFID}\\
\hline\hline
No. & $S_{1.1\/\micron}$ & RA$_{1.4 }$ & DEC$_{1.4}$ & $S_{1.4}$ & Dist. & $p$ & RA$_{24}$ & DEC$_{24}$ & $S_{24}$ & Dist. & $p$  \\
 & (mJy) & (deg) & (deg) & ($\mu$Jy) & ($''$) & & (deg) & (deg) & ($\mu$Jy) & ($''$) & \\
\hline
 
1 & $5.29^{+1.40}_{-1.68}$
 & $\cdots$ & $\cdots$ & $\cdots$ & $\cdots$ & $\cdots$ & 34.411903 & -4.559684
 & $95\pm2$ & 6.7 & 0.211 \\
2 & $3.99^{+1.10}_{-1.34}$
 & $\cdots$ & $\cdots$ & $\cdots$ & $\cdots$ & $\cdots$ & 34.442192 & -4.796477
 & $77\pm7$ & 5.3 & 0.199 \\
3 & $3.75^{+1.22}_{-1.18}$ & 34.479333 & -4.789667 & $113\pm20$ & 1.5 & 0.002 & 
34.479359 & -4.789783 & $776\pm11$ & 1.3 & 0.002 \\
4 & $4.39^{+1.67}_{-1.58}$
 & $\cdots$ & $\cdots$ & $\cdots$ & $\cdots$ & $\cdots$ & 34.630307 & -4.650681
 & $107\pm5$ & 9.5 & 0.325 \\
5 & $3.57^{+1.19}_{-1.21}$ & 34.425542 & -4.941028 & $89\pm17$ & 1.0 & 0.002 & 
$\cdots$ & $\cdots$ & $\cdots$ & $\cdots$ & $\cdots$ \\
\hline
\end{longtable}

\end{landscape}
\twocolumn

\bsp

\label{lastpage}


\begin{thebibliography}{133}
\expandafter\ifx\csname natexlab\endcsname\relax\def\natexlab#1{#1}\fi
\expandafter\ifx\csname url\endcsname\relax
  \def\url#1{{\tt #1}}\fi
\expandafter\ifx\csname urlprefix\endcsname\relax\def\urlprefix{URL }\fi

\bibitem[{{Amblard} et~al.(2010){Amblard}, {Cooray}, {Serra}
  et~al.}]{amblard10}
{Amblard} A., et~al., 2010, \aap, 518, L9

\bibitem[{{Aretxaga} et~al.(2003){Aretxaga}, {Hughes}, {Chapin}
  et~al.}]{aretxaga03}
{Aretxaga} I., {Hughes} D.H., {Chapin} E.L., {Gazta{\~n}aga} E., {Dunlop} J.S.,
  {Ivison} R.J., 2003, \mnras, 342, 759

\bibitem[{{Aretxaga} et~al.(2007){Aretxaga}, {Hughes}, {Coppin}
  et~al.}]{aretxaga07}
{Aretxaga} I., et~al., 2007, \mnras, 379, 1571

\bibitem[{{Aretxaga} et~al.(2011){Aretxaga}, {Wilson}, {Aguilar}
  et~al.}]{aretxaga11}
{Aretxaga} I., et~al., 2011, \mnras, 415, 3831

\bibitem[{{Ashby} et~al.(2006){Ashby}, {Dye}, {Huang} et~al.}]{ashby06}
{Ashby} M.L.N., et~al., 2006, \apj, 644, 778

\bibitem[{{Austermann} et~al.(2009){Austermann}, {Aretxaga}, {Hughes}
  et~al.}]{austermann09}
{Austermann} J.E., et~al., 2009, \mnras, 393, 1573

\bibitem[{{Austermann} et~al.(2010){Austermann}, {Dunlop}, {Perera}
  et~al.}]{austermann10}
{Austermann} J.E., et~al., 2010, \mnras, 401, 160

\bibitem[{{Barger} et~al.(1998){Barger}, {Cowie}, {Sanders} et~al.}]{barger98}
{Barger} A.J., {Cowie} L.L., {Sanders} D.B., {Fulton} E., {Taniguchi} Y.,
  {Sato} Y., {Kawara} K., {Okuda} H., 1998, \nat, 394, 248

\bibitem[{{Barger} et~al.(1999){Barger}, {Cowie}, \& {Sanders}}]{barger99}
{Barger} A.J., {Cowie} L.L., {Sanders} D.B., 1999, \apjl, 518, L5

\bibitem[{{Barger} et~al.(2000){Barger}, {Cowie}, \& {Richards}}]{barger00}
{Barger} A.J., {Cowie} L.L., {Richards} E.A., 2000, \aj, 119, 2092

\bibitem[{{Barger} et~al.(2001){Barger}, {Cowie}, {Steffen} et~al.}]{barger01}
{Barger} A.J., {Cowie} L.L., {Steffen} A.T., {Hornschemeier} A.E., {Brandt}
  W.N., {Garmire} G.P., 2001, \apjl, 560, L23

\bibitem[{{Barger} et~al.(2002){Barger}, {Cowie}, {Brandt} et~al.}]{barger02}
{Barger} A.J., {Cowie} L.L., {Brandt} W.N., {Capak} P., {Garmire} G.P.,
  {Hornschemeier} A.E., {Steffen} A.T., {Wehner} E.H., 2002, \aj, 124, 1839

\bibitem[{{Bertin} \& {Arnouts}(1996)}]{sextractor}
{Bertin} E., {Arnouts} S., 1996, \aaps, 117, 393

\bibitem[{{Bertoldi} et~al.(2007){Bertoldi}, {Carilli}, {Aravena}
  et~al.}]{bertoldi07}
{Bertoldi} F., et~al., 2007, \apjs, 172, 132

\bibitem[{{B{\'e}thermin} et~al.(2012){B{\'e}thermin}, {Le Floc'h}, {Ilbert}
  et~al.}]{bethermin12}
{B{\'e}thermin} M., et~al., 2012, \aap, 542, A58

\bibitem[{{Biggs} et~al.(2011){Biggs}, {Ivison}, {Ibar} et~al.}]{biggs11}
{Biggs} A.D., et~al., 2011, \mnras, 413, 2314

\bibitem[{{Bolzonella} et~al.(2000){Bolzonella}, {Miralles}, \&
  {Pell{\'o}}}]{bolzonella00}
{Bolzonella} M., {Miralles} J., {Pell{\'o}} R., 2000, \aap, 363, 476

\bibitem[{{Borys} et~al.(2005){Borys}, {Smail}, {Chapman} et~al.}]{borys05}
{Borys} C., {Smail} I., {Chapman} S.C., {Blain} A.W., {Alexander} D.M.,
  {Ivison} R.J., 2005, \apj, 635, 853

\bibitem[{{Bruzual} \& {Charlot}(2003)}]{bruzualcharlot03}
{Bruzual} G., {Charlot} S., 2003, \mnras, 344, 1000

\bibitem[{{Bussmann} et~al.(2012){Bussmann}, {Dey}, {Armus}
  et~al.}]{bussmann12}
{Bussmann} R.S., et~al., 2012, \apj, 744, 150

\bibitem[{{Calzetti} et~al.(2000){Calzetti}, {Armus}, {Bohlin}
  et~al.}]{calzetti00}
{Calzetti} D., {Armus} L., {Bohlin} R.C., {Kinney} A.L., {Koornneef} J.,
  {Storchi-Bergmann} T., 2000, \apj, 533, 682

\bibitem[{{Capak} et~al.(2008){Capak}, {Carilli}, {Lee} et~al.}]{capak08}
{Capak} P., et~al., 2008, \apjl, 681, L53

\bibitem[{{Capak} et~al.(2011){Capak}, {Riechers}, {Scoville} et~al.}]{capak11}
{Capak} P.L., et~al., 2011, \nat, 470, 233

\bibitem[{{Caputi} et~al.(2006){Caputi}, {Dole}, {Lagache} et~al.}]{caputi06b}
{Caputi} K.I., {Dole} H., {Lagache} G., {McLure} R.J., {Dunlop} J.S., {Puget}
  J., {Le Floc'h} E., {P{\'e}rez-Gonz{\'a}lez} P.G., 2006, \aap, 454, 143

\bibitem[{{Caputi} et~al.(2011){Caputi}, {Cirasuolo}, {Dunlop}
  et~al.}]{caputi11}
{Caputi} K.I., {Cirasuolo} M., {Dunlop} J.S., {McLure} R.J., {Farrah} D.,
  {Almaini} O., 2011, \mnras, 413, 162

\bibitem[{{Chabrier}(2003)}]{chabrier03}
{Chabrier} G., 2003, \apjl, 586, L133

\bibitem[{{Chapin} et~al.(2009){Chapin}, {Pope}, {Scott} et~al.}]{chapin09}
{Chapin} E.L., et~al., 2009, \mnras, 398, 1793

\bibitem[{{Chapin} et~al.(2011){Chapin}, {Chapman}, {Coppin} et~al.}]{chapin11}
{Chapin} E.L., et~al., 2011, \mnras, 411, 505

\bibitem[{{Chapman} et~al.(2001){Chapman}, {Richards}, {Lewis}, {Wilson}, \&
  {Barger}}]{chapman01}
{Chapman} S.C., {Richards} E.A., {Lewis} G.F., {Wilson} G., {Barger} A.J.,
  2001, \apjl, 548, L147

\bibitem[{{Chapman} et~al.(2003){Chapman}, {Barger}, {Cowie}
  et~al.}]{chapman03c}
{Chapman} S.C., et~al., 2003, \apj, 585, 57

\bibitem[{{Chapman} et~al.(2005){Chapman}, {Blain}, {Smail}, \&
  {Ivison}}]{chapman05}
{Chapman} S.C., {Blain} A.W., {Smail} I., {Ivison} R.J., 2005, \apj, 622, 772

\bibitem[{{Cirasuolo} et~al.(2007){Cirasuolo}, {McLure}, {Dunlop}
  et~al.}]{cirasuolo07}
{Cirasuolo} M., et~al., 2007, \mnras, 380, 585

\bibitem[{{Cirasuolo} et~al.(2010){Cirasuolo}, {McLure}, {Dunlop}
  et~al.}]{cirasuolo10}
{Cirasuolo} M., {McLure} R.J., {Dunlop} J.S., {Almaini} O., {Foucaud} S.,
  {Simpson} C., 2010, \mnras, 401, 1166

\bibitem[{{Clements} et~al.(2008){Clements}, {Vaccari}, {Babbedge}
  et~al.}]{clements08}
{Clements} D.L., et~al., 2008, \mnras, 387, 247

\bibitem[{{Combes} et~al.(2012){Combes}, {Rex}, {Rawle} et~al.}]{combes12}
{Combes} F., et~al., 2012, \aap, 538, L4

\bibitem[{{Condon}(1992)}]{condon}
{Condon} J.J., 1992, \araa, 30, 575

\bibitem[{{Coppin} et~al.(2005){Coppin}, {Halpern}, {Scott}, {Borys}, \&
  {Chapman}}]{coppin05}
{Coppin} K., {Halpern} M., {Scott} D., {Borys} C., {Chapman} S., 2005, \mnras,
  357, 1022

\bibitem[{{Coppin} et~al.(2006){Coppin}, {Chapin}, {Mortier} et~al.}]{coppin06}
{Coppin} K., et~al., 2006, \mnras, 372, 1621

\bibitem[{{Coppin} et~al.(2008){Coppin}, {Halpern}, {Scott} et~al.}]{coppin08}
{Coppin} K., et~al., 2008, \mnras, 384, 1597

\bibitem[{{Coppin} et~al.(2010){Coppin}, {Pope}, {Men{\'e}ndez-Delmestre}
  et~al.}]{coppin10}
{Coppin} K., et~al., 2010, \apj, 713, 503

\bibitem[{{Coppin} et~al.(2009){Coppin}, {Smail}, {Alexander}
  et~al.}]{coppin09}
{Coppin} K.E.K., et~al., 2009, \mnras, 395, 1905

\bibitem[{{Cox} et~al.(2011){Cox}, {Krips}, {Neri} et~al.}]{cox11}
{Cox} P., et~al., 2011, \apj, 740, 63

\bibitem[{{Croton} et~al.(2006){Croton}, {Springel}, {White} et~al.}]{croton06}
{Croton} D.J., et~al., 2006, \mnras, 365, 11

\bibitem[{{Daddi} et~al.(2009{\natexlab{a}}){Daddi}, {Dannerbauer}, {Krips}
  et~al.}]{daddi09b}
{Daddi} E., {Dannerbauer} H., {Krips} M., {Walter} F., {Dickinson} M., {Elbaz}
  D., {Morrison} G.E., 2009{\natexlab{a}}, \apjl, 695, L176

\bibitem[{{Daddi} et~al.(2009{\natexlab{b}}){Daddi}, {Dannerbauer}, {Stern}
  et~al.}]{daddi09}
{Daddi} E., et~al., 2009{\natexlab{b}}, \apj, 694, 1517

\bibitem[{{Dannerbauer} et~al.(2004){Dannerbauer}, {Lehnert}, {Lutz}
  et~al.}]{dannerbauer04}
{Dannerbauer} H., {Lehnert} M.D., {Lutz} D., {Tacconi} L., {Bertoldi} F.,
  {Carilli} C., {Genzel} R., {Menten} K.M., 2004, \apj, 606, 664

\bibitem[{{Dannerbauer} et~al.(2010){Dannerbauer}, {Daddi}, {Morrison}
  et~al.}]{dannerbauer10}
{Dannerbauer} H., et~al., 2010, \apjl, 720, L144

\bibitem[{{Downes} et~al.(1986){Downes}, {Peacock}, {Savage}, \&
  {Carrie}}]{downes86}
{Downes} A.J.B., {Peacock} J.A., {Savage} A., {Carrie} D.R., 1986, \mnras, 218,
  31

\bibitem[{{Downes} et~al.(2012){Downes}, {Welch}, {Scott} et~al.}]{downes12}
{Downes} T.P., {Welch} D., {Scott} K.S., {Austermann} J., {Wilson} G.W., {Yun}
  M.S., 2012, \mnras, 423, 529

\bibitem[{{Dunlop} et~al.(1989){Dunlop}, {Peacock}, {Savage} et~al.}]{dunlop89}
{Dunlop} J.S., {Peacock} J.A., {Savage} A., {Lilly} S.J., {Heasley} J.N.,
  {Simon} A.J.B., 1989, \mnras, 238, 1171

\bibitem[{{Dunlop} et~al.(2004){Dunlop}, {McLure}, {Yamada} et~al.}]{dunlop04}
{Dunlop} J.S., et~al., 2004, \mnras, 350, 769

\bibitem[{{Dunlop} et~al.(2010){Dunlop}, {Ade}, {Bock} et~al.}]{dunlop10}
{Dunlop} J.S., et~al., 2010, \mnras, 408, 2022

\bibitem[{{Dye} et~al.(2006){Dye}, {Warren}, {Hambly} et~al.}]{dye06}
{Dye} S., et~al., 2006, \mnras, 372, 1227

\bibitem[{{Dye} et~al.(2008){Dye}, {Eales}, {Aretxaga} et~al.}]{dye08}
{Dye} S., et~al., 2008, \mnras, 386, 1107

\bibitem[{{Eales} et~al.(2009){Eales}, {Chapin}, {Devlin} et~al.}]{eales09}
{Eales} S., et~al., 2009, \apj, 707, 1779

\bibitem[{{Egami} et~al.(2004){Egami}, {Dole}, {Huang} et~al.}]{egami04}
{Egami} E., et~al., 2004, \apjs, 154, 130

\bibitem[{{Engel} et~al.(2010){Engel}, {Tacconi}, {Davies} et~al.}]{engel10}
{Engel} H., et~al., 2010, \apj, 724, 233

\bibitem[{{Ezawa} et~al.(2004){Ezawa}, {Kawabe}, {Kohno}, \& {Yamamoto}}]{aste}
{Ezawa} H., {Kawabe} R., {Kohno} K., {Yamamoto} S., 2004, In: {J.~M.~Oschmann
  Jr.} (ed.) Society of Photo-Optical Instrumentation Engineers (SPIE)
  Conference Series, vol. 5489 of Society of Photo-Optical Instrumentation
  Engineers (SPIE) Conference Series, 763--772

\bibitem[{{Ezawa} et~al.(2008){Ezawa}, {Kohno}, {Kawabe} et~al.}]{aste2}
{Ezawa} H., et~al., 2008, In: Society of Photo-Optical Instrumentation
  Engineers (SPIE) Conference Series, vol. 7012 of Society of Photo-Optical
  Instrumentation Engineers (SPIE) Conference Series

\bibitem[{{Furusawa} et~al.(2008){Furusawa}, {Kosugi}, {Akiyama}
  et~al.}]{furusawa08}
{Furusawa} H., et~al., 2008, \apjs, 176, 1

\bibitem[{{Greve} et~al.(2004){Greve}, {Ivison}, {Bertoldi} et~al.}]{greve04}
{Greve} T.R., {Ivison} R.J., {Bertoldi} F., {Stevens} J.A., {Dunlop} J.S.,
  {Lutz} D., {Carilli} C.L., 2004, \mnras, 354, 779

\bibitem[{{Greve} et~al.(2005){Greve}, {Bertoldi}, {Smail} et~al.}]{greve05}
{Greve} T.R., et~al., 2005, \mnras, 359, 1165

\bibitem[{{Greve} et~al.(2008){Greve}, {Pope}, {Scott} et~al.}]{greve08}
{Greve} T.R., {Pope} A., {Scott} D., {Ivison} R.J., {Borys} C., {Conselice}
  C.J., {Bertoldi} F., 2008, \mnras, 389, 1489

\bibitem[{{Hainline} et~al.(2009){Hainline}, {Blain}, {Smail}
  et~al.}]{hainline09}
{Hainline} L.J., {Blain} A.W., {Smail} I., {Frayer} D.T., {Chapman} S.C.,
  {Ivison} R.J., {Alexander} D.M., 2009, \apj, 699, 1610

\bibitem[{{Hainline} et~al.(2011){Hainline}, {Blain}, {Smail}
  et~al.}]{hainline11}
{Hainline} L.J., {Blain} A.W., {Smail} I., {Alexander} D.M., {Armus} L.,
  {Chapman} S.C., {Ivison} R.J., 2011, \apj, 740, 96

\bibitem[{{Hatsukade} et~al.(2010){Hatsukade}, {Iono}, {Akiyama}
  et~al.}]{hatsukade10}
{Hatsukade} B., et~al., 2010, \apj, 711, 974

\bibitem[{{Hatsukade} et~al.(2011){Hatsukade}, {Kohno}, {Aretxaga}
  et~al.}]{hatsukade11}
{Hatsukade} B., et~al., 2011, \mnras, 411, 102

\bibitem[{{Hayward} et~al.(2011){Hayward}, {Kere{\v s}}, {Jonsson}
  et~al.}]{hayward11b}
{Hayward} C.C., {Kere{\v s}} D., {Jonsson} P., {Narayanan} D., {Cox} T.J.,
  {Hernquist} L., 2011, \apj, 743, 159

\bibitem[{{Holland} et~al.(1999){Holland}, {Robson}, {Gear}
  et~al.}]{hollandscuba}
{Holland} W.S., et~al., 1999, \mnras, 303, 659

\bibitem[{{Hughes} et~al.(1998){Hughes}, {Serjeant}, {Dunlop}
  et~al.}]{hughes98}
{Hughes} D.H., et~al., 1998, \nat, 394, 241

\bibitem[{{Ibar} et~al.(2009){Ibar}, {Ivison}, {Biggs} et~al.}]{ibar09}
{Ibar} E., {Ivison} R.J., {Biggs} A.D., {Lal} D.V., {Best} P.N., {Green} D.A.,
  2009, \mnras, 397, 281

\bibitem[{{Ibar} et~al.(2010){Ibar}, {Ivison}, {Best} et~al.}]{ibar10}
{Ibar} E., {Ivison} R.J., {Best} P.N., {Coppin} K., {Pope} A., {Smail} I.,
  {Dunlop} J.S., 2010, \mnras, 401, L53

\bibitem[{{Ivison} et~al.(2002){Ivison}, {Greve}, {Smail} et~al.}]{ivison02}
{Ivison} R.J., et~al., 2002, \mnras, 337, 1

\bibitem[{{Ivison} et~al.(2005){Ivison}, {Smail}, {Dunlop} et~al.}]{ivison05}
{Ivison} R.J., et~al., 2005, \mnras, 364, 1025

\bibitem[{{Ivison} et~al.(2007){Ivison}, {Greve}, {Dunlop} et~al.}]{ivison07}
{Ivison} R.J., et~al., 2007, \mnras, 380, 199

\bibitem[{{Joye} \& {Mandel}(2003)}]{ds9}
{Joye} W.A., {Mandel} E., 2003, In: {H.~E.~Payne, R.~I.~Jedrzejewski, \&
  R.~N.~Hook} (ed.) Astronomical Data Analysis Software and Systems XII, vol.
  295 of Astronomical Society of the Pacific Conference Series, 489

\bibitem[{{Knudsen} et~al.(2008){Knudsen}, {Kneib}, \& {Egami}}]{knudsen08b}
{Knudsen} K.K., {Kneib} J.P., {Egami} E., 2008, In: {Chary} R.R., {Teplitz}
  H.I., {Sheth} K. (eds.) Infrared Diagnostics of Galaxy Evolution, vol. 381 of
  Astronomical Society of the Pacific Conference Series, 372

\bibitem[{{Knudsen} et~al.(2010){Knudsen}, {Kneib}, {Richard}, {Petitpas}, \&
  {Egami}}]{knudsen09}
{Knudsen} K.K., {Kneib} J., {Richard} J., {Petitpas} G., {Egami} E., 2010,
  \apj, 709, 210

\bibitem[{{Lawrence} et~al.(2007){Lawrence}, {Warren}, {Almaini}
  et~al.}]{lawrence07}
{Lawrence} A., et~al., 2007, \mnras, 379, 1599

\bibitem[{{Lindner} et~al.(2011){Lindner}, {Baker}, {Omont} et~al.}]{lindner11}
{Lindner} R.R., et~al., 2011, \apj, 737, 83

\bibitem[{{Madau}(1995)}]{madau95}
{Madau} P., 1995, \apj, 441, 18

\bibitem[{{Marsden} et~al.(2011){Marsden}, {Chapin}, {Halpern}
  et~al.}]{marsden11}
{Marsden} G., et~al., 2011, \mnras, 417, 1192

\bibitem[{{McLure} et~al.(2009){McLure}, {Cirasuolo}, {Dunlop}, {Foucaud}, \&
  {Almaini}}]{mclure09}
{McLure} R.J., {Cirasuolo} M., {Dunlop} J.S., {Foucaud} S., {Almaini} O., 2009,
  \mnras, 395, 2196

\bibitem[{{Micha{\l}owski} et~al.(2010{\natexlab{a}}){Micha{\l}owski},
  {Hjorth}, \& {Watson}}]{michalowski10smg}
{Micha{\l}owski} M., {Hjorth} J., {Watson} D., 2010{\natexlab{a}}, \aap, 514,
  A67

\bibitem[{{Micha{\l}owski} et~al.(2010{\natexlab{b}}){Micha{\l}owski},
  {Watson}, \& {Hjorth}}]{michalowski10smg4}
{Micha{\l}owski} M.J., {Watson} D., {Hjorth} J., 2010{\natexlab{b}}, \apj, 712,
  942

\bibitem[{{Micha{\l}owski} et~al.(2012){Micha{\l}owski}, {Dunlop}, {Cirasuolo}
  et~al.}]{michalowski12mass}
{Micha{\l}owski} M.J., {Dunlop} J.S., {Cirasuolo} M., {Hjorth} J., {Hayward}
  C.C., {Watson} D., 2012, \aap, 541, A85

\bibitem[{{Miyazaki} et~al.(2002){Miyazaki}, {Komiyama}, {Sekiguchi}
  et~al.}]{suprimecam}
{Miyazaki} S., et~al., 2002, \pasj, 54, 833

\bibitem[{{Mortier} et~al.(2005){Mortier}, {Serjeant}, {Dunlop}
  et~al.}]{mortier05}
{Mortier} A.M.J., et~al., 2005, \mnras, 363, 563

\bibitem[{{Negrello} et~al.(2010){Negrello}, {Hopwood}, {De Zotti}
  et~al.}]{negrello10}
{Negrello} M., et~al., 2010, Science, 330, 800

\bibitem[{{Perera} et~al.(2008){Perera}, {Chapin}, {Austermann}
  et~al.}]{perera08}
{Perera} T.A., et~al., 2008, \mnras, 391, 1227

\bibitem[{{Pope} et~al.(2006){Pope}, {Scott}, {Dickinson} et~al.}]{pope06}
{Pope} A., et~al., 2006, \mnras, 370, 1185

\bibitem[{{Riechers} et~al.(2010){Riechers}, {Capak}, {Carilli}
  et~al.}]{riechers10}
{Riechers} D.A., et~al., 2010, \apjl, 720, L131

\bibitem[{{Santini} et~al.(2010){Santini}, {Maiolino}, {Magnelli}
  et~al.}]{santini10}
{Santini} P., et~al., 2010, \aap, 518, L154

\bibitem[{{Schael}(2009)}]{schael09phd}
{Schael} A.M., 2009, The Star-Formation History of Massive Galaxies, Ph.D.
  thesis, University of Edinburgh

\bibitem[{{Schinnerer} et~al.(2008){Schinnerer}, {Carilli}, {Capak}
  et~al.}]{schinnerer08}
{Schinnerer} E., et~al., 2008, \apjl, 689, L5

\bibitem[{{Scott} et~al.(2008){Scott}, {Austermann}, {Perera} et~al.}]{scott08}
{Scott} K.S., et~al., 2008, \mnras, 385, 2225

\bibitem[{{Scott} et~al.(2010){Scott}, {Yun}, {Wilson} et~al.}]{scott10}
{Scott} K.S., et~al., 2010, \mnras, 405, 2260

\bibitem[{{Scott} et~al.(2012){Scott}, {Wilson}, {Aretxaga} et~al.}]{scott12}
{Scott} K.S., et~al., 2012, \mnras, 423, 575

\bibitem[{{Scott} et~al.(2002){Scott}, {Fox}, {Dunlop} et~al.}]{scott}
{Scott} S.E., et~al., 2002, \mnras, 331, 817

\bibitem[{{Serjeant} et~al.(2008){Serjeant}, {Dye}, {Mortier}
  et~al.}]{serjeant08}
{Serjeant} S., et~al., 2008, \mnras, 386, 1907

\bibitem[{{Shang} et~al.(2011){Shang}, {Brotherton}, {Wills} et~al.}]{shang11}
{Shang} Z., et~al., 2011, \apjs, 196, 2

\bibitem[{{Shimizu} et~al.(2012){Shimizu}, {Yoshida}, \& {Okamoto}}]{shimizu12}
{Shimizu} I., {Yoshida} N., {Okamoto} T., 2012, \mnras, submitted, {\tt
  arXiv:1207.3856}

\bibitem[{{Silva} et~al.(1998){Silva}, {Granato}, {Bressan}, \&
  {Danese}}]{silva98}
{Silva} L., {Granato} G.L., {Bressan} A., {Danese} L., 1998, \apj, 509, 103

\bibitem[{{Siringo} et~al.(2009){Siringo}, {Kreysa}, {Kov{\'a}cs}
  et~al.}]{laboca}
{Siringo} G., et~al., 2009, \aap, 497, 945

\bibitem[{{Smail} et~al.(1997){Smail}, {Ivison}, \& {Blain}}]{smail97}
{Smail} I., {Ivison} R.J., {Blain} A.W., 1997, \apjl, 490, L5

\bibitem[{{Smail} et~al.(1999){Smail}, {Ivison}, {Kneib} et~al.}]{smail99}
{Smail} I., {Ivison} R.J., {Kneib} J.P., {Cowie} L.L., {Blain} A.W., {Barger}
  A.J., {Owen} F.N., {Morrison} G., 1999, \mnras, 308, 1061

\bibitem[{{Smail} et~al.(2002){Smail}, {Ivison}, {Blain}, \& {Kneib}}]{smail02}
{Smail} I., {Ivison} R.J., {Blain} A.W., {Kneib} J.P., 2002, \mnras, 331, 495

\bibitem[{{Smail} et~al.(2004){Smail}, {Chapman}, {Blain}, \&
  {Ivison}}]{smail04}
{Smail} I., {Chapman} S.C., {Blain} A.W., {Ivison} R.J., 2004, \apj, 616, 71

\bibitem[{{Smol{\v c}i{\'c}} et~al.(2011){Smol{\v c}i{\'c}}, {Capak}, {Ilbert}
  et~al.}]{smolcic11}
{Smol{\v c}i{\'c}} V., et~al., 2011, \apjl, 731, L27

\bibitem[{{Springel} et~al.(2005){Springel}, {White}, {Jenkins}
  et~al.}]{springel05}
{Springel} V., et~al., 2005, \nat, 435, 629

\bibitem[{{Swinbank} et~al.(2004){Swinbank}, {Smail}, {Chapman}
  et~al.}]{swinbank04}
{Swinbank} A.M., {Smail} I., {Chapman} S.C., {Blain} A.W., {Ivison} R.J.,
  {Keel} W.C., 2004, \apj, 617, 64

\bibitem[{{Swinbank} et~al.(2006){Swinbank}, {Chapman}, {Smail}
  et~al.}]{swinbank06}
{Swinbank} A.M., {Chapman} S.C., {Smail} I., {Lindner} C., {Borys} C., {Blain}
  A.W., {Ivison} R.J., {Lewis} G.F., 2006, \mnras, 371, 465

\bibitem[{{Tacconi} et~al.(2006){Tacconi}, {Neri}, {Chapman}
  et~al.}]{tacconi06}
{Tacconi} L.J., et~al., 2006, \apj, 640, 228

\bibitem[{{Tacconi} et~al.(2008){Tacconi}, {Genzel}, {Smail}
  et~al.}]{tacconi08}
{Tacconi} L.J., et~al., 2008, \apj, 680, 246

\bibitem[{{Takagi} et~al.(2004){Takagi}, {Hanami}, \& {Arimoto}}]{takagi04}
{Takagi} T., {Hanami} H., {Arimoto} N., 2004, \mnras, 355, 424

\bibitem[{{Takagi} et~al.(2007){Takagi}, {Mortier}, {Shimasaku}
  et~al.}]{takagi07}
{Takagi} T., et~al., 2007, \mnras, 381, 1154

\bibitem[{{Targett} et~al.(2011){Targett}, {Dunlop}, {McLure}
  et~al.}]{targett11}
{Targett} T.A., {Dunlop} J.S., {McLure} R.J., {Best} P.N., {Cirasuolo} M.,
  {Almaini} O., 2011, \mnras, 412, 295

\bibitem[{{Taylor}(2005)}]{topcat}
{Taylor} M.B., 2005, In: {P.~Shopbell, M.~Britton, \& R.~Ebert} (ed.)
  Astronomical Data Analysis Software and Systems XIV, vol. 347 of Astronomical
  Society of the Pacific Conference Series, 29

\bibitem[{{van Kampen} et~al.(2005){van Kampen}, {Percival}, {Crawford}
  et~al.}]{vankampen05}
{van Kampen} E., et~al., 2005, \mnras, 359, 469

\bibitem[{{Vega} et~al.(2008){Vega}, {Clemens}, {Bressan} et~al.}]{vega08}
{Vega} O., {Clemens} M.S., {Bressan} A., {Granato} G.L., {Silva} L., {Panuzzo}
  P., 2008, \aap, 484, 631

\bibitem[{{Vieira} et~al.(2010){Vieira}, {Crawford}, {Switzer}
  et~al.}]{viera10}
{Vieira} J.D., et~al., 2010, \apj, 719, 763

\bibitem[{{Wagg} et~al.(2009){Wagg}, {Owen}, {Bertoldi} et~al.}]{wagg09}
{Wagg} J., {Owen} F., {Bertoldi} F., {Sawitzki} M., {Carilli} C.L., {Menten}
  K.M., {Voss} H., 2009, \apj, 699, 1843

\bibitem[{{Wall} et~al.(2008){Wall}, {Pope}, \& {Scott}}]{wall08}
{Wall} J.V., {Pope} A., {Scott} D., 2008, \mnras, 383, 435

\bibitem[{{Walter} et~al.(2012){Walter}, {Decarli}, {Carilli}
  et~al.}]{walter12b}
{Walter} F., et~al., 2012, ArXiv e-prints

\bibitem[{{Wang} et~al.(2004){Wang}, {Cowie}, \& {Barger}}]{wang04}
{Wang} W.H., {Cowie} L.L., {Barger} A.J., 2004, \apj, 613, 655

\bibitem[{{Wang} et~al.(2011){Wang}, {Cowie}, {Barger}, \& {Williams}}]{wang11}
{Wang} W.H., {Cowie} L.L., {Barger} A.J., {Williams} J.P., 2011, \apjl, 726,
  L18

\bibitem[{{Wardlow} et~al.(2011){Wardlow}, {Smail}, {Coppin}
  et~al.}]{wardlow11}
{Wardlow} J.L., et~al., 2011, \mnras, 415, 1479

\bibitem[{{Warren} et~al.(2007){Warren}, {Hambly}, {Dye} et~al.}]{warren07}
{Warren} S.J., et~al., 2007, \mnras, 375, 213

\bibitem[{{Webb} et~al.(2003){Webb}, {Lilly}, {Clements} et~al.}]{webb03}
{Webb} T.M.A., {Lilly} S.J., {Clements} D.L., {Eales} S., {Yun} M., {Brodwin}
  M., {Dunne} L., {Gear} W.K., 2003, \apj, 597, 680

\bibitem[{{Wei{\ss}} et~al.(2009){Wei{\ss}}, {Kov{\'a}cs}, {Coppin}
  et~al.}]{weiss09b}
{Wei{\ss}} A., et~al., 2009, \apj, 707, 1201

\bibitem[{{Wilson} et~al.(2008){Wilson}, {Austermann}, {Perera} et~al.}]{aztec}
{Wilson} G.W., et~al., 2008, \mnras, 386, 807

\bibitem[{{Younger} et~al.(2008){Younger}, {Dunlop}, {Peck} et~al.}]{younger08}
{Younger} J.D., et~al., 2008, \mnras, 387, 707

\bibitem[{{Yun} et~al.(2012){Yun}, {Scott}, {Guo} et~al.}]{yun12}
{Yun} M.S., et~al., 2012, \mnras, 420, 957

\end{thebibliography}
\end{document}